\documentclass[12pt]{article}
\pdfoutput=1
\usepackage[DIV13]{typearea}
\usepackage{indentfirst}
\RequirePackage{amsmath,array}
\RequirePackage{amssymb,dsfont}
\RequirePackage{mathrsfs}
\RequirePackage{epsfig}
\RequirePackage{graphicx,caption}
\usepackage{subfigure}
\usepackage{multicol, multirow, bigdelim}
\usepackage[usenames,dvipsnames]{color}
\usepackage{cite}
\usepackage{slashed}
\RequirePackage[colorlinks=true
,urlcolor=blue
,anchorcolor=blue
,citecolor=blue
,filecolor=blue
,linkcolor=blue
,menucolor=blue
,pagecolor=blue
,linktocpage=true
,pdfproducer=medialab
]{hyperref}
\usepackage{cancel}

\usepackage[DIV13]{typearea}
\usepackage{amsmath}
\usepackage{amssymb}
\usepackage{graphicx}
\usepackage{cite}
\usepackage{bbold}
\usepackage[usenames]{color}
\def\cE{{\cal E}}

\def\cN{{\cal N}}

\newcommand{\mcE}{\mathcal{E}}
\newcommand{\mcN}{\mathcal{N}}

\newcommand{\nc}{\newcommand}

\newcommand{\be}{\beta}

\nc{\eps}{\varepsilon}
\nc{\vp}{\varphi}
\nc{\tvp}{\widetilde{\varphi}}
\nc{\vpj }{\mbox{${\vp^\dag i\,\raisebox{2mm}{\boldmath ${}^\leftrightarrow$}\hspace{-4mm} D_\mu\,\vp}$}}
\nc{\vpjt}{\mbox{${\vp^\dag i\,\raisebox{2mm}{\boldmath ${}^\leftrightarrow$}\hspace{-4mm} D_\mu^{\,I}\,\vp}$}}
\def\be{\begin{equation}}
\def\ee{\end{equation}}
\def\beq{\begin{equation}}
\def\eeq{\end{equation}}
\def\bc{\begin{center}}
\def\ec{\end{center}}
\def\bea{\begin{eqnarray}}
\def\eea{\end{eqnarray}}
\def\dd{\displaystyle}
\def\nn{\nonumber}

\newcommand{\hc}{\text{h.c.}}

\newcommand{\lag}[1]{{\mathcal{L}_{\textrm{#1}}}}

\newcommand{\mcL}{\mathcal{L}}

\newcommand{\elem}{\textrm{el}}
\newcommand{\comp}{\textrm{comp}}
\newcommand{\yl}{{Y^{\ast}_L}}
\newcommand{\yr}{{Y^\ast_{R}}}

\newcommand{\email}[1]{\href{mailto:#1}{\tt #1}}
\newcommand{\mt}{\tilde{m}}
\newcommand{\Xt}{\tilde{X}}
\newcommand{ \mysmall}[1]{\scriptscriptstyle #1} % a smaller #

\allowdisplaybreaks
\begin{document} 
\begin{titlepage}
\vspace*{-1cm}
\phantom{hep-ph/***} 
\flushright
\hfil{DFPD-2015/TH/20}\\
%\hfil{TUM-HEP-YYY/12}\\

\vskip 1.5cm
\begin{center}
\mathversion{bold}
{\LARGE\bf Lepton Flavour Violation}\\[3mm]
{\LARGE\bf in Composite Higgs Models}\\[3mm]
\mathversion{normal}
\vskip .3cm
\end{center}
\vskip 0.5  cm
\begin{center}
{\large Ferruccio Feruglio}~$^{a)}$,
{\large Paride Paradisi}~$^{a)}$\\[2mm]
{\large and Andrea Pattori}~$^{b)}$
\\
\vskip .7cm
{\footnotesize
$^{a)}$~Dipartimento di Fisica e Astronomia `G.~Galilei', Universit\`a di Padova
\\
INFN, Sezione di Padova, Via Marzolo~8, I-35131 Padua, Italy
\vskip .1cm
$^{b)}$~Physik-Institut, Universit\"at Z\"urich, CH-8057 Z\"urich, Switzerland
\\
\vskip .1cm
\vskip .5cm
\begin{minipage}[l]{.9\textwidth}
\begin{center} 
\textit{E-mail:} 
\email{feruglio@pd.infn.it},\\ \qquad\qquad
\email{paride.paradisi@pd.infn.it},\\
\qquad\email{pattori@physik.uzh.ch}
\end{center}
\end{minipage}
}
\end{center}
\vskip 1cm
\begin{abstract}
We discuss in detail the constraints on partial compositeness coming from flavour and CP violation in the leptonic sector. In a first part we present a formulation of partial compositeness in terms of a flavour symmetry group and a set of spurions, whose background values
specify the symmetry breaking pattern. In such a framework we construct the complete set of dimension-six operators describing
lepton flavour violation and CP violation. By exploiting the existing bounds, we derive limits on the compositeness scale in different scenarios,
characterised by increasing restrictions on the spurion properties. We confirm that in the most general case the compositeness scale should lie well-above 10 TeV.
However, if in the composite sector mass parameters and Yukawa couplings are universal, such a bound can be significantly lowered, without necessarily reproducing the case of minimal flavour violation.
The most sensitive processes are decays of charged leptons either of radiative type or into three charged leptons, $\mu\to e$ conversion in nuclei and the electric dipole moment of the electron.
In a second part we explicitly compute the Wilson coefficients of the relevant dimension-six operators in the so-called two-site model, embodying the symmetry
breaking pattern discussed in our first part, and we compare the results with those of the general spurion analysis.
\end{abstract}
\end{titlepage}
\setcounter{footnote}{0}
%%%%%%%%%%%%%%%%%%%%%%%%%%%%%%%%%%%%%%%%%%%%%%%%%%%%%%%%%%%%%%%%%%%%%%%%%%%%%%%%%%%%%%%%%%%%%
\section{Introduction}
\label{sec1}
%%%%%%%%%%%%%%%%%%%%%%%%%%%%%%%%%%%%%%%%%%%%%%%%%%%%%%%%%%%%%%%%%%%%%%%%%%%%%%%%%%%%%%%%%%%%%
If the solution to the gauge hierarchy problem is based on a new symmetry and not on antrophic considerations \cite{Schellekens:2013bpa}  or on special evolutions of the scalar sector in the early universe \cite{Graham:2015cka},
new physics at the TeV scale is expected. In most of the existing models new degrees of freedom carrying flavour quantum numbers are present at the TeV scale, representing potential sources of
flavour changing neutral currents (FCNC) and CP violation. So far both direct searches at the LHC and indirect searches in the context of precision tests and flavour physics have brought no conclusive evidence
of new physics at the TeV scale. The negative outcome of the search for new physics in the flavour sector is particularly intriguing. Indeed a scale of new physics $\Lambda_{NP}$ as large as $10^5$ TeV \cite{Isidori:2013ez} 
is required by an effective operator analysis to preserve the good agreement between observations and theory predictions,
unless the new flavour sector is highly non generic and involves specific mechanisms to suppress FCNC and CP violation at the desired level.

The latter possibility is empirically supported by the huge hierarchies among fermion masses and fermion mixing angles, which can only be explained by some special dynamics.
An effective mechanism suppressing FCNC and CP violation
can be introduced by observing that in the electroweak theory the symmetry of the flavour sector is broken only by the Yukawa interactions. 
Minimal Flavour Violation (MFV) \cite{D'Ambrosio:2002ex} is defined by the assumption that, even including new physics contributions, Yukawa couplings are the only source of such symmetry breaking.
In MFV flavour effects from new physics are controlled and damped by the smallness of the fermion masses and mixing angles. In this framework data allow $\Lambda_{NP}$ to be considerably smaller,
close to the TeV scale. MFV provides a useful benchmark for the discussion of the flavour sector, but it does not emerge as a unique framework 
from the known mechanisms aiming to explain the observed fermion spectrum \cite{Feruglio:2015jfa}, or from the known models providing a solution to the gauge hierarchy problem.

In this paper we reconsider the possibility that both the origin of fermion masses and the suppression of FCNC and CP violation are due to the mechanism of partial compositeness (PC) \cite{Kaplan:1991dc},
as realized in the context of composite Higgs models (for reviews see \cite{Bellazzini:2014yua,Panico:2015jxa}), and we perform a detailed analysis of flavour and CP violations in the leptonic sector.
According to PC there are no direct couplings between the elementary fermions and the Higgs doublet. The Higgs doublet has potentially strong couplings to
a composite sector, including, in the simplest case, a set of vector-like fermions with masses of the order of the compositeness scale. The SM fermions are
mostly elementary and get their masses through mixing terms with operators of the composite sector, often modelled by vector-like fermions. 

An appealing realization of this idea involves anarchic Yukawa couplings in the composite
sector. In this case the observed hierarchies between SM fermion masses and mixing angles are entirely due to the elementary-composite mixing terms. 
This is of particular interest, especially for the lepton sector, since the known pattern of neutrino masses and mixing angles as extracted from neutrino oscillation experiments \cite{Altarelli:2014dca}
seems to support the idea of an underlying anarchic dynamics \cite{Hall:1999sn,Haba:2000be}.
Higher dimensional operators describing low-energy FCNC and CP violations are depleted by both inverse powers of the compositeness scale 
and by the mixing terms, thus realising an efficient suppression mechanism known as RS-GIM \cite{Agashe:2004cp}.
Quantitative studies in concrete models show that in the anarchic scenario limits from CP violation in the 
quark sector lead to $\Lambda_{NP}>10$ TeV \cite{Panico:2015jxa}, while the existing bound on the rate of $\mu\to e \gamma$ results in $\Lambda_{NP}>25$ TeV \cite{Panico:2015jxa}. 
This strongly disfavors the anarchic scenario when the compositeness scale is of the order of 1 TeV. It is also known that PC at the TeV scale can satisfy the bound from 
flavour physics if Yukawa couplings in the composite sector are non-generic.
For instance if we assume that such couplings are universal and, at the same time, that the only irreducible sources of flavour symmetry breaking are proportional to the SM Yukawa couplings,
we reproduce exactly the MFV scheme \cite{Redi:2013pga}. 

In this paper we focus on the lepton sector. 
Several studies of lepton flavour violation (LFV) and CP violation in the lepton sector of composite Higgs models have been realised. Most of the analysis have been performed in the context 
of five-dimensional (5D) models with a warped space-time metric, weakly coupled duals to strongly coupled four-dimensional conformal theories, believed to provide a calculable framework for 
composite Higgs models.
Explicit computations with anarchic Yukawa couplings have been carried out in refs. \cite{Huber:2003tu,Moreau:2006np,Agashe:2006iy,Agashe:2009tu,Csaki:2010aj,Iyer:2012db,Beneke:2015lba}. 
They get a lower bound on the masses of the first Kaluza-Klein modes of order 10 TeV. These bounds can be relaxed by requiring discrete \cite{Csaki:2008qq,delAguila:2010vg,Kadosh:2010rm,Hagedorn:2011un} or continuous \cite{Perez:2008ee,Santiago:2008vq,vonGersdorff:2012tt} flavour symmetries.
The original motivation for discrete symmetries, tailored to approximately reproduce the tri-bimaximal mixing pattern in the lepton sector, has weaken after the precise measurement of the $\theta_{13}$ angle
and the models considered in \cite{Csaki:2008qq,delAguila:2010vg,Kadosh:2010rm,Hagedorn:2011un} require considerable corrections now.
Continuous non-abelian symmetries in the composite sector, broken by the elementary-composite mixing terms have been analysed in detail, especially in relation to FCNC and CP violation in the quark sector
\cite{Barbieri:2008zt,Redi:2011zi,Barbieri:2012tu,Barbieri:2012uh,Redi:2012uj}. A review on flavour physics in 5D models with warped space-time metric can be found in ref. \cite{vonGersdorff:2013rwa}.
LFV and CP violation in the lepton sector have been investigated also in SM extensions with extra vector-like heavy leptons \cite{Ishiwata:2013gma,Falkowski:2013jya}, that can mimic the PC scenario, at least 
as far as the contributions from heavy fermions is concerned.

In the present work we recast the framework of PC in terms of a generalized flavour symmetry and a suitable set of relevant spurions, much along the lines of 
refs. \cite{Agashe:2004cp,Barbieri:2008zt,Redi:2012uj,KerenZur:2012fr}. This involves a certain degree of model dependence, since both the flavour symmetry and the spurions 
are determined by the specific set of composite leptons, that we choose by following the criterium of minimality. In particular we work in the limit of vanishing neutrino masses,
turning off the potential effects related to massive neutrinos. It is well-known that within PC, LFV and CP violation in the lepton sector are present also in the 
limit of massless neutrinos. By assuming that the set of adopted spurions are the only  irreducible sources of flavour and CP violation, we can construct an exhaustive list of 
Wilson coefficients related to dimension six operators describing LFV and CP violating processes. 
At variance with the previous studies, we include for the first time all Wilson coefficients containing up to four powers of the spurions describing the elementary-composite mixing
and we discuss their role in deriving the bounds on the compositeness scale. We also provide a complete list of the LFV Wilson coefficients that can be constructed in the limit of 
vanishing ``wrong" Yukawa couplings.
``Wrong" Yukawas in the composite sector are allowed by gauge symmetry, but do not contribute to SM lepton masses, at the leading order. 
They directly contribute to the dipole operators describing radiative decays of the charged leptons and setting such Yukawas to zero can relax the bounds on the compositeness scale.

The general scope of our analysis is to check whether there are alternative solutions, beyond MFV, to reconcile PC at the TeV scale with the existing bounds on LFV and CP violation.
We also wish to verify if the anarchic scenario is completely ruled out or not.
By exploiting the effective Lagrangian of our construction and the existing experimental bounds we estimate the limits on the new physics scale in several scenarios, where our 
spurions are subjected to a series of increasingly restrictive conditions.
We confirm that in general the anarchic scenario is not compatible with new physics at the TeV scale and we provide examples of how PC can be realized at the TeV scale,
without necessarily resorting to MFV.

In the second part of our paper we consider as an explicit model realization of the flavour symmetry and its breaking pattern the so-called two-site model, first introduced in ref. \cite{Contino:2006nn}. Such realization contains explicitly vector-like leptons, 
implementing partial compositness in the lepton sector, as well as a set of spin-one resonances. By integrating out the states at the compositness scale, we evaluate the Wilson coefficients of 
the relevant LFV and CP-violating dimension six operators and we compare the results with those of the general spurion analysis.

Our paper has the following plan.
In section 2 we define the flavour symmetry and the set of spurions of our setup and we characterize the Wilson coefficients of the dimension-six operators relevant to LFV and CP violation in the lepton sector.
In section 3 we perform a phenomenological analysis and we study the bounds on the new physics scale obtained by making different types of assumptions on the available spurions.
In section 4 we recall the main aspects of the two-site model, that explicitly incorporates the features of the flavour symmetry breaking defined in section 2. In section 5
we collect our results on the dimension six operators obtained from the model by integrating out heavy fermions and heavy gauge vector bosons. In section 6 we present a phenomenological analysis of LFV
in the two-site model. Finally we draw our conclusion. In the appendix B we show the result of our computation of the full one-loop contribution to the electromagnetic dipole operator in the two-site model.
%%%%%%%%%%%%%%%%%%%%%%%%%%%%%%%%%%%%%%%%%%%%%%%%%%%%%%%%%%%%%%%%%%%%%%%%%%%%%%%%%%%%%%%%%%%%%
\section{Effective field theory for lepton flavour violation}
\label{sec2}
%%%%%%%%%%%%%%%%%%%%%%%%%%%%%%%%%%%%%%%%%%%%%%%%%%%%%%%%%%%%%%%%%%%%%%%%%%%%%%%%%%%%%%%%%%%%%
As a first step, we have to choose the flavour symmetry group of our effective theory and its breaking terms. Throughout this paper we will work in the limit
of massless neutrinos. The leptons are those of the SM, that is three copies of $SU(2)$ doublets $\ell$ and singlets $\tilde{e}$.
In MFV the flavour symmetry group of the leptonic sector is $SU(3)_\ell \times SU(3)_{\tilde{e}} $, corresponding to independent transformations made 
on $\ell$ and $\tilde{e}$. In our analysis we will instead assume a PC scenario. Charged leptons have no direct coupling to the Higgs doublet and
acquire masses via mixing with vector-like heavy fermions. In this framework it is natural to assume as flavour symmetry group (focusing only on the non-Abelian part):
\begin{align}
\label{eq4: flavor group}
G_f & = SU(3)^6  	 \nonumber \\*
			& =	SU(3)_\ell \times SU(3)_{\tilde{e}} \times SU(3)_{L_L} 
						\times SU(3)_{L_R} \times SU(3)_{\tilde{E}_L} \times SU(3)_{\tilde{E}_R} \; ,
\end{align}
under which the lepton fields rotate in generation space in the following way:
\be 
\label{eq4: CH flavor rotation}
\ell_{Li}  \rightarrow (V_\ell)_{ij}\ell_{Lj} \; , ~~~~~~~~~~~~~~~~~~~~~~~~~   \tilde{e}_{Ri} 
											 \rightarrow (V_{\tilde{e}})_{ij}\tilde{e}_{Rj} \; ,
\ee
where $V_\ell$ and $V_{\tilde{e}}$ are elements of $SU(3)_\ell$ and  $SU(3)_{\tilde{e}}$, respectively. In other words, the SM leptons only transform under the 
$SU(3)_\ell \times SU(3)_{\tilde{e}} $ component of the flavour group $G_f$, and are invariant under the remaining $SU(3)^4$ factor. Such a factor will be used 
to specify the spurions of our effective field theory. 
We introduce three sets of spurions. We need spurions $(\Delta,\tilde\Delta)$ that mix SM leptons with heavy fermions. Moreover we allow for spurions $(m,\tilde{m})$ 
describing the masses of the heavy fermions in the limit of unbroken electroweak symmetry. Finally the heavy fermion sector can interact with the Higgs doublet and 
this implies additional spurions $(\yl,\yr)$. Notice that this set of spurions is the most general one compatible with our flavour group $G_f$ and with the assumption that 
the SM Higgs doublet directly couples only to the heavy sector. From these considerations the following transformation properties for our spurions can be deduced
\begin{align}
  \begin{aligned}
    m  		& \rightarrow V_{L_L} m V_{L_R}^\dagger \; , 		\\
    \tilde{m}& \rightarrow V_{\tilde{E}_L} \tilde{m} V_{\tilde{E}_R}^\dagger \; , 
  \end{aligned}
  &&
  \begin{aligned}
    \Delta	& \rightarrow V_\ell \Delta V_{L_R}^\dagger \; , 		\\
    \tilde\Delta & \rightarrow V_{\tilde{e}} \tilde\Delta V_{\tilde{E}_L}^\dagger \; ,
  \end{aligned}
  &&
  \begin{aligned}
    \yr		& \rightarrow V_{L_L} \yr V_{\tilde{E}_R}^\dagger \; , 	\\
    \yl		& \rightarrow V_{L_R} \yl V_{\tilde{E}_L}^\dagger \; ,
  \end{aligned}
\end{align}
\label{tr1}
where it is evident from our notation which are the $SU(3)$ factors of $G_f$ that are involved in each transformation. An equivalent set of spurions is obtained by replacing 
the dimensionful quantities $(\Delta,\tilde\Delta)$ with dimensionless combinations $(X,\tilde X)=(\Delta m^{-1},\tilde\Delta \tilde m^{-1\dagger})$ transforming as
\be
 \begin{aligned}
    X	& \rightarrow V_\ell X V_{L_L}^\dagger \; , 		\\
    \tilde X & \rightarrow V_{\tilde{e}} \tilde X V_{\tilde{E}_R}^\dagger \; .
  \end{aligned}
\label{tr2}
\ee
In such a case at the leading order (LO) SM charged lepton masses are generated by the operator
\be
{\cal O}^{(0)}_M=(\overline{\ell}_L \varphi) X~ \yr~ \tilde X^\dagger \tilde{e}_{R} ~~~,
\label{massLO}
\ee
where $\varphi$ is the Higgs electroweak doublet. 

It is well-know that, if all the spurions $(X,\tilde X)$, $(\yl,\yr)$ and $(m,\tilde{m})$ are present at the same time and if the Yukawa couplings $(\yl,\yr)$ are anarchic, 
a severe bound on the new physics scale applies. Indeed in concrete models belonging to the general class we are considering, the one-loop exchange of Higgs and 
heavy fermions leads to the following electromagnetic dipole operator:
\be
\label{Dcom}
 \frac{1}{16 \pi^2}(\overline{\ell}_L \varphi)(\sigma\cdot F) X~ \yr \tilde m^{-1} \yl^\dagger m^{-1} \yr~  \tilde X^\dagger \tilde{e}_{R}~~~.
\ee
If such a contribution is present, and the Yukawa couplings $(\yl,\yr)$ are assumed to be anarchic of ${\cal O}(1)$,
then the electromagnetic dipole operator and the mass operator are not aligned in flavour space and 
the heavy fermion scale is bounded to be heavier than about 30 TeV, to respect the bound on ${\rm BR}(\mu\to e \gamma)$.
One way to eliminate this dangerous contribution, while maintaining non-vanishing lepton masses, is to assume $\yl=0$. We will come back to this assumption
later in this section. For the moment we will adopt it as a working hypothesis. Our purpose is to analyse the flavour violating contributions surviving in this limit
and to estimate the corresponding bounds on the new physics scale.

At variance with MFV, spurions with the dimension of a mass are present  in our setup and some additional
prescriptions are needed:
\begin{itemize}
\item
First of all we require that our operators are local in the spurions $(X,\tilde X)$. These are mixing parameters
that are generically treated as small and provide one set of expansion parameters for our spurion analysis. 
\item
We also assume that the operators
are local in the Yukawa couplings $\yr$, that we restrict in the range $1\le|\yr|\le4\pi$. 
\item
Each power of $\yr$ occurs accompanied either
by an Higgs electroweak doublet $\varphi$ or by a factor $1/4 \pi$.
\item
Masses of the composite sector are described by the spurions $(m,\tilde{m})$ 
to which we add an additional parameter $M$, singlet under the flavour symmetry group, to describe masses of other composite particles, 
such as for instance a new set of vector boson resonances. These additional states are coupled to the Higgs doublet and to the heavy fermions
with a strong coupling constant $g^*$, in the range $1\le g^*\le4\pi$. 
\item
In our operators masses will always appear in negative powers,
to allow for decoupling of all the operators with the scale of new physics. 
\end{itemize}
With this set of assumptions, the LO operator describing charged lepton masses is still ${\cal O}_M$ in eq.~(\ref{massLO}).
We stress that this operator provides the definition of the spurion $\yr$. Any polynomial of the type
\be
\yr \left[
1+\frac{c_1}{16\pi^2}\left(\yr^\dagger \yr\right)+\frac{c_2}{(16 \pi^2)^2}\left(\yr^\dagger \yr\right)^2+....
\right]
\label{yrp}
\ee
could replace $\yr$ in ${\cal O}_M$. With no loss of generality we can redefine as $\yr$ the particular combination of eq.~(\ref{yrp}) occurring in ${\cal O}_M$.

The physical processes we are interested in concern lepton flavour violation in the charged lepton sector as well as the magnetic dipole moments and the electric dipole moments (EDM)
of the charged leptons. To this purpose it is convenient to adopt an effective field theory description where the SM Lagrangian is extended
by an appropriate set of gauge invariant operators depending on the SM fields \cite{Buchmuller:1985jz}:
\be
\mathcal{L} = \mathcal{L}_{\rm SM} + 
\frac{1}{\Lambda^2} \sum_i C_i\, Q_i +...
\label{Lag6}
\ee
where we have restricted our attention to the lowest dimensional operators relevant to the processes we are interested in, namely those of dimension six. 
Dots denote higher-dimensional operators. We list below a complete set of dimension six operators depending on lepton fields and on the scalar electroweak 
doublet $\vp$ \cite{Grzadkowski:2010es}. We start with the dipole operators
\begin{align}
\label{DD}
(Q_{eW})_{ij}&=(\bar \ell_{Li} \sigma^{\mu\nu} \tilde e_{Rj}) \tau^I \vp W_{\mu\nu}^I~,
\nn\\
(Q_{eB})_{ij}&=(\bar \ell_{Li}  \sigma^{\mu\nu} \tilde e_{Rj}) \vp B_{\mu\nu}~~.
\end{align}
$W_{\mu\nu}^I$ and $B_{\mu\nu}$ are the field strengths for the gauge vector bosons of $SU(2)$ and $U(1)$, respectively.
The flavour structure of these two operators is the same and we will focus on the combinations 
\begin{align}
(Q_{e\gamma})_{ij}&=\cos\theta_W~(Q_{eB})_{ij}-\sin\theta_W~(Q_{eW3})_{ij}~~,\nn\\
(Q_{eZ})_{ij}&=\sin\theta_W~(Q_{eB})_{ij}+\cos\theta_W~(Q_{eW3})_{ij}~~,
\end{align}
where $\theta_W$ is the weak mixing angle and $Q_{eW3}$ denotes the contribution to $Q_{eW}$ obtained setting to zero $W_{\mu\nu}^{1,2}$. 
The electromagnetic dipole operator $Q_{e\gamma}$
is the only operator that gives a tree-level contribution to the radiative decays of charged leptons, when $i\ne j$. The diagonal elements, $i=j$, contribute
to the anomalous magnetic moments and to the EDM of the charged leptons. We have operators bilinear in the Higgs doublet
\begin{align}
\label{Vope}
(Q_{\vp l}^{(1)})_{ij}&=(\vpj)( \bar \ell_{Li}\gamma^\mu \ell_{Lj})~,\nn \\
(Q_{\vp l}^{(3)})_{ij}&=(\vpjt)(\bar \ell_{Li} \tau^I \gamma^\mu \ell_{Lj})~,\\
(Q_{\vp e})_{ij}&=(\vpj)(\overline{\tilde e}_{Ri} \gamma^\mu \tilde e_{Rj})\nn~~.
\end{align}
After the breaking of the electroweak symmetry these operators modify the couplings of the $Z$ boson to leptons, potentially violating both universality, 
through the diagonal terms, and lepton flavour, through the non-diagonal ones. There is a unique operator trilinear in the Higgs doublet
\be
(Q_{e\vp})_{ij}=(\vp^\dag \vp)(\bar \ell_{Li} \tilde e_{Rj} \vp)~~,
\ee
which contributes, with a different weight, to both the masses and the Higgs couplings of the charged leptons. Finally we have four-lepton operators
\begin{align}
\label{Cope}
(Q_{ll})_{ijmn}&=(\bar \ell_{Li} \gamma_\mu \ell_{Lj})(\bar \ell_{Lm} \gamma^\mu \ell_{Ln})~,\nn\\
(Q_{ee})_{ijmn}&=(\overline{\tilde e}_{Ri} \gamma_\mu \tilde e_{Rj})(\overline{\tilde e}_{Rm} \gamma^\mu \tilde e_{Rn})~,\\
(Q_{le})_{ijmn}&=(\bar \ell_{Li} \gamma_\mu \ell_{Lj})(\overline {\tilde e}_{Rm} \gamma^\mu \tilde e_{Rn})\nn~~,
\end{align}
which can contribute to muon and tau decays into three charged leptons.
We also consider dimension-six operators of the type $llqq$ that can contribute to $\mu\to e$ conversion in nuclei
\footnote{The operators $(Q_{\ell q}^{(u)})_{ij}$ and $(Q_{\ell q}^{(d)})_{ij}$ are linear combinations of the operators $(Q_{\ell q}^{(1)})_{ij}$ and $(Q_{\ell q}^{(3)})_{ij}$ of ref. \cite{Grzadkowski:2010es}.}:
\begin{align}
\label{mutoe}
(Q_{\ell q}^{(u)})_{ij}&=(\bar \ell_{Li} \gamma_\mu \ell_{Lj})(\bar u_{Lm} \gamma^\mu u_{Ln})~~~~~~~~~~~~~~~\nn\\
(Q_{\ell q}^{(d)})_{ij}&=(\bar \ell_{Li} \gamma_\mu \ell_{Lj})(\bar d_{Lm} \gamma^\mu d_{Ln})~~~~~~~~~~~~~~~\nn\\
(Q_{\ell u,d})_{ij}&=(\bar \ell_{Li} \gamma_\mu \ell_{Lj})(\bar u_{Rm} \gamma^\mu u_{Rn})~,(\bar \ell_{Li} \gamma_\mu \ell_{Lj})(\bar d_{Rm} \gamma^\mu d_{Rn})~,\nn\\
(Q_{eq})_{ij}&=(\overline{\tilde e}_{Ri} \gamma_\mu \tilde e_{Rj})(\bar q_{Lm} \gamma^\mu q_{Ln})~~~~~~~~~~~~~~~~q=(u,d),\\
(Q_{e u,d})_{ij}&=(\overline{\tilde e}_{Ri} \gamma_\mu \tilde e_{Rj})(\bar u_{Rm} \gamma^\mu u_{Rn})~,(\overline{\tilde e}_{Ri} \gamma_\mu \tilde e_{Rj})(\bar d_{Rm} \gamma^\mu d_{Rn})~.\nn
\end{align}
There are other 3 independent operators of this type \cite{Grzadkowski:2010es}, but the chosen subset is sufficiently general for the purposes of the present discussion.
Notice that each operator carries flavour indices. Hermiticity of the effective Lagrangian is guaranteed either by appropriate symmetry properties of the Wilson 
coefficients under transposition of the family indices or by addition of the hermitian conjugate operator. 
%[\textcolor{red}{Should we discuss the relation to other basis?}] 
\begin{table}[h] 
\centering
\begin{tabular}{| c | c | l | l |}
\hline
$N_Y$ & $N_X$ & HF & HB \\
\hline
1 & 2 & $  X   \yr   \tilde{X}^\dagger $
			  &  $ X   \yr   (\tilde{c}^\dagger \tilde{c})^{-1}   \tilde{X}^\dagger $ \\
   &    &  &  $ X   (c c^\dagger)^{-1}   \yr   \tilde{X}^\dagger $ \\
\hline
1 & 4 & $ X   \yr   \tilde{X}^\dagger   \tilde{X}   \tilde{X}^\dagger $
			  & $ X   (c c^\dagger)^{-1}   \yr   \tilde{X}^\dagger   \tilde{X}   \tilde{X}^\dagger $  \\
   &    & $  X   X^\dagger   X   \yr   \tilde{X}^\dagger $
   			  & $ X   \yr   (\tilde{c}^\dagger \tilde{c})^{-1}   \tilde{X}^\dagger   \tilde{X}   \tilde{X}^\dagger $ \\
   &    &  & $ X   \yr   \tilde{X}^\dagger   \tilde{X}   (\tilde{c}^\dagger \tilde{c})^{-1}   \tilde{X}^\dagger $ \\
   &    &  & $ X   (c c^\dagger)^{-1}   X^\dagger   X   \yr   \tilde{X}^\dagger $ \\
   &    &  & $ X   X^\dagger   X   (c c^\dagger)^{-1}   \yr   \tilde{X}^\dagger $ \\
   &    &  & $ X   X^\dagger   X   \yr   (\tilde{c}^\dagger \tilde{c})^{-1}   \tilde{X}^\dagger $ \\
\hline
3 & 2 & $  X   \yr   \yr^\dagger   \yr   \tilde{X}^\dagger $
			  & $ X   (c c^\dagger)^{-1}   \yr   \yr^\dagger   \yr   \tilde{X}^\dagger  $ \\
	&   &  & $ X   \yr   (\tilde{c}^\dagger \tilde{c})^{-1}   \yr^\dagger   \yr   \tilde{X}^\dagger  $ \\
   &    &  & $ X   \yr   \yr^\dagger   (c c^\dagger)^{-1}   \yr   \tilde{X}^\dagger  $ \\
   &    &  & $ X   \yr   \yr^\dagger   \yr   (\tilde{c}^\dagger \tilde{c})^{-1}   \tilde{X}^\dagger  $ \\
\hline
3 & 4 & $  X   \yr   \tilde{X}^\dagger   \tilde{X}   \yr^\dagger   \yr   \tilde{X}^\dagger  $
			  & $  X   (c c^\dagger)^{-1}   \yr   \tilde{X}^\dagger   \tilde{X}   \yr^\dagger   \yr   \tilde{X}^\dagger  $ \\
  &   & $   X   \yr   \yr^\dagger   X^\dagger   X   \yr   \tilde{X}^\dagger  $
			  &  ... \\
  &   & $   X   \yr   \yr^\dagger   \yr   \tilde{X}^\dagger   \tilde{X}   \tilde{X}^\dagger  $
			  &  $   X   (c c^\dagger)^{-1}   \yr   \yr^\dagger   X^\dagger   X   \yr   \tilde{X}^\dagger  $ \\
  &   & $  X   X^\dagger   X   \yr   \yr^\dagger   \yr   \tilde{X}^\dagger  $
			  &  ... \\
\hline
\end{tabular}
\caption{Spurion combination $C_{\bar L R}$, in a matrix notation, for the lepton bilinear $\bar \ell_{Li} (C_{\bar L R})_{ij} \tilde e_{Rj}$. $N_Y$ and $N_X$ are the orders of the expansion in $\yr$ and $(X,\tilde X)$, respectively.
We restrict the list to $N_Y\le 3$ and $N_X\le 4$.
For convenience we distinguish spurion combinations depending on composite fermion matrices $c$ and $\tilde{c}$ (column HB), from combinations not involving $c$ and $\tilde{c}$ (column HF).}
\end{table}

Our aim is to estimate the Wilson coefficients of these operators, by expressing them in terms of the spurions using the set of rules described above.
We expand each Wilson coefficient in powers of the mixings $(X,\tilde X)$ and the Yukawa coupling $\yr$. Since the spurions $(X,\tilde X)$ control lepton masses, they are expected to be small, of order $(0.1/\yr)$ at most. We will stop the expansion 
in $(X,\tilde X)$ at the fourth order. In the expansion we will go up to the third order in $\yr$, since in the anarchic scenario
trilinear combinations of Yukawa couplings are in general misaligned with respect to linear ones. Higher orders in Yukawa couplings do not bring any new qualitative feature in our analysis.
Finally the correct dimension is provided by negative powers of $M$ or $(m,\tilde{m})$.
Since the operators under study have dimension six, in practice we have two classes of operators, those suppressed by $1/M^2$ and those suppressed by the heavy fermion masses.
Each of these two classes refers to a specific decoupling limit. When $M\gg |m|,|\tilde{m}|$, the heavy bosons decouple first and the operators are suppressed by negative powers of $(m,\tilde{m})$.
We call this heavy boson (HB) case.
In the opposite limit, $M\ll |m|,|\tilde{m}|$, we have a fast heavy fermion decoupling and the operators are suppressed by the smaller scale $M$. We call this heavy fermion (HF) case.
When the two scales $M$ and $(m,\tilde{m})$
are comparable, the Wilson coefficient can be a generic function of the ratio of the two scales. For the present discussion the two limiting cases are sufficient to capture the behaviour of the system.
For our spurion analysis it is convenient to rewrite the mass matrices $m$, $\tilde{m}$ in this way
\be
m=m_0~ c~~~,~~~~~~~~~~~~~~~~~~~~~\tilde{m}=m_0~ \tilde{c}~~~,
\ee
where $m_0$ is a flavour-independent mass parameter, while the flavour dependence is carried by the dimensionless matrices $c$, $\tilde{c}$. 
To facilitate the identification of the relevant Wilson coefficients, it is useful to identify the combinations of spurions that fit lepton bilinears
\be
\bar \ell_{Li} (C_{\bar L R})_{ij} \tilde e_{Rj}~~~,~~~~~~~~~~~~~~~~~~~~\bar \ell_{Li}  (C_{\bar L L})_{ij} \ell_{Lj}~~~,~~~~~~~~~~~~~~~~~~~~\overline{\tilde e}_{Ri} (C_{\bar R R})_{ij}  \tilde e_{Rj}~~~.
\ee
In tables 1, 2 and 3 we collect such combinations, constructed with the rules outlined above.
\begin{table}[t]
\centering
\begin{tabular}{| c | c | l | l |}
\hline
$N_Y$ & $N_X$ & HF & HB \\
\hline
0 &0 & \mbox{l\hspace{-0.55em}1} &\mbox{l\hspace{-0.55em}1}\\
\hline
0 & 2 	& $ X   X^\dagger  $
			& 	$X   (c c^\dagger)^{-1}   X^\dagger $ \\
\hline
0 & 4 	& $  X   X^\dagger   X   X^\dagger $
			& 	$ X   (c c^\dagger)^{-1}   X^\dagger   X   X^\dagger $ \\
   &  &  & 	$ X   X^\dagger   X   (c c^\dagger)^{-1}   X^\dagger $ \\
\hline
2 & 2   & $  X   \yr   \yr^\dagger   X^\dagger $
		    & 	$ X   (c c^\dagger)^{-1}   \yr   \yr^\dagger   X^\dagger $ \\
	&  &  & 	$ X   \yr   (\tilde{c}^\dagger \tilde{c})^{-1}   \yr^\dagger   X^\dagger $ \\
   &  &  & 	$ X   \yr   \yr^\dagger   (c c^\dagger)^{-1}   X^\dagger  $ \\

\hline
2 & 4 	& $  X   \yr   \tilde{X}^\dagger   \tilde{X}   \yr^\dagger   X^\dagger $
			&  $  X   (c c^\dagger)^{-1}   \yr   \tilde{X}^\dagger   \tilde{X}   \yr^\dagger   X^\dagger $ \\
  &   	& $  X   \yr   \yr^\dagger   X^\dagger   X   X^\dagger $ 
			&  ... \\
  &  	& $  X   X^\dagger   X   \yr   \yr^\dagger   X^\dagger $ 
			&  $  X   (c c^\dagger)^{-1}   \yr   \yr^\dagger   X^\dagger   X   X^\dagger $  \\
  &   &  & ... \\
\hline
\end{tabular}
\caption{Spurion combination $C_{\bar L L}$, in a matrix notation, for the lepton bilinear $\bar \ell_{Li} (C_{\bar L L})_{ij} \ell_{Lj}$. $N_Y$ and $N_X$ are the orders of the expansion in $\yr$ and $(X,\tilde X)$, respectively.
We restrict the list to $N_Y\le 2$ and $N_X\le 4$. For convenience we distinguish spurion combinations depending on composite fermion matrices $c$ and $\tilde{c}$ (column HB), from combinations not involving $c$ and $\tilde{c}$ (column HF).}
\end{table}
\begin{table}[h!]
\centering
\begin{tabular}{| c | c | l | l |}
\hline
$N_Y$ & $N_X$ & HF & HB \\
\hline
0 &0 & \mbox{l\hspace{-0.55em}1} &\mbox{l\hspace{-0.55em}1}\\
\hline
0 & 2 	& $   \tilde{X}   \tilde{X}^\dagger  $
			& 	$\tilde{X}   (\tilde{c}^\dagger \tilde{c})^{-1}   \tilde{X}^\dagger $ \\
\hline
0 & 4 	& $  \tilde{X}   \tilde{X}^\dagger   \tilde{X}   \tilde{X}^\dagger $
			& 	$ \tilde{X}   (\tilde{c}^\dagger \tilde{c})^{-1}   \tilde{X}^\dagger   \tilde{X}   \tilde{X}^\dagger $ \\
   &  &  & 	$ \tilde{X}   \tilde{X}^\dagger   \tilde{X}   (\tilde{c}^\dagger \tilde{c})^{-1}   \tilde{X}^\dagger $ \\
\hline
2 & 2   & $  \tilde{X}   \yr^\dagger   \yr   \tilde{X}^\dagger $
			& 	$ \tilde{X}   (\tilde{c}^\dagger \tilde{c})^{-1}   \yr^\dagger   \yr   \tilde{X}^\dagger $ \\
   &	&  & 	$ \tilde{X}   \yr^\dagger   (c c^\dagger)^{-1}   \yr   \tilde{X}^\dagger $ \\
   &  &  & 	$ \tilde{X}   \yr^\dagger   \yr   (\tilde{c}^\dagger \tilde{c})^{-1}   \tilde{X}^\dagger  $ \\
\hline
2 & 4 	& $  \tilde{X}   \yr^\dagger   X^\dagger   X   \yr   \tilde{X}^\dagger $
			& $  \tilde{X}   (\tilde{c}^\dagger \tilde{c})^{-1}   \yr^\dagger   X^\dagger   X   \yr   \tilde{X}^\dagger $  \\
 &  	& $  \tilde{X}   \yr^\dagger   \yr   \tilde{X}^\dagger   \tilde{X}   \tilde{X}^\dagger $ 
			&  ... \\
 &  	& $  \tilde{X}   \tilde{X}^\dagger   \tilde{X}   \yr^\dagger   \yr   \tilde{X}^\dagger $ 
			&  $  \tilde{X}   (\tilde{c}^\dagger \tilde{c})^{-1}   \yr^\dagger   \yr   \tilde{X}^\dagger   \tilde{X}   \tilde{X}^\dagger $  \\
 &    &  & ... \\
\hline
\end{tabular}
\caption{Spurion combination $C_{\bar R R}$, in a matrix notation, for the lepton bilinear $\overline{\tilde e}_{Ri} (C_{\bar R R})_{ij} \tilde e_{Rj}$. $N_Y$ and $N_X$ are the orders of the expansion in $\yr$ and $(X,\tilde X)$, respectively.
We restrict the list to $N_Y\le 2$ and $N_X\le 4$. For convenience we distinguish spurion combinations depending on composite fermion matrices $c$ and $\tilde{c}$ (column HB), from combinations not involving $c$ and $\tilde{c}$ (column HF).}
\end{table}
%%%%%%%%%%%%%%%%%%%%%%%%%%%%%%%%%%%%%%%%%%%%%%%%%%%%%%%%%%%%%%%%%%%%%%%%%%%%%%%%%%%%%%%%%%%%%
\subsection{Dipole operators}
%%%%%%%%%%%%%%%%%%%%%%%%%%%%%%%%%%%%%%%%%%%%%%%%%%%%%%%%%%%%%%%%%%%%%%%%%%%%%%%%%%%%%%%%%%%%%
The dipole operators in eq.~(\ref{DD}) involve the lepton bilinear $\bar \ell_{Li} (C_{\bar L R})_{ij} \tilde e_{Rj}$ and their possible Wilson coefficients are the $(C_{\bar L R})_{ij}$ combinations
listed in table 1.
We expect that the dipole operators are loop-generated in perturbation theory. The naive loop suppression factor $1/(16\pi^2)$ is not present in table 1 and should 
be included in the Wilson coefficient 
\be
(C_{e\gamma,Z})_{ij}=\frac{1}{16\pi^2}(C_{\bar L R})_{ij}~~~. 
\ee
The new physics scale $\Lambda$ associated to dipole operators is $M$ in the HF case ($M\ll |m|,|\tilde{m}|$)
and $m_0$ in the HB one ($M\gg |m|,|\tilde{m}|$).

The first important outcome of our analysis is that there is a large set of potentially lepton flavour violating combinations, beyond that of eq.~(\ref{Dcom}).
The actual appearance of these combinations in concrete models containing our set of spurions will depend on the specific dynamics of the model under consideration.
We will discuss the phenomenological implications of the new structures for the Wilson coefficients $(C_{e\gamma,Z})_{ij}$ in section 3.
%%%%%%%%%%%%%%%%%%%%%%%%%%%%%%%%%%%%%%%%%%%%%%%%%%%%%%%%%%%%%%%%%%%%%%%%%%%%%%%%%%%%%%%%%%%%%
\subsection{Scalar operator}
%%%%%%%%%%%%%%%%%%%%%%%%%%%%%%%%%%%%%%%%%%%%%%%%%%%%%%%%%%%%%%%%%%%%%%%%%%%%%%%%%%%%%%%%%%%%%
The scalar operator
$(Q_{e\vp})_{ij}=(\vp^\dag \vp)(\bar \ell_{Li} \tilde e_{Rj} \vp)$
has exactly the same flavour structure of the dipole operators and we can directly read from table 1 the list of possible Wilson coefficients, provided we 
have at least three Yukawa couplings, $N_Y\ge 3$, since this operator contains three Higgs doublets. Up to an overall flavour-independent coefficient,
which is expected to be of order one, we have: 
\be
(C_{e\vp})_{ij}=(C_{\bar L R})_{ij}~~~~~~~~~~~~(N_Y\ge 3)~~~.
\ee
Notice that there is no loop suppression in this case, since this operator can be generated at tree level.         
As for the dipole operators, the scale of new physics $\Lambda$ is $M$ in the HF case and $m_0$ in the HB one.

This operator, together with the mass operator of eq.~(\ref{massLO}), contributes to both masses and Yukawa couplings of charged leptons.
If the Wilson coefficients are not exactly aligned in flavour space, we have flavour violating decays of the Higgs. Even in the case where there is a perfect alignment among all Wilson
coefficients, the overall strength of Yukawa couplings is altered compared to the case of the SM, where it is completely fixed by the fermion mass and by the electroweak VEV.
We will discuss the impact of these modifications in section \ref{pheno}.
%%%%%%%%%%%%%%%%%%%%%%%%%%%%%%%%%%%%%%%%%%%%%%%%%%%%%%%%%%%%%%%%%%%%%%%%%%%%%%%%%%%%%%%%%%%%%
\subsection{Vector operators}
%%%%%%%%%%%%%%%%%%%%%%%%%%%%%%%%%%%%%%%%%%%%%%%%%%%%%%%%%%%%%%%%%%%%%%%%%%%%%%%%%%%%%%%%%%%%%
In the operators $(Q_{\vp l}^{(1,3)})_{ij}$, $(Q_{\vp e})_{ij}$ of eq.~(\ref{Vope}) new flavour structures arise. The corresponding Wilson coefficients can be read from table 2, for $(Q_{\vp l}^{(1,3)})_{ij}$
and table 3, for $(Q_{\vp e})_{ij}$, but excluding the case $N_Y=0$, since the vector operators are bilinear in the Higgs doublets and this requires at least two powers of the Yukawa couplings.
We have 
\be
(C_{\vp l}^{1,3})_{ij}=(C_{\bar L L})_{ij}~~~,~~~~~~~~~~~~~~~~~~~~(C_{\vp e})_{ij}=(C_{\bar R R})_{ij}~~~,~~~~~~~~~~~(N_Y\ge 2)~~~.
\ee
There is no loop suppression, in general, and the new physics scale is identified as for the previous operators.
We see that, in general, the Wilson coefficients for the vector operators are not diagonal in the mass basis.
This leads to violation of flavour in the couplings of the $Z$ boson, with consequences that we discuss in section \ref{pheno}. 
%%%%%%%%%%%%%%%%%%%%%%%%%%%%%%%%%%%%%%%%%%%%%%%%%%%%%%%%%%%%%%%%%%%%%%%%%%%%%%%%%%%%%%%%%%%%%
\subsection{Contact operators}
%%%%%%%%%%%%%%%%%%%%%%%%%%%%%%%%%%%%%%%%%%%%%%%%%%%%%%%%%%%%%%%%%%%%%%%%%%%%%%%%%%%%%%%%%%%%%
Finally we consider the contact operators $(Q_{ll})_{ijmn}$, $(Q_{ee})_{ijmn}$ and $(Q_{le})_{ijmn}$ given in eq.~(\ref{Cope}). In these operators we recognise combinations of the flavour structures
already discussed. We expect the following possible factorisations:
\be
\label{eq:C_ll}
(C_{ll})_{ijmn}=\left\{
\begin{array}{l}
(C_{\bar L L})_{ij}\times (C_{\bar L L})_{mn}\\
(C_{\bar L L})_{in}\times (C_{\bar L L})_{mj}
\end{array}~,
\right.
\ee
\be
\label{eq:C_ee}
(C_{ee})_{ijmn}=\left\{
\begin{array}{l}
(C_{\bar R R})_{ij}\times (C_{\bar R R})_{mn}\\
(C_{\bar R R})_{in}\times (C_{\bar R R})_{mj}
\end{array}~,
\right.
\ee
\be
\label{eq:C_le}
(C_{le})_{ijmn}=\left\{
\begin{array}{l}
(C_{\bar L L})_{ij}\times (C_{\bar R R})_{mn}\\
(C_{\bar L R})_{in}\times (C_{\bar L R}^\dagger)_{mj}
\end{array}~.
\right.
\ee
Since the contact terms contain no Higgs doublets, we have no restrictions on $N_Y$. For the same reason, each bilinear in the Yukawa coupling 
should be accompanied by a loop suppression factor $1/(16\pi^2)$ \footnote{If contact terms originate from dim-8 operators, the loop factor can be effectively replaced by $v^2/\Lambda^2$.}.
Concerning the new physics scale $\Lambda$, when the two factors $C$ come from the HF column we have $\Lambda=M$.
When one the factors come from the HF column and the other from the HB column, we have $\Lambda=m_0$. 
Similarly, for the $llqq$ operators we have:
\be
(C_{\ell q}^{(u,d)})_{ij}=(C_{\bar L L})_{ij}~~~,~~~~~~~(C_{\ell u,d})_{ij}=(C_{\bar L L})_{ij}~~~,
\ee
\be
(C_{eq})_{ij}=(C_{\bar R R})_{ij}~~~,~~~~~~~(Q_{e u,d})_{ij}=(C_{\bar R R})_{ij}~~~.
\ee
with no restrictions on $N_Y$.
%%%%%%%%%%%%%%%%%%%%%%%%%%%%%%%%%%%%%%%%%%%%%%%%%
\subsection{Stability of the solution $\yl=0$}
%%%%%%%%%%%%%%%%%%%%%%%%%%%%%%%%%%%%%%%%%%%%%%%%%

Since we are interested in the scenario where $\yl$ is nearly vanishing, 
a legitimate question is whether and to which extent such a limit 
is stable under quantum corrections. To this purpose it is better to distinguish the two regimes $v\ll m_0\ll M$ and $v\ll M\ll m_0$.
When $v\ll m_0\ll M$, we can consider the following spurion combinations
\be
\label{th}
\dd\frac{m_0^2}{M^2}~c^\dagger  \yr  \tilde c^\dagger~~~,~~~~~~~~~~\frac{m_0^2}{M^2}~c^\dagger \yr\yr^\dagger\yr ~ \tilde c^\dagger~~~,
\ee
that behave as effective Yukawas of type $\yl$ since they have the same transformation properties as $\yl$. We expect that terms like those in (\ref{th})
arise, in perturbation theory, through threshold corrections induced by the one loop exchange of heavy gauge bosons and heavy fermions, from which 
we estimate
 \be
\label{threshold1}
\yl\approx \frac{g^2_*}{16\pi^2}\dd\frac{m_0^2}{M^2}~c^\dagger  \yr  \tilde c^\dagger~~~~,~~~~~~~~~~\yl\approx\dd k~\frac{g^2_*}{16\pi^2}\frac{m_0^2}{M^2}~c^\dagger \yr\yr^\dagger\yr ~ \tilde c^\dagger~~~~,
\ee
where $g_*$ is a coupling constant of the strong sector and $k$ is either an additional loop factor $1/16\pi^2$ stemming from a Higgs loop, or a factor coming from the electroweak VEV, $v^2/M^2$, if the two Yukawas are attached to external Higgs legs.
The contributions of eq.~(\ref{threshold1}) are small if there is an hierarchy between heavy fermions and heavy gauge boson masses. They are also suppressed in the semi-perturbative regime $ g_*, \yr \approx 1$.
Similarly, exchange of ordinary gauge bosons and heavy fermions lead to effective $\yl$ couplings of the type
\be
\label{aa}
\yl\approx \frac{g^2}{16\pi^2}\dd\frac{m_{W,Z}^2}{m_0^2}~c^{-1}  \yr  \tilde c^{-1}~~~,~~~~~~~~~~\yl\approx\dd k~\frac{g^2}{16\pi^2}\frac{m_{W,Z}^2}{m_0^2}~c^{-1}  \yr\yr^\dagger\yr ~ \tilde c^{-1}~~~.
\ee
These contributions are naturally suppressed by the ratio $m_{W,Z}^2/m_0^2$ and can be easily kept at the percent level.

In the other regime, $v\ll M\ll m_0$, beyond the combinations of eq.~(\ref{aa}), we can also consider 
\be
\label{threshold2}
\frac{M^2}{m_0^2}~c^{-1}  \yr  \tilde c^{-1}~~~,~~~~~~~~~~\frac{M^2}{m_0^2}~c^{-1}  \yr\yr^\dagger\yr ~ \tilde c^{-1}~~~,
\ee
which also transform as the spurion $\yl$. These combinations formally decouple in the limit of infinitely large $m_0$. There are however also combinations that do not decouple, such as
\be
c^{-1}  \yr  \tilde c^\dagger~~~,~~~~~~~~~~~c^{-1}  \yr\yr^\dagger\yr ~ \tilde c^\dagger~~~.
\ee
We have checked, through a one-loop computation in the two-site model that we will consider in section 4, that this kind of contributions are generated.
They cannot be parametrically suppressed by mass ratios and the corresponding $\yl$ can be depleted only in the semi-perturbative regime $ g_*, \yr \approx 1$.

Notice that for $m_0 \approx M$, the effective Yukawa couplings arising from eqs.~(\ref{threshold1},\ref{threshold2}),  are insensitive to the overall mass scale.
In summary, in the regime of heavy gauge bosons much heavier than heavy fermions, the effective $\yl$ can remain close to the percent level even in a strongly 
coupled regime, while in the opposite regime the solution $\yl \approx0$ is typically stable under quantum corrections (up to the \% level) only if we assume the
semi-perturbative regime $ g_*, \yr \approx 1$.
%%%%%%%%%%%%%%%%%%%%%%%%%%%%%%%%%%%%%%%%%%%%%%%%%%%%%%%%%%%%%%%%%%%%%%%%%%%%%%%%%%%%%%%%
\section{Model independent bounds on the scale of new physics}
\label{pheno}
%%%%%%%%%%%%%%%%%%%%%%%%%%%%%%%%%%%%%%%%%%%%%%%%%%%%%%%%%%%%%%%%%%%%%%%%%%%%%%%%%%%%%%
Model-independent studies of LFV processes have already been done in the literature \cite{Brignole:2004ah,Crivellin:2013hpa,Pruna:2014asa}.
In this section we collect the bounds on the Wilson coefficients of the Lagrangian of eq.~(\ref{Lag6}) and discuss their impact on our spurion analysis.
The main bounds on the coefficients of the dipole operators $O_{e\gamma}$ and $O_{eZ}$ are given in table 5, showing the results obtained in 
ref. \cite{Crivellin:2013hpa,Pruna:2014asa}  and derived from ref. \cite{Kitano:2002mt} in the case of $\mu^- {\rm Au} \to e^- {\rm Au}$.
They come from the present limits reported in table~\ref{nextgenexp}.
\begin{table}[t!]
\centering
\begin{tabular}{|c|c|c|}
\hline
LFV Process & Present Bound & Future Sensitivity  \\
\hline
$\mu \to e \gamma$ & $5.7 \times 10^{-13}$ \cite{Adam:2013mnn} & $\approx 6 \times 10^{-14}$ \cite{Baldini:2013ke}  \\
$\mu \to 3 e$ & $1.0 \times 10^{-12}$\cite{Bellgardt:1987du} & $\approx 10^{-16}$ \cite{Blondel:2013ia}\\
$\mu^- {\rm Au} \to e^- {\rm Au}$ & $7.0 \times 10^{-13}$ \cite{Bertl:2006up} & $ ? $  \\
$\mu^- {\rm Ti} \to e^- {\rm Ti}$ & $4.3 \times 10^{-12}$ \cite{Dohmen:1993mp} & $?$ \\
$\mu^- {\rm Al} \to e^- {\rm Al}$ & $-$  & $\approx 10^{-16}$ \cite{comet,mu2e} \\
$\tau \to e \gamma$ & $3.3 \times 10^{-8}$ \cite{Aubert:2009ag}& $\sim 10^{-8}-10^{-9}$ \cite{Hayasaka:2013dsa} \\
$\tau \to \mu \gamma$ & $4.4 \times 10^{-8}$ \cite{Aubert:2009ag}& $\sim 10^{-8}-10^{-9}$ \cite{Hayasaka:2013dsa} \\
$\tau \to 3 e$ & $2.7\times10^{-8}$\cite{Hayasaka:2010np} & $\sim 10^{-9}-10^{-10}$ \cite{Hayasaka:2013dsa}  \\
$\tau \to 3 \mu$ & $2.1\times10^{-8}$\cite{Hayasaka:2010np} & $\sim 10^{-9}-10^{-10}$ \cite{Hayasaka:2013dsa}  \\
\hline
Lepton EDM & Present Bound & Future Sensitivity  \\
\hline
$d_e ({\rm e~cm})$ & $8.7 \times 10^{-29}$ \cite{Baron:2013eja} & $?$ \\
$d_\mu ({\rm e~cm})$ & $1.9 \times 10^{-19}$ \cite{Bennett:2008dy} & $?$ \\
\hline
\end{tabular}
\caption{Present and future experimental sensitivities for relevant low-energy observables.}
\label{nextgenexp}
\end{table}
%
%\be
%BR(\mu\to e \gamma)<5.7\times 10^{-13}~~~~BR(\tau\to e \gamma)<3.3\times 10^{-8}~~~~BR(\tau\to \mu\gamma)<4.4\times 10^{-8}~~~.
%\ee
%
The amplitudes for these processes get a tree-level contribution from the off-diagonal elements of the electromagnetic dipole operator $O_{e\gamma}^{ij}$, while $O_{eZ}^{ij}$ 
contributes at one loop. Each bound has been derived by assuming a single non-vanishing Wilson coefficient at the time. This also applies to all the bounds discussed in this section.
Bounds on the coefficients $C_{e\gamma,Z}^{ji}$ are equal to the bounds on the coefficients $C_{e\gamma,Z}^{ij}$. 
\begin{table}[h!]
\centering
\begin{tabular}{| c | c | c | c |}
\hline
 & $|C|$ ($\Lambda=1$ TeV) & $\Lambda$ (TeV) ($|C| =1$) & LFV Process \\
 \hline
 $C_{e\gamma}^{\mu e}$& $2.5\times 10^{-10}$ & $6.3\times 10^{4}$ & $\mu\to e \gamma$\\
 \hline
 $C_{e\gamma}^{\mu e}$& $4.0\times 10^{-9}$ & $1.6\times 10^{4}$ & $\mu\to 3e$\\
 \hline
$C_{e\gamma}^{\mu e}$& $5.2\times 10^{-9}$ & $1.4\times 10^{4}$ & $\mu^- {\rm Au} \to e^- {\rm Au}$ \\
 \hline
$C_{e\gamma}^{\tau e}$& $2.4\times 10^{-6}$ & $6.5\times 10^{2}$ & $\tau\to e \gamma$\\
 \hline
 $C_{e\gamma}^{\tau\mu}$& $2.7\times 10^{-6}$ & $6.1\times 10^{2}$ & $\tau\to \mu \gamma$\\
 \hline
$C_{eZ}^{\mu e}$& $1.4\times 10^{-7}$ & $2.7\times 10^{3}$ & $\mu\to e \gamma$ {[\tt 1-loop]}\\
 \hline
$C_{eZ}^{\tau e}$& $1.3\times 10^{-3}$ & $28$ & $\tau\to e \gamma$ {[\tt 1-loop]}\\
 \hline
 $C_{eZ}^{\tau\mu}$& $1.5\times 10^{-3}$ & $26$ & $\tau\to \mu \gamma$ {[\tt 1-loop]}\\
 \hline
\end{tabular}
\label{tdip}
\caption{Bounds on off-diagonal Wilson coefficients $C_{e\gamma}^{ij}/\Lambda^2$ and $C_{eZ}^{ij}/\Lambda^2$ from ref. \cite{Crivellin:2013hpa,Pruna:2014asa}. 
The bounds from $\mu^- {\rm Au} \to e^- {\rm Au}$ have been derived from ref. \cite{Kitano:2002mt}.
In the second column we list the upper bound on $|C_{e\gamma,Z}^{ij}|$
assuming $\Lambda=1$ TeV, while in the third column we fix $|C_{e\gamma,Z}^{ij}|=1$ and we list the corresponding lower bound on $\Lambda$, in TeV. Bounds on the coefficients $C_{e\gamma,Z}^{ji}$ are equal to the bounds on the coefficients $C_{e\gamma,Z}^{ij}$. 
}
\end{table}
The diagonal elements of the dipole operators contribute to electric and magnetic dipole moments of the charged leptons. From the present bounds reported in table~\ref{nextgenexp}
%
%\be
%d_e<8.7\times 10^{-29}~e~cm~~~,~~~~~~~~~~~~~~d_\mu<1.8\times 10^{-19}~e~cm~~~,
%\ee 
%
we have  \cite{Crivellin:2013hpa}
\be
Im(C_{e\gamma}^{ee})\left(\frac{1~{\rm TeV}}{\Lambda}\right)^2<3.9 \times 10^{-12}~~~,~~~~~~~
Im(C_{e\gamma}^{\mu\mu})\left(\frac{1~{\rm TeV}}{\Lambda}\right)^2<8.4 \times 10^{-3}~~~.
\label{edm}
\ee
Given the current deviation $\Delta a_\mu=a_\mu^{EXP}-a_\mu^{SM}$ in the muon anomalous magnetic moment  $a_\mu = (g-2)_\mu /2$ \cite{Passera:2004bj,Jegerlehner:2009ry}
\be
\Delta a_\mu=(29\pm 9)\times 10^{-10}~~~,
\ee
we would need
\be
Re(C_{e\gamma}^{\mu\mu})\left(\frac{1~{\rm TeV}}{\Lambda}\right)^2=1.2\times 10^{-5}~~~
\label{gm2}
\ee
to account for the central value of the discrepancy.

%%%%%%%%%%%%%%%%%%%%%%%%%%%%%%%%%%%%%%%%%%%%%%%%%%%%%%%%%%%%%%%%%%%%%%%%%%%%%%%%%%%%%%%%%%%%%
% SCALAR OPERATORS
%%%%%%%%%%%%%%%%%%%%%%%%%%%%%%%%%%%%%%%%%%%%%%%%%%%%%%%%%%%%%%%%%%%%%%%%%%%%%%%%%%%%%%%%%%%%%
\begin{table}[h!]
\centering
\begin{tabular}{| c | c | c | c |}
\hline
 & $|C|$ ($\Lambda=1$ TeV) & $\Lambda$ (TeV) ( $|C| =1$) & LFV Process \\
  \hline
$C_{e\varphi}^{\mu e}$& $8.4\times 10^{-5}$ & $109$ & $\mu\to e \gamma$ {[\tt 2-loop]}\\
 \hline
$C_{e\varphi}^{\tau e}$& $0.33$ & $1.7$ & $\tau\to e \gamma$ {[\tt 2-loop]}\\
 \hline
 $C_{e\varphi}^{\tau\mu}$& $0.37$ & $1.6$ & $\tau\to \mu \gamma$ {[\tt 2-loop]}\\
 \hline
\end{tabular}
\label{tdip}
\caption{Bounds on off-diagonal Wilson coefficients $(C_{e\varphi})_{ij}$ from ref. \cite{Goudelis:2011un,Harnik:2012pb,Blankenburg:2012ex}. 
In the second column we list the upper bound on $|C_{e\varphi}^{ij}|$ assuming $\Lambda=1$ TeV, while in the third column we fix $|C_{e\varphi}^{ij}|=1$ 
and we list the corresponding lower bound on $\Lambda$, in TeV. Bounds on the coefficients $C_{e\varphi}^{ji}$ are equal to the bounds on the coefficients 
$C_{e\varphi}^{ij}$.}
\end{table}

Also the scalar operator $O_{e\varphi}$ is mostly bounded by the limits on radiative lepton decays \cite{Goudelis:2011un,Harnik:2012pb,Blankenburg:2012ex,Crivellin:2013hpa,Pruna:2014asa}. The scalar operator contributes to lepton masses and to higgs couplings with a different weight:
\be
{\cal L}_Y={\cal M}_{ij}
~\bar{e}_{Li}\tilde e_{Rj}+
{\cal Y}_{ij}
~h~\bar{e}_{Li}\tilde e_{Rj}+
h.c.~,
\ee
where \be
{\cal M}_{ij}=\left[
-y_{ij}^{SM}+\frac{v^2}{2\Lambda^2}(C_{e\varphi})_{ij}
\right]
~\frac{v}{\sqrt{2}}~~~,~~~~~~~~{\cal Y}_{ij}=\frac{1}{\sqrt{2}}\left[
-y_{ij}^{SM}+\frac{3v^2}{2\Lambda^2}(C_{e\varphi})_{ij}
\right]~~~,
\ee
and $y_{ij}^{SM}$ are the Standard Model Yukawa couplings. Radiative charged lepton decays constrain the off-diagonal elements of ${\cal Y}_{ij}$ in the basis where the mass matrix ${\cal M}_{ij}$ is diagonal.
To convert these constraints in bounds on the Wilson coefficients $(C_{e\varphi})_{ij}$  it is convenient to work in the basis where the SM couplings $y_{ij}^{SM}$ are diagonal and expand the unitary matrices that 
diagonalise ${\cal M}$ in powers of $v^2/\Lambda^2$. We found that, to first order in this parameter, the off-diagonal elements ${\cal Y}_{ij}$ $(i\ne j)$ in the lepton mass basis are given by:
\be
{\cal Y}_{ij}=\frac{v^2}{\sqrt{2}\Lambda^2}(C_{e\varphi})_{ij}~~~~~~~~~~~~~~(i\ne j)~~~. 
\ee
By translating the bounds on ${\cal Y}_{ij}$ $(i\ne j)$ given in ref. \cite{Goudelis:2011un,Harnik:2012pb,Blankenburg:2012ex} into bounds on $(C_{e\varphi})_{ij}$, we get the results shown in table 6.
The bounds on the coefficients $(C_{e\varphi})_{ji}$ are equal those on the coefficients $(C_{e\varphi})_{ij}$. These bounds are dominated by two-loop contributions of the corresponding operator to the radiative lepton decay, through Barr-Zee type diagrams, assuming that the top Yukawa coupling is as in the SM. One-loop contributions to charged lepton radiative decays and tree-level contributions to $\ell \to 3\ell^\prime$ decays lead to less severe bounds than the ones given in table 6.

%%%%%%%%%%%%%%%%%%%%%%%%%%%%%%%%%%%%%%%%%%%%%%%%%%%%%%%%%%%%%%%%%%%%%%%%%%%%%%%%%%%%%%%%%%%%%
% VECTOR OPERATORS
%%%%%%%%%%%%%%%%%%%%%%%%%%%%%%%%%%%%%%%%%%%%%%%%%%%%%%%%%%%%%%%%%%%%%%%%%%%%%%%%%%%%%%%%%%%%%
\begin{table}[h!]
\centering
\begin{tabular}{| c | c | c | c |}
\hline
 & $|C|$ ($\Lambda=1$ TeV) & $\Lambda$ (TeV) ($|C| =1$) & LFV Process \\
  \hline
$(C_{\vp l}^{(1,3)})_{\mu e}$& $3.7\times 10^{-5}$ & $164$ & $\mu\to 3 e$ \\
  \hline
$(C_{\vp l}^{(1,3)})_{\mu e}$& $5.0\times 10^{-6}$ & $447$ & $\mu^- {\rm Au} \to e^- {\rm Au}$ \\
  \hline
$(C_{\vp l}^{(1,3)})_{\tau e}$& $1.5\times 10^{-2}$ & $8.3$ & $\tau\to 3 e$\\
  \hline
$(C_{\vp l}^{(1,3)})_{\tau\mu}$& $1.2\times 10^{-2}$ & $9.0$ & $\tau \to 3\mu$\\
  \hline

$C_{\vp e}^{\mu e}$& $3.9\times 10^{-5}$ & $160$ & $\mu\to 3 e$\\
  \hline
  $C_{\vp e}^{\mu e}$& $5.0\times 10^{-6}$ & $447$ & $\mu^- {\rm Au} \to e^- {\rm Au}$ \\
  \hline
$C_{\vp e}^{\tau e}$& $1.5\times 10^{-2}$ & $8.1$ & $\tau\to 3 e$\\
  \hline
$C_{\vp e}^{\tau\mu}$& $1.3\times 10^{-2}$ & $8.8$ & $\tau \to 3\mu$\\
 \hline
\end{tabular}
\label{tdip}
\caption{Bounds on off-diagonal Wilson coefficients $(C_{\vp l}^{(1,3)})_{ij}$ and $C_{\vp e}^{ij}$ from ref. \cite{Crivellin:2013hpa,Pruna:2014asa}. 
The bounds from $\mu^- {\rm Au} \to e^- {\rm Au}$ have been derived from ref. \cite{Kitano:2002mt}.
In the second column we list the upper bound on the Wilson coefficients assuming $\Lambda=1$ TeV, while in the third column we set to unity the 
coefficients and we list the corresponding lower bound on $\Lambda$, in TeV.}
\end{table}

\begin{table}[h!]
\centering
\begin{tabular}{| c | c | c | c |}
\hline
 & $|C|$ ($\Lambda=1$ TeV) & $\Lambda$ (TeV) ($|C| =1$) & LFV Process \\
  \hline
$C_{ll,ee}^{\mu eee}$& $2.3\times 10^{-5}$ & $207$ & $\mu\to 3 e$\\
  \hline
$C_{ll,ee}^{e\tau ee}$& $9.2\times 10^{-3}$ & $10.4$ & $\tau\to 3 e$\\
  \hline
$C_{ll,ee}^{\mu\tau \mu\mu}$& $7.8\times 10^{-3}$ & $11.3$ & $\tau \to 3\mu$\\
  \hline
$C_{le}^{\mu eee, e e\mu e}$& $3.3\times 10^{-5}$ & $174$ & $\mu\to 3 e$\\
  \hline
  $C_{le}^{\mu \mu\mu e, e \mu \mu\mu}$& $2.1\times 10^{-4}$ & $69$ & $\mu\to e\gamma$ {[\tt 1-loop]} \\
  \hline
  $C_{le}^{\mu \tau\tau e, e \tau\tau\mu}$& $1.2\times 10^{-5}$ & $289$ & $\mu\to e\gamma$ {[\tt 1-loop]}\\
  \hline
$C_{le}^{e\tau ee,e e e\tau}$& $1.3\times 10^{-2}$ & $8.8$ & $\tau\to 3 e$\\
  \hline
$C_{le}^{\mu\tau\mu\mu,\mu\mu\mu\tau}$& $1.1\times 10^{-2}$ & $9.5$ & $\tau \to 3\mu$\\
  \hline
${C^{(u)}}_{\ell q}^{e\mu}$& $2.0 \times 10^{-6}$ & $707$ & $\mu^- {\rm Au} \to e^- {\rm Au}$ \\
  \hline
${C^{(d)}}_{\ell q}^{e\mu}$& $1.8 \times 10^{-6}$ & $745$ & $\mu^- {\rm Au} \to e^- {\rm Au}$ \\
  \hline
 $C_{eq}^{e\mu}$& $9.2 \times 10^{-7}$ & $1.0 \times 10^3$ & $\mu^- {\rm Au} \to e^- {\rm Au}$ \\
  \hline
$C_{\ell u,eu}^{e\mu}$& $2.0\times 10^{-6}$ & $707$ & $\mu^- {\rm Au} \to e^- {\rm Au}$ \\
  \hline
$C_{\ell d,ed}^{e\mu}$& $1.8\times 10^{-6}$ & $745$ & $\mu^- {\rm Au} \to e^- {\rm Au}$ \\
  \hline
%$C_{ll,le,ee}^{e\mu uu}$& $2.0\times 10^{-6}$ & $707$ & $\mu^- {\rm Au} \to e^- {\rm Au}$ \\
% \hline
%$C_{le}^{uue\mu}$& $2.0\times 10^{-6}$ & $707$ & $\mu^- {\rm Au} \to e^- {\rm Au}$ \\
%  \hline
%$C_{ll,le,ee}^{e\mu dd}$& $1.8\times 10^{-6}$ & $745$ & $\mu^- {\rm Au} \to e^- {\rm Au}$ \\
%  \hline
%  $C_{le}^{dd e\mu}$& $1.8\times 10^{-6}$ & $745$ & $\mu^- {\rm Au} \to e^- {\rm Au}$ \\
%  \hline
\end{tabular}
\label{tdip}
\caption{
Bounds on  coefficients $C_{ll,ee,le}^{ijkl}$ from ref. \cite{Crivellin:2013hpa}. 
The bounds from $\mu^- {\rm Au} \to e^- {\rm Au}$ have been derived from ref. \cite{Kitano:2002mt}.
In the second column we list the upper bound on the Wilson coefficients assuming $\Lambda=1$ TeV, 
while in the third column we set to unity the coefficients and we list the corresponding lower bound on $\Lambda$, in TeV.}
\end{table}

%%%%%%%%%%%%%%%%%%%%%%%%%%%%%%%%%%%%%%%%%%%%%%%%%%%%%%%%%%%%%
%                                                       Dominance of contact operators: (C_ll,le)^emu uu
%%%%%%%%%%%%%%%%%%%%%%%%%%%%%%%%%%%%%%%%%%%%%%%%%%%%%%%%%%%%%
%
%   R_conv = (2*0.0974 + 0.146)^2*(0.107^5/1000^4) / (13.07*6.6*10^-19) [ |C|^2 / Lambda^4 ] < 7*10^(-13)
%
%   |C|^2 < 7*10^(-13) / (2*0.0974 + 0.146)^2*(0.107^5/1000^4) / (13.07*6.6*10^-(19))
%
%  |C| < 3.1*10^-6 
%  Lambda > 565 TeV
%
%%%%%%%%%%%%%%%%%%%%%%%%%%%%%%%%%%%%%%%%%%%%%%%%%%%%%%%%%%%%%

%%%%%%%%%%%%%%%%%%%%%%%%%%%%%%%%%%%%%%%%%%%%%%%%%%%%%%%%%%%%%
%                                                       Dominance of contact operators: (C_ll,le)^emu dd
%%%%%%%%%%%%%%%%%%%%%%%%%%%%%%%%%%%%%%%%%%%%%%%%%%%%%%%%%%%%%
%
%   R_conv = (0.0974 + 2*0.146)^2*(0.107^5/1000^4) / (13.07*6.6*10^-19) [ |C|^2 / Lambda^4 ] < 7*10^(-13)
%
%   |C|^2 < 7*10^(-13) / (0.0974 + 2*0.146)^2*(0.107^5/1000^4) / (13.07*6.6*10^-(19))
%
%  |C| < 2.7*10^-6 
%  Lambda > 604 TeV
%
%%%%%%%%%%%%%%%%%%%%%%%%%%%%%%%%%%%%%%%%%%%%%%%%%%%%%%%%%%%%%

Coming to the vector operators $(Q_{\vp l}^{(1,3)})_{ij}$, $(Q_{\vp e})_{ij}$, they lead to lepton flavour violating $Z$ decays, but the corresponding limits on the 
Wilson coefficients, assuming $\Lambda=1$ TeV, are of order 10\% \cite{Crivellin:2013hpa}. Through one-loop diagrams they also contribute to radiative decays 
of the charged leptons \cite{Crivellin:2013hpa,Pruna:2014asa}.
It turns out that the most restrictive bounds come from the processes $\mu^- {\rm Au} \to e^- {\rm Au}$ and $\ell \to 3\ell^\prime$ whose branching ratios satisfy 
the experimental limits of table~\ref{nextgenexp}.
%
%\be
%BR(\mu\to 3 e)<1.0\times 10^{-12}~~~~BR(\tau\to 3 e)<2.7\times 10^{-8}~~~~BR(\tau\to 3\mu)<2.1\times 10^{-8}~~~.
%\ee
%
We collect the corresponding bounds in table 7.
%%%%%%%%%%%%%%%%%%%%%%%%%%%%%%%%%%%%%%%%%%%%%%%%%%%%%%%%%%%%%%%%%%%%%%%%%%%%%%%%%%%%%%%%%%%%%
% CONTACT OPERATORS
%%%%%%%%%%%%%%%%%%%%%%%%%%%%%%%%%%%%%%%%%%%%%%%%%%%%%%%%%%%%%%%%%%%%%%%%%%%%%%%%%%%%%%%%%%%%%
Also the contact operators can contribute to $\mu^- {\rm Au} \to e^- {\rm Au}$, $\ell \to 3\ell^\prime$ and, through one-loop diagrams, 
to the radiative decays of the charged leptons. The most significant bounds are given in table 8.

We can translate the bounds collected in tables 5-8 into limits on the masses $M$ and $m_0$ of our spurion analysis. To do this we should determine or 
make some assumptions on the parameters $c$, $\tilde c$, $X$, $\tilde X$, $Y_R^*$ and $Y_L^*$, through which we can express all the Wilson coefficients, 
as explained in section 2. By exploiting the flavour symmetry of our setup, we see that it is not restrictive to work in the basis where the mixing matrices $X$ 
and $\tilde X$ are diagonal, real and non-negative and we will adopt this choice, unless otherwise stated. The LO mass matrix of the charged fermions 
\be
m_l= X~ \yr~ \tilde X^\dagger \frac{v}{\sqrt{2}} ~~~,
\label{massLO_1}
\ee
is diagonalized by a bi-unitary transformation
\be
\left( L~ m_l~ R^\dagger \right)_{ij} = m_i \delta_{ij}\,.
\ee
Assuming that $\yr$ is anarchic, it is straightforward to find that
\be
L\approx
\left(
\begin{array}{ccc}
1&X_1/X_2& X_1/X_3\\
X_1/X_2&1&X_2/X_3\\
X_1/X_3&X_2/X_3&1
\end{array}
\right)~,~~~~~~~~~~~~~~
R\approx
\left(
\begin{array}{ccc}
1&\tilde{X}_1/\tilde{X}_2& \tilde{X}_1/\tilde{X}_3\\
\tilde{X}_1/\tilde{X}_2&1&\tilde{X}_2/\tilde{X}_3\\
\tilde{X}_1/\tilde{X}_3&\tilde{X}_2/\tilde{X}_3&1
\end{array}
\right)~,
\label{rotations}
\ee
where factors of order one have been omitted from the matrix elements. Similarly the lepton masses are approximately given by
\be
m_i\approx X_i (Y_R^*)_{ii} \tilde{X}_i \frac{v}{\sqrt{2}}~~~.
\label{mmm}
\ee
The most favorable scenario to minimize FCNC 
effects is realized when $X_i= \tilde X_i$, which we assume, for the time being. We have
\be
X_i=\tilde{X}_i=\left[\frac{\sqrt{2} m_i}{v (Y_R^*)_{ii}}\right]^{1/2}~~~,
\label{XXX}
\ee
which will be used in our estimates. 

If $c$, $\tilde c$, $Y_R^*$ and $Y_L^*$ are all generic matrices,  the dipole operator, eq.~(\ref{Dcom}), leads to the well-known limit
\be
 m_0~\frac{\langle c\rangle}{\langle Y\rangle}>33~ {\rm TeV},
 \label{bound21}
\ee
where $ \langle Y\rangle$ denotes a suitable average of the $Y_{L,R}^*$ couplings and similarly for $\langle c\rangle$.
Such a bound can be evaded only if special features of the flavour parameters $c$, $\tilde c$, $X$, $\tilde X$, $Y_R^*$ and $Y_L^*$ are adopted.
One possibility is to recover the framework of minimal flavour violation (MFV), as proposed in ref. \cite{Redi:2013pga}. This can be done by assuming 
$c$, $\tilde c$, $X$, $\tilde X$, $Y_R^*$ and $Y_L^*$ all proportional to the identity matrix, except either $X$ (right-handed compositeness) or $\tilde X$ (left-handed compositeness).
In this case, when neutrino masses are neglected, the only spurion that breaks the flavour symmetry becomes proportional to the charged lepton Yukawa couplings, exactly as in MFV.
A choice of basis where such a spurion is diagonal is always possible and LFV is only present when neutrino masses are turned on. 
In the latter case one can reconcile LFV with a scale of new physics close to the TeV scale, however we think it is interesting to explore other options allowing for a TeV scale $\Lambda$.

We start by taking $Y_L^*=0$. To guarantee the stability of this condition we should also assume either a perturbative regime, where $|\yr|\le 1$, or the hierarchy $M\gg m_0$, 
as explained in section 2.5. 
In this case the dipole operator is dominated by the contribution $(N_Y,N_X)=(1,2)$ (see table 1)
\be
\frac{C_{e\gamma}}{\Lambda^2}=\left\{
\begin{array}{c}
\dd\frac{X   \yr   (\tilde{c}^\dagger \tilde{c})^{-1}   \tilde{X}^\dagger}{m_0^2}\\
\dd\frac{X   (c c^\dagger)^{-1}   \yr   \tilde{X}^\dagger}{m_0^2}
\end{array}
\right.
\ee 
and the bound on $\mu\to e \gamma$ leads to $m_0\langle c\rangle>33$ TeV. As a next step we consider the case of $Y_L^*=0$ and universal heavy fermion masses, namely 
$c$ and $\tilde c$ proportional to the unit matrix. With this assumption the Wilson coefficients $(N_Y,N_X)=(1,2)$ of the dipole operator are diagonal in the mass basis, to LO.
This however is not the case, in general, for the coefficients $(N_Y,N_X)=(3,2)$, which are comparable in size to those of eq.~(\ref{Dcom}).
Therefore, even working in the limit of vanishing ``wrong'' Yukawa coupling $\yl$, and universal heavy fermion masses, we expect that with anarchic $Y_R^*$
the bound of eq.~(\ref{bound21}) still applies. 
To allow for a lower new physics scale $\Lambda$, we are lead to postulate that also the Yukawa couplings $\yr$ are universal. More precisely
we define a new scenario, that we call intermediate flavour violation (IFV), by the following assumption:
\begin{enumerate}
\item[] {\bf IFV} scenario:\hfill\break
{\em The Yukawa couplings $\yl$ are negligible and a  choice of basis exists where $c$, $\tilde c$ and $\yr$ are simultaneously proportional to the identity matrix.}
\end{enumerate}

From the discussion in section 2.5, we know that a non-vanishing $\yl$ can be generated. Since such a $\yl$ will only depend on  $c$, $\tilde c$ and $\yr$, it will be universal too. 
Admitting a universal $Y_L^*$ would not change our conclusions and, for simplicity, we keep the assumption $\yl=0$. Moreover, from now on we set to unity the matrices $c$ and 
$\tilde c$, absorbing their effect in the overall scale $m_0$ of the heavy fermion masses. The remaining coupling $\yr$ is described by a single parameter $y$:
\be
\yr=y~\mbox{l\hspace{-0.55em}1}~~~.
\ee
At variance with MFV, in IFV the sources of flavour breaking are both $X$ and $\tilde X$. In the basis defining IFV the mixing matrices $X$ and $\tilde X$ are not diagonal,  in general. 
They are generic complex matrices. At the LO, the lepton mass matrix is proportional to the product $X \tilde X^\dagger$. When we move to the charged lepton mass basis, in general, 
only the product $X\tilde X^\dagger$ becomes diagonal, not $X$ and $\tilde X$ individually and they both can lead to LFV. 

By exploiting the freedom related to our flavour symmetry, we can provide an alternative, but equivalent, description of the IFV scenario.  We can choose a basis where $X$ and $\tilde X$ 
are diagonal, real and non-negative. 
In this case by means of the symmetry transformations of eqs.~(\ref{tr1}-\ref{tr2}) we can still maintain $c$ and $\tilde c$ proportional to the identity, but the matrix $\yr$ becomes a 
generic unitary matrix. In such a basis the lepton mass matrix $X \yr \tilde X^\dagger$ is non-diagonal and LFV is now ascribed to the interplay between $\yr$ and $X$, $\tilde X$. 

Within IFV the special case $X=\tilde X$,
which was previously assumed to minimize FCNC effects, forbids any LFV effect. Indeed, if $X=\tilde X$, there exist a basis where both $X$ and the mass matrix $X \yr \tilde X^\dagger$ are
diagonal at the same time. As long as neutrino masses are neglected there is no source of LFV and all the Wilson coefficients are diagonal in flavour space. The only difference with respect 
to MFV is that the spurion $X$ is not proportional to the charged lepton Yukawa couplings, but to their square root. Therefore in discussing IFV we consider the general case where $X$ and 
$\tilde X$ are not equal.

In IFV no LFV is generated from the Wilson coefficients of the dipole and scalar operators with $N_X=2$, to LO. 
These coefficients are of the type $XA \tilde X^\dagger$, where $A$ is any combination of $c$, $\tilde c$ and $Y_R^*$, and they are automatically aligned to the charged lepton mass matrix, 
to LO. To estimate the other Wilson coefficients we should specify the choice of $X$ and $\tilde X$. Rather than scanning for the most general possibility, here we provide an example.
We assume that, in the basis where $X$ and $\tilde X$ are diagonal, their elements are given, up to coefficients of order one, by
\be
X=\frac{1}{\sqrt{y}}{\tt diag}(\lambda^4,\lambda^3,\lambda^3)~~~,~~~~~~~\tilde X=\frac{1}{\sqrt{y}}{\tt diag}(\lambda^5,\lambda^2,1)~~~,
\label{XX}
\ee
where $\lambda\approx 0.22$ is a small parameter of the order of the Cabibbo angle. The choice of elements of  nearly the same order of magnitude for $X$ is motivated by a possible role that
the matrix $X$ can play in describing large lepton mixing angles, once neutrino masses are turned on \cite{Feruglio:2015jfa}. The value $\lambda^3\div \lambda^4$
is chosen here for convenience, to adequately suppress
LFV. The choice for the elements of $\tilde X$ is fixed, at the level of orders of magnitude, by the relation (\ref{mmm}). The transformations  needed
to diagonalize the lepton mass matrix are given in eq.~(\ref{rotations}). 

Now we can complete our discussion concerning the Wilson  coefficients that are bilinear in the spurions $X$, $\tilde X$. Coefficients $(N_X,N_Y)=(2,2)$ of the vector operators are
of the type $X A X^\dagger$ or $\tilde X\tilde A \tilde X^\dagger$ where $A$ depends on $c$, $\tilde c$ and $Y_R^*$. From tables 2 and 3, we see that $\yr$ enters $A$ only in the combinations 
$\yr \yr^\dagger$ or $\yr^\dagger \yr$.
Therefore, up to an overall coefficient of order one, in the basis where $X$ and $\tilde X$ are diagonal and real we have
\be
(C_{\varphi l}^{(1,3)})_{ij}= y^2 X^2_i~\delta_{ij}~~~,~~~~~~~~~~~~~~~~~(C_{\varphi e})_{ij}=y^2 \tilde X^2_i~\delta_{ij}~~~.
\ee 
When we move to the charged lepton mass basis, we get:
\be
(C_{\varphi l}^{(1,3)})_{ij}\approx y^2 X_i X_j~~~,~~~~~~~~~~~~~~~~~(C_{\varphi e})_{ij}\approx y^2 \tilde X_i \tilde X_j~~~.
\ee 
The most stringent limits of table 7 arise from $\mu^- {\rm Au} \to e^- {\rm Au}$ and require
 \be
 y^2\frac{X_\mu X_e}{\Lambda^2}\approx y^2\frac{\tilde X_\mu \tilde X_e}{\Lambda^2}\approx y\frac{\lambda^7}{\Lambda^2} < 5.0\times 10^{-6}~{\rm TeV}^{-2}~~~,
\ee
which translate, respectively, into $\Lambda>\sqrt{y}~2.2$ TeV.
%\be
%y^2\frac{X_\mu X_e}{\Lambda^2}\approx y\frac{\lambda^6}{\Lambda^2} < 3.7\times 10^{-5}~{\rm TeV}^{-2}~~~,~~~~~~~y^2\frac{\tilde X_\mu \tilde X_e}{\Lambda^2}\approx y%\frac{\lambda^7}{\Lambda^2} < 3.9\times 10^{-5}~{\rm TeV}^{-2}~~~,
%\ee
%which translate, respectively, into $\Lambda>\sqrt{y}~1.8$ TeV and $\Lambda>\sqrt{y}~0.8$ TeV. 
Here and in the rest of this section $\Lambda$ stands for either $M$ or $m_0$. The limit on the decay $\tau\to 3 \mu$ gives rise to a similar bound:
\be
y^2\frac{\tilde X_\tau \tilde X_\mu}{\Lambda^2}\approx y\frac{\lambda^2}{\Lambda^2} < 1.3\times 10^{-2}~{\rm TeV}^{-2}~~~,
\ee
resulting in $\Lambda>\sqrt{y}~1.9$ TeV. Coefficients of the type $(N_X,N_Y)=(2,0)$ can arise for the contact operators of the type $llqq$ and we expect, in the mass basis: 
\be
(C_{\ell q}^{(u,d)})_{ij},(C_{\ell u,d})_{ij}\approx  X_i X_j~~~,~~~~~~~~~~~~~~~~~(C_{eq})_{ij},(C_{eu,d})_{ij}\approx \tilde X_i \tilde X_j~~~.
\ee 
The limits of table 8 from $\mu^- {\rm Au} \to e^- {\rm Au}$ require
\be
\frac{X_\mu X_e}{\Lambda^2}\approx \frac{\tilde X_\mu \tilde X_e}{\Lambda^2}\approx \frac{\lambda^7}{y\Lambda^2} < 1.0 \times 10^{-6}~{\rm TeV}^{-2}~~~,
\ee
and we get $\Lambda>5.0/\sqrt{y}~$ TeV, respectively.

We are left with the coefficients that are quadrilinear in $X$, $\tilde X$. In the IFV scenario the off-diagonal elements of the dipole operator $(Q_{e\gamma})_{ij}$ are dominated by
the coefficients  
\be
(C_{e\gamma})_{ij}=\frac{1}{16\pi^2}~(X \yr   \tilde{X}^\dagger   \tilde{X}   \tilde{X}^\dagger)_{ij}~~~,~~~~~~~~(C_{e\gamma})_{ij}=\frac{1}{16\pi^2}~(X   X^\dagger   X   \yr   \tilde{X}^\dagger)_{ij}~~~,
\label{quadr}
\ee
which, in the lepton mass basis are approximately given by
\be
(C_{e\gamma})_{ij}\approx\frac{1}{16\pi^2}~\frac{\sqrt{2} m_i}{v}  \tilde{X}_i   \tilde{X}_j~~~,~~~~~~~~(C_{e\gamma})_{ij} \approx \frac{1}{16\pi^2}~X_i   X_j   \frac{\sqrt{2} m_j}{v}~~~.
\ee
The limit
\be
\frac{(C_{e\gamma})_{e\mu,\mu e}}{\Lambda^2}<2.5\times 10^{-10}~ {\rm TeV}^{-2}  
\ee
corresponds to a bound $\Lambda>(0.6/\sqrt{y})$ TeV. Similarly, from
\be
\frac{(C_{e\gamma})_{\tau\mu,\mu\tau}}{\Lambda^2}<2.7\times 10^{-6}~ {\rm TeV}^{-2}  
\ee
we get $\Lambda>(1.1/\sqrt{y})$ TeV. We have checked that all the other limits listed in table 5-8 do not lead to more restrictive bounds on $\Lambda$. 
For instance, the bound on the contact operators
\be
\frac{C_{le}^{e\tau\tau\mu}}{\Lambda^2}<1.2\times 10^{-5}~ {\rm TeV}^{-2} ~,
\ee
produces the bound $\Lambda>(0.3/y)$ TeV. The bounds from the vector operators scale as $\sqrt{y}$, while those from the dipole or contact operators scale as $1/y$ or $1/\sqrt{y}$. 
Therefore a new physics scale $\Lambda$ around few TeV is still acceptable provided the Yukawa coupling $y$ is close to one.

A new physics scale $\Lambda\approx 1$ TeV is too large to allow, in the present framework, for an explanation of the central value of the $\Delta a _\mu$ anomaly. 
Indeed, in the lepton mass basis we have
\be
Re(C_{e\gamma}^{\mu\mu})\approx \frac{1}{16\pi^2}\frac{\sqrt{2} m_\mu}{v}=3.8\times 10^{-6}
\ee
and to match the required value, eq.~(\ref{gm2}), we need $\Lambda=0.56$ TeV. Since the contribution to $\Delta a _\mu$ scales with the inverse square of $\Lambda$, 
choosing $\Lambda=1$ TeV gives $\Delta a _\mu=9\times 10^{-10}$, less than one third of the central value of the current anomaly.
Concerning the electron EDM, we assume as in MFV that the sources of CP violation
and LFV are the same. Since the Wilson coefficients of the dipole operators with $N_X=2$ are aligned in flavour space with the mass operator, we identify in (\ref{quadr})
the dominant coefficients that can contain non trivial phases. We estimate
\be
Im(C_{e\gamma}^{ee})\approx \frac{1}{16\pi^2}~X_e X_e~\frac{\sqrt{2} m_e}{v} ~,
\ee
which, for $\Lambda>(0.16/\sqrt{y})$ TeV, respects the bound of eq.~(\ref{edm}).
 
In summary, even in the limit of ``wrong'' Yukawa coupling negligibly small, an anarchic $\yr$ requires a scale of new physics well above 10 TeV. 
One way to lower this bound consists in mimicking  the case of MFV, where there is a single non-universal spurion, either $X$ or $\tilde X$. In this case, 
when neutrino masses are neglected, LFV is absent. By discussing a large class of possible LFV effects, we have shown that MFV is not the only possibility 
to reconcile LFV bounds with a new physics scale around few TeV. In the example analyzed here, both $X$ and $\tilde X$ are non-universal and represent 
potential sources of LFV, which is allowed also in the limit of vanishing neutrino masses. The corresponding bounds on $\Lambda$ approach the TeV scale, 
provided the Yukawa couplings $\yr$ are close to one.

%%%%%%%%%%%%%%%%%%%%%%%%%%%%%%%%%%%%%%%%%%%%%%%%%%%%%%%%%%%%%%%%%%%%%%%%%%%%%%%%%
\section{Two-site model}
\label{sec4}
%%%%%%%%%%%%%%%%%%%%%%%%%%%%%%%%%%%%%%%%%%%%%%%%%%%%%%%%%%%%%%%%%%%%%%%%%%%%%%%%%

In the previous sections, we focused on PC scenarios for charged leptons from a general perspective, exploiting a spurionic analysis. 
Now, instead, we consider a specific simplified composite Higgs model, the so-called two-site model~\cite{Contino:2006nn}. Its relevant features are:
\begin{itemize}
\item[(i)] The gauge group is $G^{\rm{gauge}} =  G^{\elem}\times G^{\comp}$ where
\begin{align}
G^\elem =  [SU(2)_L \times U(1)_Y]^{\elem} \; ,\qquad G^\comp =  [SU(2)_L \times SU(2)_R \times U(1)_X]^{\comp} \; .
\end{align}
The group for $G^\comp$ has been chosen in order to provide a custodial symmetry.
 In table \ref{tab: CH gauge field} is summarized our notation for generators, boson fields and coupling constants associated with each simple subgroup.
 
PC is assumed to arise from an unknown dynamical spontaneous symmetry breaking mechanism, that take place at higher energies and that breaks $G^{\elem}\times G^{\comp}$ into the diagonal group (which can be recognized as the electroweak gauge group). This spontaneous symmetry breaking mechanism, that in turns triggers the PC scenario, will be effectively described by linear couplings between elementary and composite bosons.
    \begin{table}[h!]  % gauge fields in CH
    \centering
    \setlength{\extrarowheight}{5pt}
    \begin{tabular} {| c | c | c | c | c |} 
    \hline
    & Subgroup & Generator(s) & Field(s) & Coupling \\
    \hline
    \multirow{2}{*}{el} & $SU(2)_L$ & $T^{aL,\;\elem}, \quad a=1,2,3$ & $W^{a,\;\elem}_\mu$ & $g_1^{\elem}$ \\
    \cline{2-5}
    & $U(1)_Y$ & $Y$ & $B^\elem_\mu$ & $g_2^{\elem}$ \\
    \hline
    \multirow{4}{*}{comp} &  $SU(2)_L$ & $T^{aL,\;\comp}, \quad a=1,2,3$ & $W^{a,\;\comp}_\mu$ & $g_1^{\comp}$ \\
    \cline{2-5}
    & \multirow{3}{*}{$SU(2)_R \times U(1)_X$} & $\sqrt{\frac{3}{5}}(T^{3R,\;\comp}+\sqrt{\frac{2}{3}}X)$ & $B^\comp_\mu$ & $g_2^{\comp}$ \\
    \cline{3-5}
    & & $T^{1R,\;\comp},\; T^{2R,\;\comp}$ & $\tilde{W}^{1,2,\;\comp}_\mu$ & $g_2^{\comp}$ \\
    \cline{3-5}
    & & $\sqrt{\frac{3}{5}}(T^{3R,\;\comp}-\sqrt{\frac{2}{3}}X)$ & $\tilde{B}^\comp_\mu$ & $g_2^{\comp}$ \\
    \hline
    \end{tabular} 
    \caption{ Gauge subgroups and their associated generators, boson fields and couplings. The normalization of the $B^\comp$ and $\tilde{B}^\comp$ generators has been chosen to match the $SO(10)$ GUT normalization of the hypercharge, $Y_{GUT}=\sqrt{\frac{3}{5}}Y$.  \label{tab: CH gauge field}}
    \end{table}

\item[(ii)] The fermionic sector of the model includes three families of chiral fermions charged under $G^\elem$
and three families of vector-like fermions charged under $G^\comp$. 
PC for fermions is realized through linear mass-mixing terms between elementary and composite fermions.

\item[(iii)] The Higgs sector consists of a real bidoublet $(\tilde{\varphi},\varphi)$ charged under $[SU(2)_L \times SU(2)_R]^\comp$, 
that will be identified with the composite Higgs field, interacting only with the composite fermions.

Table \ref{tab: CH particle content} summarises the quantum numbers for fermions and Higgs doublet.

    \begin{table}[t]  % particle content in CH
    \centering
    \setlength{\extrarowheight}{5pt}
    \begin{tabular} {|c| c | c | c | c | c |} 
    \hline
    & \multicolumn{2}{c|}{Elementary} & \multicolumn{3}{c|}{Composite} \\
    \cline{2-6}
     & $SU(2)_L$ & $U(1)_Y$ & $SU(2)_L$ & $SU(2)_R$ & $U(1)_X$ \\
    \hline
    $\ell_{Li}$	 & {\bf 2} & $-\frac{1}{2}$ & {\bf 1} & {\bf 1} & $0$ \\
    \hline
    $\tilde{e}_{Ri}$& {\bf 1} & $-1$ 		& {\bf 1} & {\bf 1} & 0 \\
    \hline
    $L_i$			 & {\bf 1} & 0 			& {\bf 2} & {\bf 1} & $-\frac{1}{2} \cdot \sqrt{\frac{3}{2}}$ \\
    \hline
    $\tilde{E}_i$	 & {\bf 1} & 0 			& {\bf 1} & {\bf 1} & $-1 \cdot \sqrt{\frac{3}{2}}$ \\
    \hline
    $(\tilde{\varphi},\varphi)$ & {\bf 1} & 0 & {\bf 2} & {\bf 2} & 0 \\
    \hline
    \end{tabular} 
    \caption{ Particle content and quantum numbers of the two-site minimal model. The index $i=1,2,3$ runs over three families for each representation. 
    Lower case letters denote elementary fields, capital letters denote composite fields. The `tilde' apex denotes $SU(2)_L$ singlets, in order to distinguish 
    them from the doublets. \label{tab: CH particle content}} 
    \end{table}

\end{itemize}
We are ready now to introduce the Lagrangian of the two-site model:
\begin{equation}
\lag{}  = \lag\elem + \lag\comp + \lag{mix} \; ,
\label{eq: CH lag tot}
\end{equation}
with
\begin{align}
\lag\elem  = {} &  -\frac{1}{4}(F^a_{\mu\nu})^2 + \sum\limits_{i=1}^{3} 
\left( \bar{\ell}_{Li}i\slashed{D}\ell_{Li} + \bar{\tilde{e}}_{Ri}i\slashed{D}\tilde{e}_{Ri} \right) \; ,  
\label{eq: CH el lag} \\
\lag\comp  = {} & -\frac{1}{4}(\rho^a_{\mu\nu})^2  + 
      |D_\mu \varphi|^2 - V(\varphi) \nonumber \\*      
   & + \sum\limits_{i=1}^{3} \left( \bar{L}_{i}(i\slashed{D}-m_i)L_{i}+ 
      \bar{\tilde{E}}_{i}(i\slashed{D}-\tilde{m}_i)\tilde{E}_{i} \right) \nonumber \\*
   & - \sum\limits_{i,j=1}^{3} \left( \yl_{ij}\bar{L}_{Ri}\varphi\tilde{E}_{Lj} +
      \yr_{ij}\bar{L}_{Li}\varphi\tilde{E}_{Rj} \right) + \hc \; , 
      \label{eq: CH comp lag} \\
\lag{\textrm{mix}}  = {}  & \frac{M_\ast^2}{2}(\rho^a_\mu)^2- M_{*}^2 \frac{g^\elem}{g^\comp} A^{\mu} \rho^{*}_{\mu} + 
      \frac{M_{*}^2}{2} \left( \frac{g^\elem}{g^\comp} A_{\mu} \right)^2 \nonumber \\*
   & - \sum\limits_{i,j=1}^{3} \left( \Delta_{ij}\bar\ell_{Li} L_{Rj} + 
      \tilde{\Delta}_{ij} \bar{\tilde{e}}_{Ri} \tilde{E}_{Lj} \right) + \hc \; , 
      \label{eq: CH mix lag}
\end{align}
where lower (upper) case letters denote elementary (composite) fields and the `tilde' denotes $SU(2)_L$ singlets.
Moreover, among the heavy vector bosons, we distinguish between those mixing with the SM gauge bosons, 
$\rho^*_\mu = \{ W^*_\mu, B^*_\mu \}$, and those that do not, $\tilde\rho_\mu = \{ \tilde W_\mu, \tilde B_\mu \}$.

In $\lag{\textrm{el}}$ we can recognize a SM-like Lagrangian (with $F^a_{\mu \nu}$ collectively denoting the field strength tensors for the elementary gauge bosons)
without the Higgs doublet, that is now a composite particle. 
$\lag{\textrm{comp}}$ contains mass terms for composite leptons, kinetic terms for composite leptons and bosons (we have collectively denoted the field strength tensors with $\rho^a_{\mu\nu}$), 
their interactions, the Higgs sector and Yukawa interactions. 
Finally, $\lag{\textrm{mix}}$ contains linear mass-mixing terms among elementary and composite particles. We have also included in $\lag{\textrm{mix}}$ a mass terms for composite vector bosons.
$\lag{\textrm{mix}}$ explicitly breaks $G^{\elem}\times G^{\comp}$ down to the diagonal subgroup, and effectively reproduces the mechanism of PC.

From the above Lagrangian we can easily recognize the flavour symmetry group of eq.~(\ref{eq4: flavor group}) and convince ourselves that promoting 
$\yl$, $\yr$, $m$, $\tilde m$, $\Delta$ and $\tilde\Delta$ to spurions with the transformation properties of eq.~(3)  we actually restore $G_f$.

The Lagrangians of eqs.~(\ref{eq: CH el lag}), (\ref{eq: CH comp lag}), (\ref{eq: CH mix lag}) are expressed in the elementary/composite basis. 
For the bosons, in order to switch to the mass basis (before EWSB), we diagonalize the mass-mixings in $\lag{\textrm{mix}}$ by the following field transformations:
\begin{equation}
\label{rotations_mix_bos}
\begin{aligned}
   \left( \begin{array}{c} A_\mu \\ \rho^*_\mu \end{array} \right) & \rightarrow
      \left( \begin{array}{c c}
               \cos\theta & -\sin\theta \\
               \sin\theta & \cos\theta
      \end{array} \right)
      \left( \begin{array}{c} A_\mu \\ \rho^*_\mu \end{array} \right)  &~~~~~~
       \tan\theta = \frac{g^\elem}{g^\comp}\; ,
\\
      \tilde\rho_\mu  & \rightarrow  \tilde\rho_\mu  \; ,
\end{aligned}
\end{equation}
At this stage the fields $ A_\mu$ are massless, while $\rho^*_\mu$ and $\tilde\rho_\mu$ have masses of order $M^*$.
Taking into account EWSB, the diagonalization of the mass terms becomes rather involved. To a good approximation, eq.~(\ref{rotations_mix_bos}) still diagonalizes the boson fields, while the mass matrices for the fermions read:
\begin{align} \label{eq: mass matrix leptons}
  \begin{array}{r c c c l l}
    & \tilde{e}_R 			& E_R 								& \tilde{E}_R				&& 			\\
    & & & & \\ [-8pt]
    $\ldelim({3}{45pt}[$M_\cE = $]$
    & 0						&\Delta								& 0							&
    $\rdelim){3}{5pt}$
    & e_L		\\
    & 0						& m									& \frac{v}{\sqrt{2}} \yr 	&& E_L	 \;   \\
    & \tilde{\Delta}^\dagger 	& \frac{v}{\sqrt{2}} \yl^\dagger 	& \tilde{m} 			&& \tilde{E}_L\\
  \end{array}
  ,
&&
  \begin{array}{r c l l}
    & N_R 			&& 			\\
    & & & \\ [-8pt]
    $\ldelim({2}{45pt}[$M_\cN = $]$
    & \Delta	&
    $\rdelim){2}{5pt}$
    & \nu_L		\\
    & m && N_L	 \;
  \end{array}    .
\end{align}
With this notation, one can then perform a suitable rotation that brings to the mass basis:
\begin{equation}
\label{eq: compact lepton basis}
\begin{aligned}
\left( 
  \begin{array}{c}
    e_L \\ E_L \\ \tilde E_L 
  \end{array}
\right)  & \rightarrow
V_L^\dagger \left( 
  \begin{array}{c}
    e_L \\ E_L \\ \tilde E_L 
  \end{array}
\right) \; ,&~~~~
\left( 
  \begin{array}{c}
    \tilde e_R \\ E_R \\ \tilde E_R 
  \end{array}
\right) & \rightarrow 
V_R^\dagger \left( 
  \begin{array}{c}
    \tilde e_R \\ E_R \\ \tilde E_R 
  \end{array}
\right) \; ,
\\
\left( 
  \begin{array}{c}
    \nu_L \\ N_L 
  \end{array}
\right) & \rightarrow 
U_L^\dagger \left( 
  \begin{array}{c}
    \nu_L \\ N_L 
  \end{array}
\right) \; ,&
N_R & \rightarrow U_R N_R \; ,
\end{aligned}
\end{equation}
where $V_{L,R}, U_{L,R}$ are defined in order to have $V_L^\dagger M_\cE V_R$ and $U_L^\dagger M_\cN U_R$  diagonal.

Using rotations (\ref{rotations_mix_bos}), (\ref{eq: compact lepton basis}) one can recast the Lagrangian (\ref{eq: CH lag tot}) in terms of lepton mass eigenstates. 
Using an approximate expression for the rotation matrices $V_{L,R}, U_{L,R}$, assuming universal masses for the heavy leptons, ($m_i=m,\tilde m_i=\tilde m$),
and retaining the leading terms relevant for FC neutral-current (FCNC), one gets:
\begin{align} \label{eq: Lagrangian explicit}
\lag{FCNC} = {} &
				\frac{h}{\sqrt{2}} \left(  (X \yr)_{ij} \bar e_{Li} \tilde E_{Rj} 
				+ (\yr \Xt^\dagger)_{ij} \bar E_{Li} \tilde e_{Rj}  
				\right) + \hc 
				\nonumber \\
&
				\frac{h}{\sqrt{2}} 
				\left( 
				\frac{v^2}{2\mt^2} X \yr \yr^\dagger X^\dagger y^{\mysmall SM}_{\ell} \!+\! 
				\frac{v^2}{2m^2} y^{\mysmall SM}_{\ell} \Xt \yr^\dagger \yr \Xt^\dagger \!-\! 
				\frac{v^2}{m \tilde m} X \yr \yl^\dagger \yr \Xt^\dagger\!
				\right)_{\!\! ij} \!\!
				\bar e_{Li} \tilde e_{Rj} + \hc 
				\nonumber \\
&
				+ \frac{1}{2\sqrt{2}}\frac{g}{c_W} Z_\mu \left(
				\frac{v}{\tilde m} (X \yr)_{ij} \bar e_{Li} \gamma^\mu \tilde E_{Lj} 
				- \frac{v}{m} (\yr \Xt^\dagger)_{ij} \bar{ E}_{Ri} \gamma^\mu \tilde e_{Rj} 
				\right) + \hc
				\nonumber \\
&
				+ \frac{1}{4} \frac{g}{c_W} Z_\mu \left(
				\frac{v^2}{\tilde m^2} (X \yr \yr^\dagger X^\dagger)_{ij} \bar e_{Li} \gamma^\mu  e_{Lj}
				- \frac{v^2}{m^2} (\Xt \yr^\dagger \yr \Xt^\dagger)_{ij} \bar{\tilde e}_{Ri} \gamma^\mu \tilde e_{Rj}
				\right)
				\nonumber \\
&				
				- \frac{g_*}{2}  (B^*_\mu -\tilde B_\mu + W_\mu^{*, \, 3}) \,
				 (X X^\dagger)_{ij} \bar e_{Li} \gamma^\mu e_{Lj}
				+ g_*  (B^*_\mu -\tilde B_\mu) \,
				(\Xt \Xt^\dagger)_{ij} \bar{\tilde e} _{Ri}\gamma^\mu \tilde e_{Rj}
				\nonumber \\
&
				+ \frac{g_*}{2}  (B^*_\mu -\tilde B_\mu +W_\mu^{*, \, 3}) \,				
				 X_{ij} \bar e_{Li} \gamma^\mu E_{Lj}
				+ g_*  (B^*_\mu -\tilde B_\mu ) \, \Xt_{ij} \bar{\tilde e}_{Ri} \gamma^\mu \tilde E_{Rj}  +\hc 
\end{align}

From Lagrangian (\ref{eq: Lagrangian explicit}) one can easily read some of the most relevant features of FCNC in this class of models:
\begin{itemize}
\item The first line of eq.~(\ref{eq: Lagrangian explicit}) accounts for FCNC interactions among light- and heavy-leptons and the Higgs.
They contribute to low-energy processes such as $\ell_i\to\ell_j\gamma$ at loop-level. 
\item The second line of eq.~(\ref{eq: Lagrangian explicit}) refers to FCNC interactions among light-leptons and the Higgs.
As we know from previous sections, FCNC effects can arise in this case through the dim-6 operator 
$(Q_{e\vp})_{ij}=(\vp^\dag \vp)(\bar \ell_{Li} \tilde e_{Rj} \vp)$ and this explains the appearance of the factors $v^2/\tilde m^2$, 
$v^2/m^2$ and  $v^2/m\tilde m$ after EW symmetry breaking.
Such Yukawa interactions generate FCNC Higgs decays like $\varphi\to \tau\mu$, tree level contributions to processes sensitive 
to four fermion operators such as $\mu\to 3e$ and $\mu N \to e N$ and loop induced effects to $\ell_i \to \ell_j\gamma$.
\item The third line of eq.~(\ref{eq: Lagrangian explicit}) describes the interactions among light- and heavy-leptons and the Z boson. 
Since the operators $\bar e_{Li} \slashed{Z} \tilde E_{Lj}$ and $\bar{E}_{Ri} \slashed{Z} \tilde e_{Rj}$ are not $SU(2)_L$ invariant, 
they can be generated only after the EW symmetry breaking through the $SU(2)_L$ invariant dim-5 operators 
$\bar\ell_{Li} \slashed{Z} \tilde E_{Lj}\varphi$ and $\bar{L}_{Ri} \slashed{Z} \tilde e_{Rj}\varphi$. This explains the factors of $v/\tilde m$ 
and $v/m$. These interactions will contribute to $\ell_i\to\ell_j\gamma$ at loop-level.
\item The fourth line of eq.~(\ref{eq: Lagrangian explicit}) contains FCNC interactions among light-leptons and the Z bosons. 
As discussed in previous sections, such effects can arise by means of dim-6 operators bilinear in the Higgs doublet like
$(Q_{\vp e})_{ij}=(\vpj)(\bar{\tilde e}_{Ri} \gamma^\mu \tilde e_{Rj})$. As a result, after EW symmetry breaking, we generate the operators 
$\bar e_{Li} \slashed{Z} e_{Lj}$ and $\bar{\tilde e}_{Ri} \slashed{Z} \tilde e_{Rj}$ which are suppressed by $v^2/m^2$ and $v^2/\tilde m^2$ 
factors, respectively.
The leading effects induced by these terms are the tree level FCNC decay modes $Z\to \ell_i \ell_j$ as well as 
$\ell_i\to 3\ell_j$ and $\mu N \to e N$. 
\item The fifth line of eq.~(\ref{eq: Lagrangian explicit}) describes $SU(2)_L$ invariant interactions between heavy gauge 
bosons and light-fermions. Also in this case we can induce FCNC processes at tree level such as $\ell_i\to 3\ell_j$ and
$\mu N \to e N$ and loop induced effects to $\ell_i \to \ell_j\gamma$.
\item The sixth line of eq.~(\ref{eq: Lagrangian explicit}) refers to FC interactions among heavy gauge bosons, 
heavy- and light-leptons. Effects to low-energy observables are induced by the loop exchange of
heavy gauge bosons and leptons. If the heavy leptons of different generations were degenerate, 
FCNC effects would vanish according to the GIM-mechanism. However, the GIM cancellation is 
broken by non-universal mass splittings of order $X^\dagger X$ and ${\tilde X}^\dagger {\tilde X}$.
The latter point has been overlooked in the literature so far.
\end{itemize}
Concerning the flavor structure of the various interaction terms, we remember that each 
Higgs electroweak doublet $\varphi$ or its vacuum expectation value $v$ occur accompanied by 
a power of $\yr$ or $\yl$, while the lepton fields $e_{L}$ and $\tilde e_{R}$ come with $X$ and 
$\tilde X$, respectively.

Before discussing LFV in this model we comment on the issues of renormalizability, gauge-invariance and UV sensitivity.
The two-site model is a non-renormalizable effective theory and therefore it is valid only up to energies of the order of an UV cut-off $\Lambda_{UV}$.
Such a cut-off has been estimated in ref. \cite{Contino:2006nn}, where it has been found that $\Lambda_{UV}\approx 8 \pi M^*/g^*$, by analysing self-interactions in the composite sector.
The non-renormalizability is a consequence of the explicit breaking of the gauge symmetry $G^{\elem}\times G^{\comp}$, by the heavy vector boson masses, by the Yukawa couplings
$\yl$ and $\yr$ in the composite sector and by the mixing between elementary and composite fermions $\Delta$ and $\tilde \Delta$.

This explicit breaking raises the question of the reliability and consistency of our results. We can always 
promote our effective Lagrangian to a gauge-invariant theory by interpreting the sources of gauge symmetry breaking as spurions. Providing the spurions with suitable transformation properties
under the gauge group we can recover the gauge invariance under the full local group $G^{\elem}\times G^{\comp}$. Once we treat the spurions as dynamical fields, this procedure defines a (non-unique)  embedding of the effective theory
in a possible UV completion. The spurions include two types of degrees of freedom: the would-be Goldstone bosons,
eaten by the heavy gauge vector bosons of the composite sector through the Higgs mechanism, and physical scalars $\Phi_i$.
By working in the unitary gauge, which we adopt in our computations, a generic amplitude comprises two separate contributions: 
one coming only from the exchange of the physical polarizations of the gauge vector bosons and one including the exchange of some physical scalar degrees of freedom $\Phi_i$.
Clearly our computation retains only the first one, while the second is missing and makes our results sensitive to the details of the UV completion.
If the masses of the extra scalars $\Phi_i$ are close the cut-off scale $\Lambda_{UV}$, we expect that the contributions we are
neglecting in our estimates of the Wilson coefficients are generically  of order $(M/\Lambda_{UV})^2$ and/or $(m,\tilde m/\Lambda_{UV})^2$.

The embedding of the model in a UV completion also shows that it is not consistent to deal with the HF case, within the effective theory. Indeed to keep the masses of the elementary
leptons non-vanishing in the HF limit, we have to consider at the same time large $(\Delta,\tilde \Delta)$, such that the ratios $(X,\tilde X)$ remain constant.
In a UV completion $\Delta,\tilde \Delta$ are proportional to VEVs that break the gauge symmetry $G^{\elem}\times G^{\comp}$ 
and contribute to the masses of the heavy gauge vector bosons of the composite sector. Therefore, barring tuning of the parameters, we cannot
make $m,\tilde m\gg M$.

%%%%%%%%%%%%%%%%%%%%%%%%%%%%%%%%%%%%%%%%%%%%%%%%%%%%%%%%%%%%%
\section{Lepton flavour violation in the two-site model}
\label{sec.5}
%%%%%%%%%%%%%%%%%%%%%%%%%%%%%%%%%%%%%%%%%%%%%%%%%%%%%%%%%%%%%

In this section we will present our results for the Wilson coefficients of the various LFV operators considered in sec.~\ref{sec2}, 
in the context of the two-site model introduced above.
We work at the leading order in the loop expansion and we pay particular attention to the spurionic structure of the coefficients.
Throughout this section we assume universal masses for the heavy leptons: $m_i=m$ and $\tilde m_i=\tilde m$.

\subsection{Dipole operators}

\begin{figure}[t]
\centering
\includegraphics[scale=0.67]{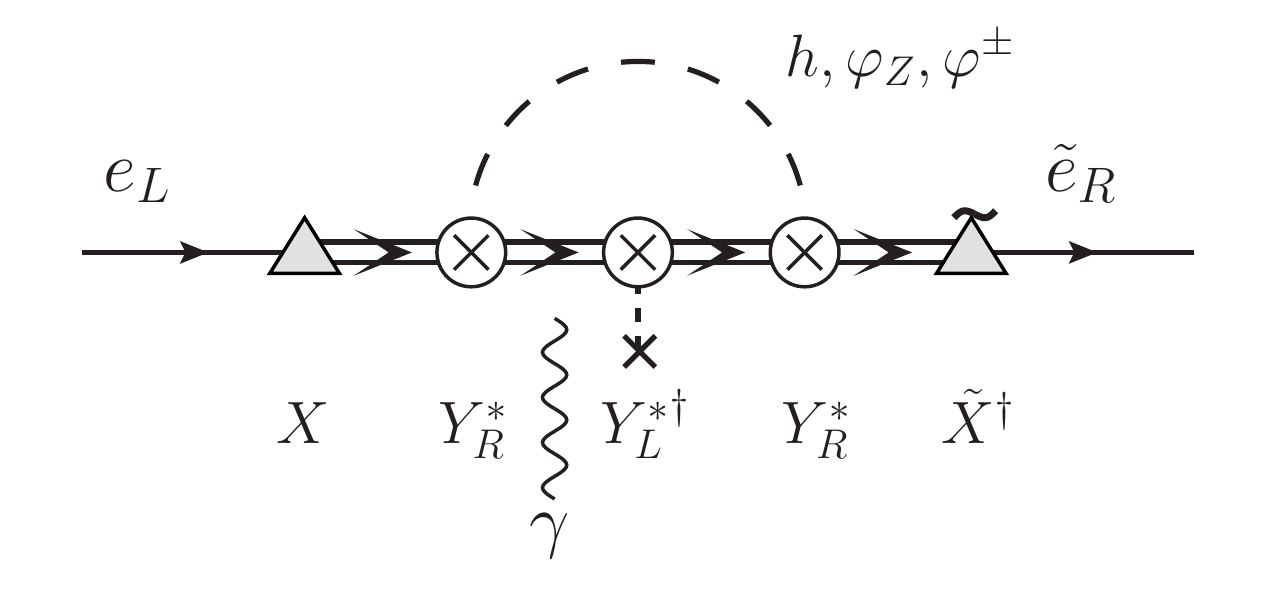}
\includegraphics[scale=0.67]{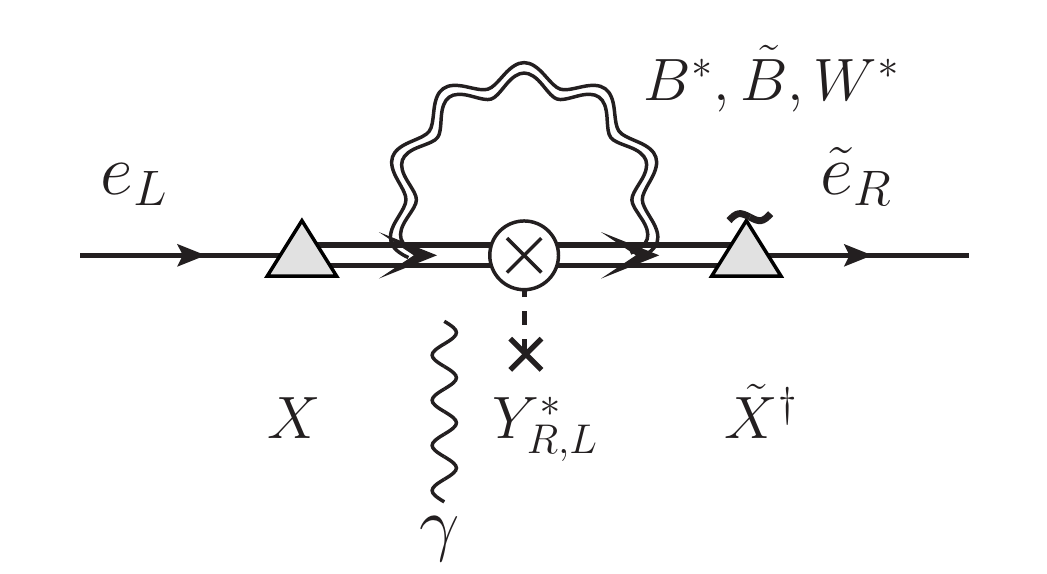}
\caption{Feynman diagrams for LO contributions to $\ell_i\to \ell_j\gamma$ with loop-exchange of 
SM bosons $h,Z,W$ and heavy fermions (left) and heavy gauge bosons and heavy fermions (right).}
\label{fig:LO_diagram}
\end{figure}

Dipole operators are among the most important operators in the leptonic sector as they generate 
LFV radiative decays such as $\mu\to e\gamma$, leptonic EDMs and $(g-2)$s. We will 
consider the one-loop contribution to dipole operators in the two-site model, assuming the validity of the 
perturbative expansion. 
The result of our computation of the full one-loop contribution to the electromagnetic dipole operator is shown in the appendix B.
Given the fact that the composite Higgs scenario is naturally 
characterised by a strongly interacting regime, the use of perturbation theory can be 
questioned. Here we will restrict ourselves to the portion of the parameter space that is compatible with
perturbation theory. This allows us to consider moderately large coupling constants of the composite sector, 
with loop factors still providing a sizeable suppression. The same considerations apply to processes
receiving non-vanishing contributions already at the tree level.
Within the two-site model, dipole amplitudes receive leading contributions from the virtual exchange 
of i) light SM bosons ($h,Z,W$) and heavy fermions and ii) heavy gauge bosons and heavy fermions.   
Examples of diagrams for the above classes of contributions are shown in fig.~\ref{fig:LO_diagram}.
In particular, summing over the $h,Z$ and $W$ amplitudes, we find that the dominant effects for class i) 
are given by
\begin{align}
\frac{\left(C_{e\gamma}\right)_{h+Z+W}}{\Lambda^2} &= \frac{e}{64\pi^2} \frac{1}{m \mt} X \yr \yl^\dagger \yr \Xt^\dagger\;, 
\label{class_i_LO}  
\end{align}
in agreement with the result of refs.\cite{Kannike:2011ng,Falkowski:2013jya}.
As we can see, the amplitude of eq.~(\ref{class_i_LO}) has precisely the same spurionic structure as
the one of eq.~(\ref{Dcom}). In particular, according to our spurionic classification of sec.~\ref{sec2}, 
it turns out that $\left(C_{e\gamma}\right)_{h+Z+W}$ is of order $(N_Y, N_X)=(3,2)$, where $N_Y$ and $N_X$ are the 
orders of the expansion in $(\yr,\yl)$ and $(X,\tilde X)$, respectively.
Interestingly, we observe that contributions of order $(N_Y, N_X)=(3,2)$ containing only the Yukawa 
$\yr$, which would be allowed by the flavor symmetries of our model, are absent, to one loop order.

\medskip

The leading effects for class ii) start from the order $(N_Y, N_X)=(1,2)$ and read  
\begin{align}
\frac{\left(C_{e\gamma}\right)_{\tilde B + B^*}}{\Lambda^2}  &=
%\left(C_{e\gamma}\right)_{\tilde B + B^*} &=
\frac{e}{256 \pi^2} \frac{g_*^2}{M^2}  \,  
\left[ X  \yr \Xt^\dagger  f^{B}_1(y,z)  + X  \yl \Xt^\dagger f^{B}_2(y,z)  \right]\;,  \\
\frac{\left(C_{e\gamma}\right)_{W^*}}{\Lambda^2}  &=
%\left(C_{e\gamma}\right)_{W^*} &=
-\frac{e}{256 \pi^2} \frac{g_*^2}{M^2}  \, X  \yr \Xt^\dagger  f^{W^*}_1(y)\;,
\label{class_ii_LO}
\end{align}
where $M$ stands for a common heavy boson mass, $y = m^2/M^2$, $z = \mt^2/M^2$ and 
the loop functions are defined in the appendix. 
Notice that the first term of $\left(C_{e\gamma}\right)_{\tilde B+B^*}$ and $\left(C_{e\gamma}\right)_{W^*}$ are aligned with the mass operator
as long as $m,\tilde m\propto \mbox{l\hspace{-0.55em}1}$, at least.
Therefore, as already pointed out in the literature \cite{Agashe:2004cp}, they cannot induce neither flavor nor CP violating 
effects after switching to the mass basis for the SM leptons. Yet, we stress that they contribute to the 
leptonic $g-2$. However, we point out here that the second term of $\left(C_{e\gamma}\right)_{\tilde B+B^*}$, not discussed in the literature to our knowledge, 
is not aligned with the mass operator and therefore generates flavor and CP violating effects.  

From eq.~(\ref{class_i_LO}) one can easily find that, in the anarchic scenario, the bound from 
${\rm BR}(\mu\to e\gamma)$ 
%$\mathcal{B}(\mu\to e\gamma)$ 
imposes that $\sqrt{{m \mt}}/\langle Y\rangle\gtrsim 10~$TeV. 
In order to relax such a strong bound while keeping $\yr$ anarchic, we analyze here in great detail 
the solution with $\yl=0$, as already done in the model-independent analysis of section 2.

\medskip

\begin{figure}[t]
\centering
\includegraphics[scale=0.54]{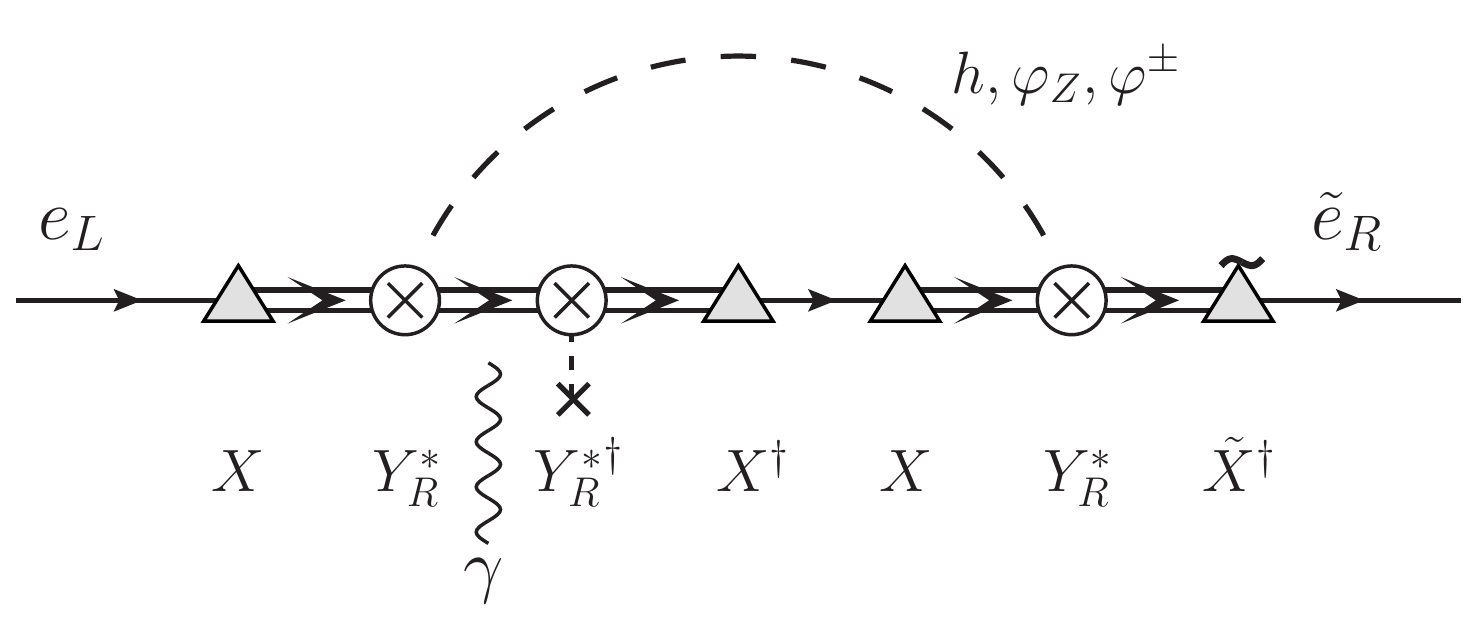}~~
\includegraphics[scale=0.54]{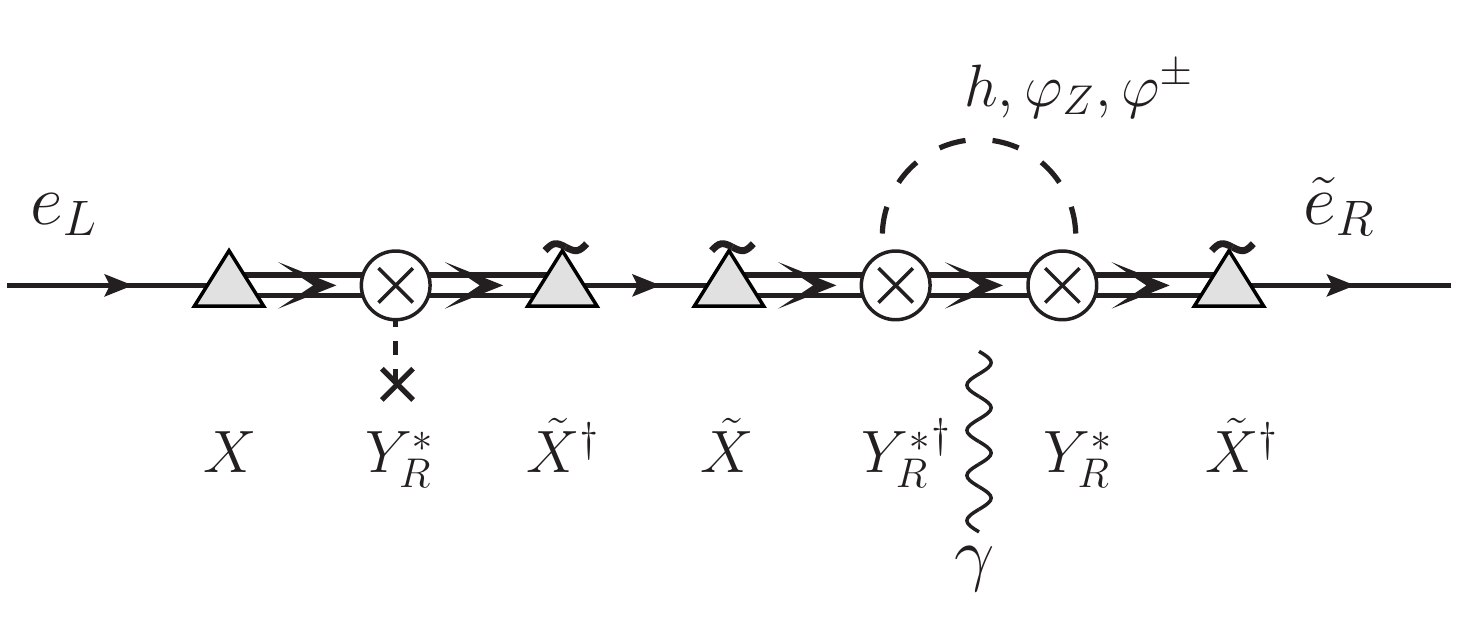}
\caption{NLO Feynman diagrams for $\ell_i\to \ell_j\gamma$ of order $(N_Y, N_X)=(3,4)$ with loop-exchange of $h,Z,W$ and heavy fermions.}
\label{fig:LO_diagram_Higgs_12}
\end{figure}

\begin{figure}[t]
\centering
\includegraphics[scale=0.55]{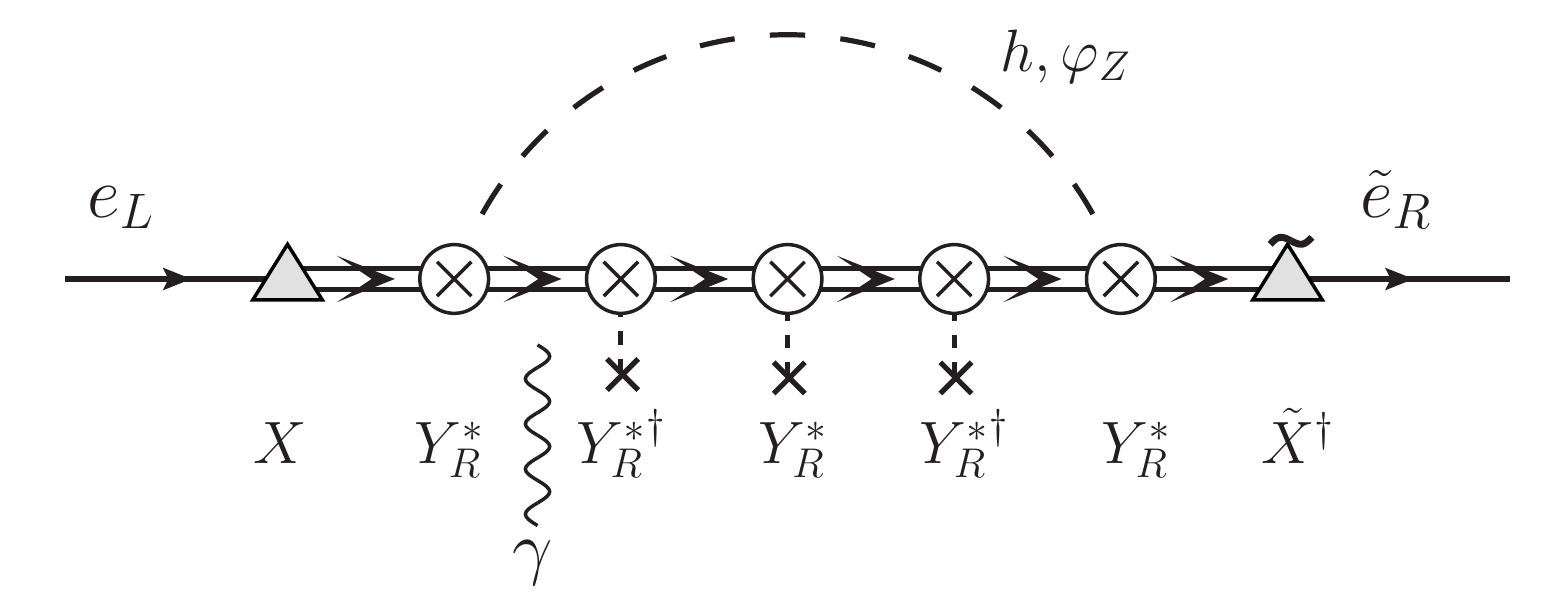}
\includegraphics[scale=0.55]{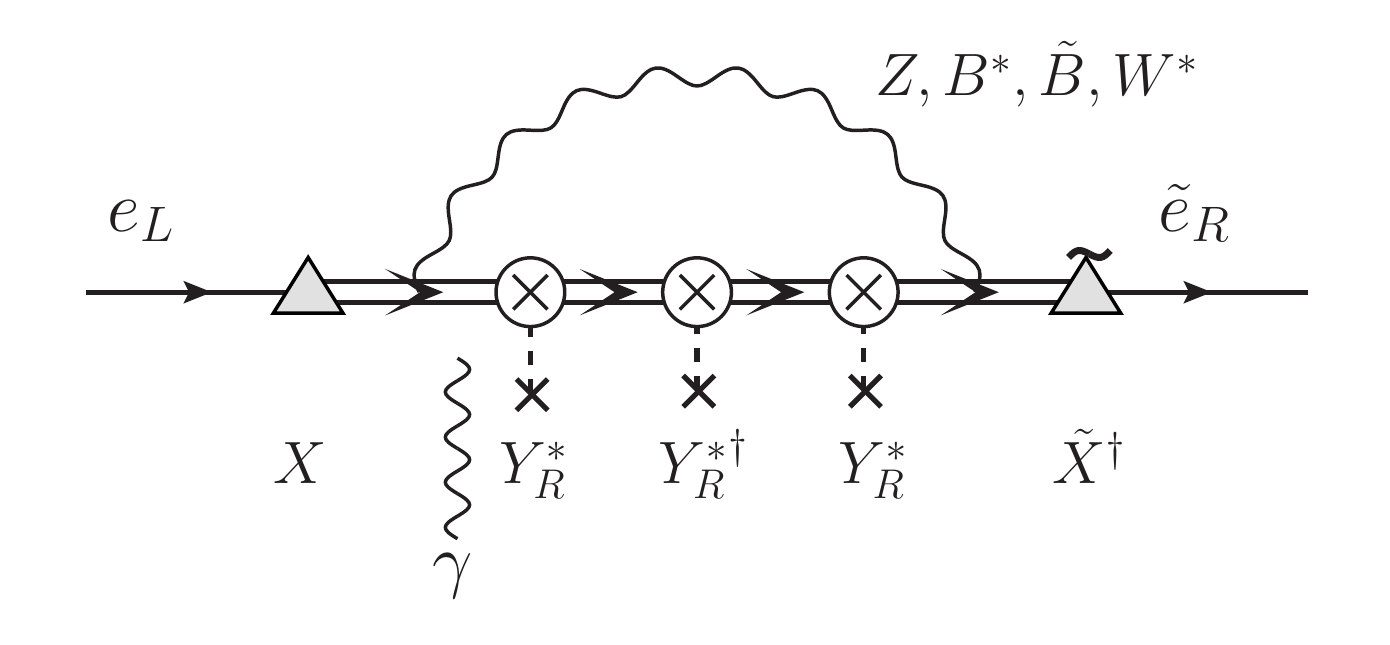}
\caption{NLO Feynman diagrams for $\ell_i\to \ell_j\gamma$ arising from $dim-8$ operators. The diagram on the left is of order
$(N_Y, N_X)=(5,2)$, the one on the right of order $(N_Y, N_X)=(3,2)$.}
\label{fig:dim8-diagrams}
\end{figure}

Setting $\yl=0$, the NLO effects stemming from class i) are given by:
\begin{align}
\frac{\left(C_{e\gamma}\right)_{h}}{\Lambda^2}  ={}& -
\frac{e}{256\pi^2} 
\Bigg[
			 \frac{f^h_1(x)}{\mt^2} X \yr \yr^\dagger X^\dagger  X \yr \Xt^\dagger
			+ \frac{f^h_1(x^{-1})}{m^2}  X \yr \Xt^\dagger \Xt \yr^\dagger \yr \Xt^\dagger 
\label{eq:NLO_HB1}			
 \nonumber \\*
&
+  \frac{v^2}{(m^2 - \mt^2)^2} X \yr \yr^\dagger \yr \yr^\dagger \yr \Xt^\dagger  \Bigg] \;,
			\\
\frac{\left(C_{e\gamma}\right)_{Z}}{\Lambda^2}  ={}& -
\frac{e}{256\pi^2} 
\Bigg[ 
\frac{f^Z_1(x)}{\mt^2} X \yr \yr^\dagger X^\dagger X \yr \Xt^\dagger + \frac{f^Z_2(x)}{m^2}  X \yr \Xt^\dagger \Xt \yr^\dagger \yr \Xt^\dagger  
\label{eq:NLO_HB2}
 \nonumber \\*
&
			+  \frac{v^2}{(m^2 - \mt^2)^2} X \yr \yr^\dagger \yr \yr^\dagger \yr \Xt^\dagger
			+ 8 \, X \yr \yr^\dagger \yr \Xt^\dagger \frac{M_Z^2}{m^2 \mt^2}  \Bigg] \;,
			 \\
\frac{\left(C_{e\gamma}\right)_{W}}{\Lambda^2}  ={}& 
\frac{e}{256\pi^2} 
			\left(
			 \frac{20}{3 \mt^2} X \yr \yr^\dagger X^\dagger X \yr \Xt^\dagger
			+ \frac{4}{3 m^2}  X \yr \Xt^\dagger \Xt \yr^\dagger \yr \Xt^\dagger
		        \right) \;.
\label{eq:NLO_HB3}
\end{align}						
where  $x = m^2/\mt^2$.
The first (second) terms of $\left(C_{e\gamma}\right)_{h}, \left(C_{e\gamma}\right)_{Z}$ and $\left(C_{e\gamma}\right)_{W}$ arise from diagrams such as those shown in the left (right) plot of 
fig.~\ref{fig:LO_diagram_Higgs_12}. These contributions are of order $(N_Y, N_X)=(3,4)$ and occur through a chirality flip implemented in 
the internal heavy-fermion line (left plot) and external light-fermion line (right plot).
On the other hand, the last term of $\left(C_{e\gamma}\right)_{h}$ as well as the third terms of $\left(C_{e\gamma}\right)_{Z}$ stem from dim-8 operators such as 
\be
\frac{\varphi^\dagger \varphi}{16 \pi^2}(\overline{\ell}_L \varphi)(\sigma\cdot F) 
X~ \yr \left({\tilde m}^\dagger \tilde m \right)^{-2} \yr^\dagger \yr \yr^\dagger \yr~  \tilde X^\dagger \tilde{e}_R~,
\label{eq:dim8_operators}
\ee
and the corresponding diagrams are shown in the left plot of fig.~\ref{fig:dim8-diagrams}. 
Such contributions are of order $(N_Y, N_X)=(5,2)$ and turn out to be suppressed by a 
factor of $v^2/\tilde m^2$ compared to dim-6 contributions. 
As a result, the parametric ratio between dim-6 and dim-8 operators is of order $X^2 \tilde m^2/\yr^2 v^2$
showing that both kind of operators might provide the dominant effects depending on the model parameters.
Let us stress that contributions of order $(N_Y, N_X)=(5,2)$ can arise also at dim-6 level if the extra Higgses
of eq.~(\ref{eq:dim8_operators}) close in a loop instead of getting a vacuum expectation value. 
Therefore, we expect that one-loop induced dim-8 operators with $(N_Y, N_X)=(5,2)$ dominate over
two-loop induced dim-6 operators with $(N_Y, N_X)=(5,2)$ provided $v^2 /\tilde m^2 \gtrsim 1/16\pi^2$.
\begin{figure}[t]
\centering
\includegraphics[scale=0.56]{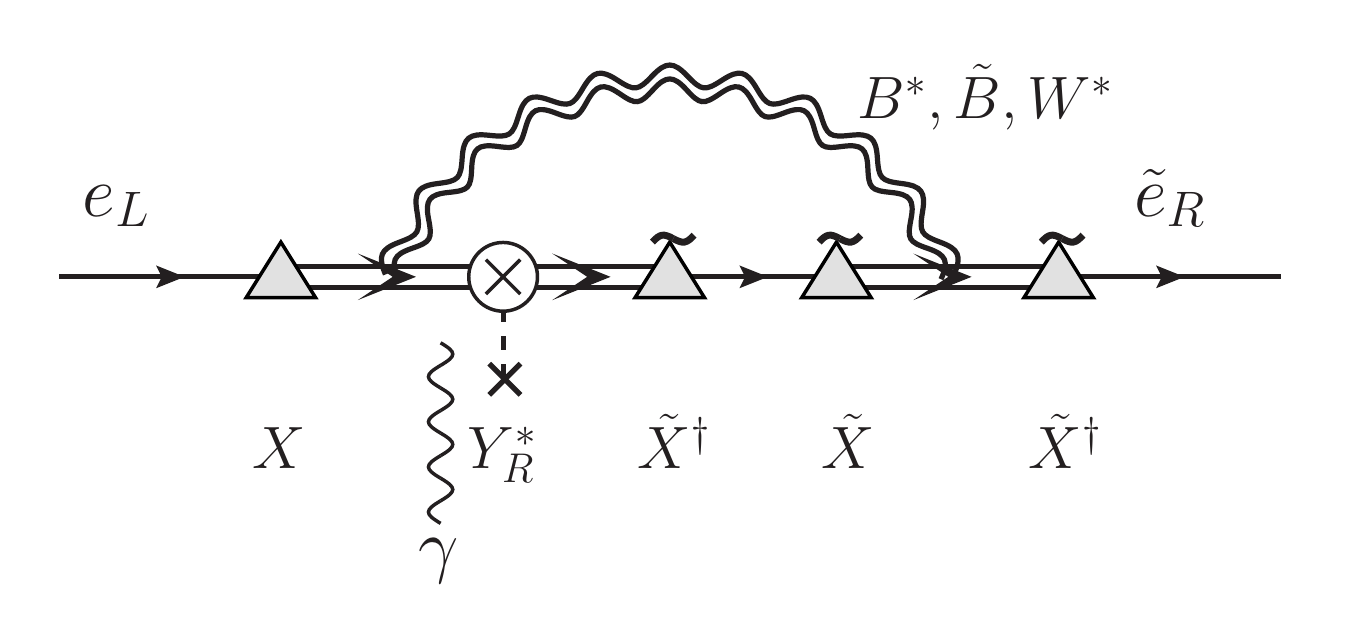}
\includegraphics[scale=0.56]{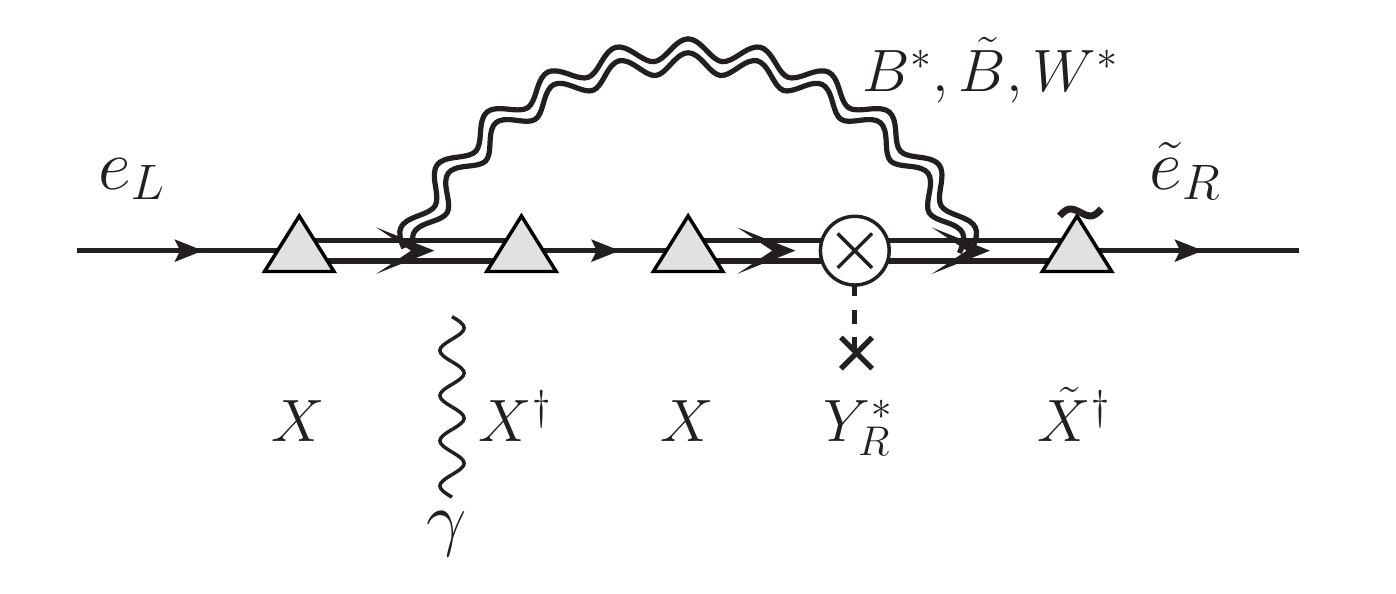}
\caption{NLO Feynman diagrams for $\ell_i\to \ell_j\gamma$ of order $(N_Y, N_X)=(1,4)$ with loop-exchange of heavy gauge bosons and heavy fermions.}
\label{fig:heavy_heavy}
\end{figure}
The last term of $\left(C_{e\gamma}\right)_{Z}$ stems from dim-8 operators of the form
\be
g^2
\frac{\varphi^\dagger \varphi}{16 \pi^2}(\overline{\ell}_L \varphi)(\sigma\cdot F) 
X~ \yr \left({\tilde m}^\dagger \tilde m \right)^{-2} \yr^\dagger \yr~  \tilde X^\dagger \tilde{e}_R~,
\label{eq:dim8_operators2}
\ee
and the relevant diagram is shown in the right plot of fig.~\ref{fig:dim8-diagrams}. 
Finally, we observe that there are not effects for $\left(C_{e\gamma}\right)_{W}$ from dim-8 operators 
since our model doesn't include heavy right-handed neutrinos.
 
We discuss now NLO effects arising from class ii) which are given by:	
\begin{align}	
\frac{\left(C_{e\gamma}\right)_{\tilde B + B^*}}{\Lambda^2}  ={}& -\frac{e}{256\pi^2} 
\frac{g^2_*}{M^2} 
\Bigg[ 
f^B_3(y,z) \, X  X^\dagger X \yr \Xt^\dagger +  f^B_4(y,z) \, X \yr \Xt^\dagger \Xt \Xt^\dagger   
\label{eq:NLO_HF1}
\nonumber \\*
&
+ f^B_5(y,z)  \frac{v^2}{M^2} X \yr \yr^\dagger \yr \Xt^\dagger   \Bigg] \; , \\
\frac{\left(C_{e\gamma}\right)_{W^*}}{\Lambda^2}  ={}& -\frac{e}{256\pi^2} 
\frac{g^2_*}{M^2}  
\Bigg[ f^{W^*}_2(y,z)   X  X^\dagger X \yr \Xt^\dagger +  f^{W^*}_3(y,z)  X \yr \Xt^\dagger \Xt \Xt^\dagger 
 \nonumber \\
& 
			+ f^{W^*}_4(y,z)  \frac{v^2}{M^2} X \yr \yr^\dagger \yr \Xt^\dagger   
\Bigg] \; ,
\label{eq:NLO_HF2}
\end{align}
The first two terms of $\left(C_{e\gamma}\right)_{\tilde B+B^*}$ and $\left(C_{e\gamma}\right)_{W^*}$ are of order $(N_Y, N_X)=(1,4)$ and arise
from the one loop-exchange of heavy fermions and bosons. The relevant Feynman diagrams are
shown in fig.~\ref{fig:heavy_heavy}. On the other hand, the last contributions to  $\left(C_{e\gamma}\right)_{\tilde B+B^*}$ and 
$\left(C_{e\gamma}\right)_{W^*}$ are of order $(N_Y, N_X)=(3,2)$ and stem from dim-8 operators such as, for example, 
\be
\frac{g_*^2}{M^2}
\frac{\varphi^\dagger \varphi}{16 \pi^2}(\overline{\ell}_L \varphi)(\sigma\cdot F) 
X\, \yr \left({\tilde m}^\dagger \tilde m \right)^{-1} \yr^\dagger \yr\, \tilde X^\dagger \tilde{e}_R~.
\label{eq:dim8_operators3}
\ee
The relevant Feynman diagrams are shown in the right plot of fig.~\ref{fig:dim8-diagrams}.

%%%%%%%%%%%%%%%%%%%%%%%%%%%%%%%%%%%%%%%%%%%%%%%%%
\subsection{Scalar operator}
%%%%%%%%%%%%%%%%%%%%%%%%%%%%%%%%%%%%%%%%%%%%%%%%%

The Yukawa interactions for charged leptons are modified after we integrate out at tree level the heavy fermions. 
Setting $\yl = 0$, we find that:
\begin{align}
-\mcL_Y &= \frac{h}{\sqrt{2}} \, \bar e_L \,y_\ell\,  e_R + \hc \;,
\end{align}
where in the mass basis for the charged leptons we have
%
%\begin{align}
%y_\ell  = \frac{v}{\sqrt{2}} X \yr \Xt^\dagger - \left( \frac{v}{\sqrt{2}} \right)^3  
%\left( \frac{1}{\mt^2} X \yr \yr^\dagger X^\dagger X \yr \Xt^\dagger + \frac{1}{m^2} X \yr \Xt^\dagger \Xt \yr^\dagger \yr \Xt^\dagger \right)
%\end{align}
%
%
\begin{align}
y_\ell  = y^{\mysmall SM}_\ell - \frac{v^2}{2\mt^2} \, X \yr \yr^\dagger X^\dagger y^{\mysmall SM}_\ell - \frac{v^2}{2m^2} \, y^{\mysmall SM}_\ell \Xt \yr^\dagger \yr \Xt^\dagger \;,
\end{align}
with
\be
y^{\mysmall SM}_\ell =X \yr \tilde X^\dagger~~~.
\ee
%
%\begin{align}
%\left( m_{SM}-\frac{v}{\sqrt{2}}Y_{SM} \right)  = 
%			\left( \frac{v}{\sqrt{2}} \right)^3 X \yr \Xt^\dagger 
%			\left( \frac{1}{\mt^2} X \yr \yr^\dagger X^\dagger + \frac{1}{m^2} \Xt \yr^\dagger \yr \Xt^\dagger \right)
%\end{align}
%
Therefore, the coefficient $C_{e\vp}$ of the dim-6 operator $Q_{e\vp}=C_{e\vp}(\vp^\dag \vp)(\bar \ell_{L} e_{R} \vp)$ reads
%
%\begin{align}
%\left( m_{SM}-\frac{v}{\sqrt{2}}Y_{SM} \right)  = -2 \left( \frac{v}{\sqrt{2}} \right)^3 C_{e\vp}
%\end{align}
%
%thus:
%
\begin{align}
\frac{C_{e\vp}}{\Lambda^2} =  \frac{1}{2\mt^2} X \yr \yr^\dagger X^\dagger y^{\mysmall SM}_\ell + y^{\mysmall SM}_\ell \frac{1}{2 m^2} \Xt \yr^\dagger \yr \Xt^\dagger \;.
\end{align}
Notice that for $\yl=0$ the corrections to $y^{\mysmall SM}_\ell$ are proportional to $y^{\mysmall SM}_\ell$ itself, as
in scenarios where the Higgs is a pseudo-Goldstone boson~\cite{Agashe:2009di,Giudice:2007fh,Buras:2011ph}.
%%%%%%%%%%%%%%%%%%%%%%%%%%%%%%%%%%%%%%%%%%%%%%%%%%%%%%%%%%
\subsection{Vector operators}
%%%%%%%%%%%%%%%%%%%%%%%%%%%%%%%%%%%%%%%%%%%%%%%%%%%%%%%%%%

The $Z$ boson interactions with charged leptons are also modified after we integrate out at tree level the heavy fermions. 
In particular, we find that 
%In addition to the SM coupling, once integrated out the heavy fields, in this model (at NLO in the spurions and at tree level) 
%the $Z$ boson has the additional interaction term:
%
\begin{align}
\mcL_Z &= \frac{g}{c_W} \,
\bar e  \left( \, g_L P_L  + g_R P_R \, \right)  \slashed{Z}  e\;,
\end{align}
where $g_L$ and $g_R$ are defined as follow
\begin{align} 
g_L &=   -\frac{1}{2} + s^2_W  + \frac{v^2}{4 \mt^2} \, X \yr \yr^\dagger X^\dagger \;, &
g_R &=  s^2_W -  \frac{v^2}{4 m^2} \, \Xt \yr^\dagger \yr \Xt^\dagger \;.
\end{align}
%
%Starting from the operators $Q_{\vp l}^{(1)}, Q_{\vp l}^{(3)}, Q_{\vp e}$ one can derive the additional interaction term between 
%the Z boson and the charged leptons:
%
%\begin{align}
%\mcL_Z &=  - \frac{g}{2 c_W} \, Z^\mu \, v^2  \left(
%			 (C^{(1)}_{\vp l} + \frac{1}{4} C^{(3)}_{\vp l} )  \,   \bar e_L \gamma_\mu e_L  
%			+ C_{\vp e}  \, \bar e_R \gamma_\mu e_R  \right) \;.
%\end{align}
%
Switching to the coefficients $C^{(1)}_{\vp l}$, $C^{(3)}_{\vp l}$ and $C_{\vp e}$ we find that:
 \begin{align}
\frac{C^{(1)}_{\vp l}}{\Lambda^2}  &= - \frac{1}{2 \mt^2} X \yr \yr^\dagger X^\dagger\;, &
 C^{(3)}_{\vp l}& = 0\;,&
\frac{C_{\vp e}}{\Lambda^2} &= \frac{1}{2 m^2} \Xt \yr^\dagger \yr \Xt^\dagger \;.
 \end{align}
 %

%%%%%%%%%%%%%%%%%%%%%%%%%%%%%%%%%%%%%%%%%%%%%%%%%%%%%%%%%%
\subsection{Contact operators}
%%%%%%%%%%%%%%%%%%%%%%%%%%%%%%%%%%%%%%%%%%%%%%%%%%%%%%%%%%

Contact operators induced by the exchange of heavy gauge bosons can be easily derived starting from the coefficients 
$C^{(\rho)}_{L L}$ and $C^{(\rho)}_{R R}$ entering the interaction Lagrangian of heavy gauge bosons and SM leptons which
we report in the appendix. Integrating out the heavy gauge bosons at tree level, we find that
\begin{equation}
\begin{aligned}
\frac{(C_{ll})_{e\mu ee}}{\Lambda^2} & = ( X X^\dagger )_{e\mu} \left( \frac{g_1^2}{4 M^2_{B^*}} + \frac{g^2_2}{4M^2_{W^{*}_3}}  \right)\;,
\\
\frac{(C_{ee})_{e\mu ee}}{\Lambda^2} & = ( \tilde X \tilde X^\dagger )_{e\mu} \left( \frac{g^2_1}{M^2_{B^*}}  \right)\;, \\
\frac{(C_{le})_{e\mu ee}}{\Lambda^2} & = ( X X^\dagger )_{e\mu} \left( \frac{g^2_1}{2M^2_{B^*}}  \right)\;, 
\\
\frac{(C_{le})_{eee\mu}}{\Lambda^2} & = ( \tilde X \tilde X^\dagger )_{e\mu} \left( \frac{g^2_1}{2M^2_{B^*}}  \right)\;, 
\end{aligned}
\end{equation}
and
\begin{equation}
\begin{aligned}
\frac{(C_{\ell q}^{(u)})_{e\mu}}{\Lambda^2} & = - \frac{1}{4}( X X^\dagger )_{e\mu} \left(\frac{g_1^2}{3 M^2_{B^*}} + \frac{g^2_2}{M^2_{W^{*}_3}}  \right)\;,
&
\frac{(C_{\ell q}^{(d)})_{e\mu}}{\Lambda^2} & = -\frac{1}{4}( X X^\dagger )_{e\mu} \left(\frac{g_1^2}{3 M^2_{B^*}} - \frac{g^2_2}{M^2_{W^{*}_3}}  \right)\;,
\\
\frac{(C_{eu})_{e\mu}}{\Lambda^2} & = -( \tilde X \tilde X^\dagger )_{e\mu} \left(\frac{2g^2_1}{3M^2_{B^*}}  \right)\;, 
&
\frac{(C_{ed})_{e\mu}}{\Lambda^2} & = ( \tilde X \tilde X^\dagger )_{e\mu} \left(\frac{g^2_1}{3M^2_{B^*}}  \right)\;, 
\\
\frac{(C_{\ell u})_{e\mu}}{\Lambda^2} & = - ( X X^\dagger )_{e\mu} \left( \frac{g^2_1}{3M^2_{B^*}}  \right)\;, 
&
\frac{(C_{\ell d})_{e\mu}}{\Lambda^2} & =  ( X X^\dagger )_{e\mu} \left( \frac{g^2_1}{6M^2_{B^*}}  \right)\;, 
\\
\frac{(C_{eq})_{e\mu}}{\Lambda^2} & = -( \tilde X \tilde X^\dagger )_{e\mu} \left( \frac{g^2_1}{6M^2_{B^*}}  \right)\;, 
&
%\frac{(C_{eq})_{e\mu}}{\Lambda^2} & = -( \tilde X \tilde X^\dagger )_{e\mu} \left( \frac{g^2_1}{6M^2_{B^*}}  \right)\;,
\end{aligned}
\end{equation}
where higher-order effects suppressed by additional factors of $(g_{SM}/g_*)^2$, $v^2/m^2(\tilde m^2)$ and $X^2(\tilde X^2)$ have been neglected.
%%%%%%%%%%%%%%%%%%%%%%%%%%%%%%%%%%%%%%%%%%%%%%%%%%%%%%%%%%%%%%%%%%%%%%%%%
\section{Phenomenological analysis}
%%%%%%%%%%%%%%%%%%%%%%%%%%%%%%%%%%%%%%%%%%%%%%%%%%%%%%%%%%%%%%%%%%%%%%%%%

In this section, we evaluate the most relevant low-energy processes in the charged lepton sector in the context of the two-site 
model by making use of the Wilson coefficients derived in the previous section.
One of the most interesting features of the two-site model is the absence of the Wilson coefficients $(N_Y, N_X)=(3,2)$ from dimension six dipole operators,
at least in the one-loop approximation. Such coefficients are particularly dangerous for LFV. As we saw in section 3, their presence
is not compatible with anarchic Yukawas $\yr$ if the compositeness scale is close to 1 TeV.
We do not know if this feature of the model is an accident of the one-loop approximation or if it persists at higher loop orders.
In what follows we will neglect the $(N_Y, N_X)=(3,2)$ contribution to dimension six dipole operators,
assuming that higher loops provide a sufficient suppression. Our purpose is to check if under these conditions
anarchic Yukawas are still viable or not when we consider a compositeness scale around 1 TeV.

Among the most interesting LFV channels are $\mu\to e\gamma$, $\mu \to 3e$, $\mu \to e$ conversion in Nuclei as well as 
$\tau$ LFV processes. 
However, hereafter, we focus on processes with an underlying $\mu\to e$ transition since they are the best probes of composite Higgs models with 
anarchic $\yr$.
Concerning flavour conserving processes, we are interested in the electron EDM and the muon $g-2$.
The current status and future experimental sensitivities for the above processes are collected in table~\ref{nextgenexp}.
%
%\begin{table}[t!]
%\centering
%\begin{tabular}{|c|c|c|}
%\hline
%LFV Process & Present Bound & Future Sensitivity  \\
%\hline
%$\mu \to e \gamma$ & $5.7 \times 10^{-13}$ \cite{Adam:2013mnn} & $\approx 6 \times 10^{-14}$ \cite{Baldini:2013ke}  \\
%$\mu \to 3 e$ & $1.0 \times 10^{-12}$\cite{Bellgardt:1987du} & $\approx 10^{-16}$ \cite{Blondel:2013ia}\\
%$\mu^-$ Au $\to$ $e^-$ Au & $7.0 \times 10^{-13}$ \cite{Bertl:2006up} & $ ? $  \\
%$\mu^-$ Ti $\to$ $e^-$ Ti & $4.3 \times 10^{-12}$ \cite{Dohmen:1993mp} & $?$ \\
%$\mu^-$ Al $\to$ $e^-$ Al & $-$  & $\approx 10^{-16}$ \cite{comet,mu2e} \\
%$\tau \to \mu \gamma$ & $4.4 \times 10^{-8}$ \cite{Aubert:2009ag}& $\sim 10^{-8}-10^{-9}$ \cite{Hayasaka:2013dsa} \\
%$\tau \to 3 \mu$ & $2.1\times10^{-8}$\cite{Hayasaka:2010np} & $\sim 10^{-9}-10^{-10}$ \cite{Hayasaka:2013dsa}  \\
%\hline
%Electron EDM & Present Bound & Future Sensitivity  \\
%\hline
%$d_e ({\rm e~cm})$ & $8.7 \times 10^{-29}$ \cite{Baron:2013eja} & $?$ \\
%\hline
%\end{tabular}
%\caption{Present and future experimental sensitivities for relevant low-energy observables.}
%\label{nextgenexp}
%\end{table}
%
As recalled in section 3, there is a $\sim 3.5 \sigma$ discrepancy between the {\small SM} prediction and the
experimental value of the muon $g-2$.

%%%%%%%%%%%%%%%%%%%%%%%%%%%%%%%%%%%%%%%%%%%%%%%%
\boldmath
\subsection{$\ell_i\to\ell_j\gamma$}
\unboldmath
%%%%%%%%%%%%%%%%%%%%%%%%%%%%%%%%%%%%%%%%%%%%%%%%

The dipole transition $\ell_i \to \ell_j\gamma$ is responsible for both LFV radiative decays (when $ i\neq j$)
like $\mu\to e\gamma$ and flavor conserving processes like the electron EDM and the muon $g-2$ when 
$ i = j = e$ or $\mu$, respectively. The branching ratio for the process $\mu\to e\gamma$ can be written as
\begin{equation}
{\rm BR} (\mu \to e  \gamma)=
\dfrac{24\pi^2}{G^2_F\Lambda^4} 
\left(\dfrac{v^2}{m_{\mu}^2}\right) 
\left( \left|C^{e\mu}_{e\gamma} \right|^2 + \left|C^{\mu e}_{e\gamma} \right|^2 \right ) \;.
\label{Brmuegamma}
\end{equation}
Radiative LFV transitions are tightly related to the magnetic and electric leptonic dipole moments
which are also extremely sensitive probes of new physics. In particular, one can find that
\begin{equation}
d_{e} = -
\sqrt{2} \, \frac{v}{\Lambda^2} \, {\rm Im}\left(C^{ee}_{e\gamma} \right) \;,\qquad
a_{\mu} = 
\dfrac{2 \sqrt{2}}{e} \,
\frac{m_{\mu}v}{\Lambda^2} \, {\rm Re}\left(C^{\mu\mu}_{e\gamma} \right)\; .   
\ee
In concrete scenarios as our two-site model, $\Delta a_{\ell}$, $d_{\ell}$ and
${\rm BR}(\ell\to \ell^{\prime}\gamma)$ are expected to be correlated. However, such 
correlations depend on the flavor and CP structure of the couplings which are unknown.
In our discussion, we assume order one CP-violating phases and anarchic $\yr$.  

Here we provide the dominant contribution 
to $\mu \to e \gamma$ which arises from dim-8 operators. 
Focusing on the HB scenario ($M\gg m,\tilde m$) with anarchic $\yr$ and $X=\tilde X$, it turns out that 
%
%\beq
%BR(\mu\to e \gamma) \approx
%\frac{3\alpha_{em}}{128\pi} \frac{v^8 |\yr|^8}{(m^2 - \mt^2)^4}  
%\left( 
%\left|\frac{X_e}{X_\mu}\right|^2 + \left|\frac{\tilde X_e }{ \tilde X_\mu}\right|^2 
%\right)
%\eeq
%
%
\beq
{\rm BR}(\mu\to e \gamma) \approx
\frac{3\alpha_{em}}{64\pi} \frac{v^8 |\yr|^8}{(m^2 - \mt^2)^4} \,\frac{m_e}{m_\mu}\,,
\label{eq:mueg_approx}
\eeq
where we made use of eq.~(\ref{XXX}) to eliminate $X,\tilde X$ and where $\yr$ now stands for an average element of the anarchic matrix $\yr_{ij}$.
Notice that the above expression is valid only in the mass insertion approximation which requires 
that $m, \tilde m, |m-\tilde m|\gg v \yr$. As an example, if $|m^2-\tilde m^2|\approx m^2$, one can find
\be
{\rm BR}(\mu\to e \gamma) \approx 3 \times 10^{-13} \left(\frac{1.5\,{\rm TeV}}{m}\right)^8 |\yr|^8~,
\label{eq:mueg_approx2}
\ee
implying that $\mu \to e\gamma$ saturates its current experimental bound for $m\approx 1.5~$TeV and $\yr \approx 1$.
This is confirmed by our numerical results shown in fig.~\ref{fig:mueg_LO_vs_NLO} where we have reported the predictions 
for ${\rm BR}(\mu\to e \gamma)$ as a function of the heavy fermion mass $m$. 
In fig.~\ref{fig:mueg_LO_vs_NLO}, as well as in all other plots,
we have assumed anarchic $\yr$ and $\yl$ and $X=\tilde X$ so that the relevant flavor mixing angles entering $\mu\to e$
transitions are $X_e/X_\mu =\tilde X_e/ \tilde X_\mu \sim \sqrt{m_e /m_\mu}$.  
The lower (upper) red line in the left plot
refers to the case where $\yr=\yl=1$ ($\yr=\yl=2$) while the lower (upper) black line corresponds to $\yl=0$ and $\yr=1$ ($\yr=2$).
The most prominent features emerging by this plot are: i) in the case of $\yl=0$ ($\yl \neq 0$), the bound on $\mu\to e\gamma$
is satisfied for $m\ge 1~$TeV ($m\ge10~$TeV), ii) for $\yl=0$, ${\rm BR}(\mu\to e\gamma)\sim \yr^8/m^8$ while for  
$\yl \neq 0$, ${\rm BR}(\mu\to e\gamma)\sim \yl^2\yr^2/m^4$ and this explains the different growth with $\yr,\yl$ 
and the decoupling properties with $m$ of the various curves. In the right plot of fig.~\ref{fig:mueg_LO_vs_NLO}
we show the branching ratio of $\mu \to e\gamma$ as a function of the heavy fermion mass $m$ for $\yl=0$ and 
various values of $\yr$.
\begin{figure}[t]
\centering
\includegraphics[scale=0.4]{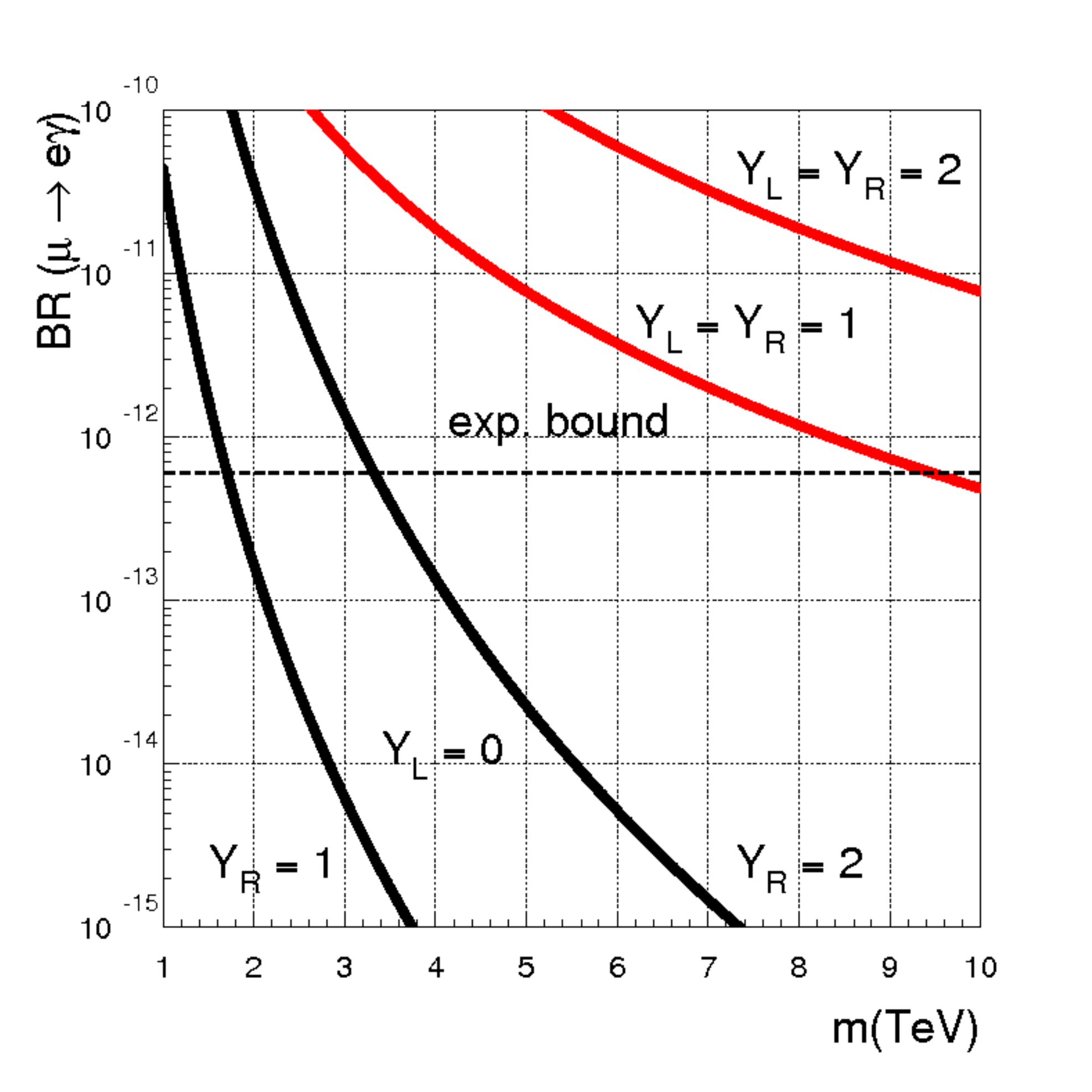}
\includegraphics[scale=0.4]{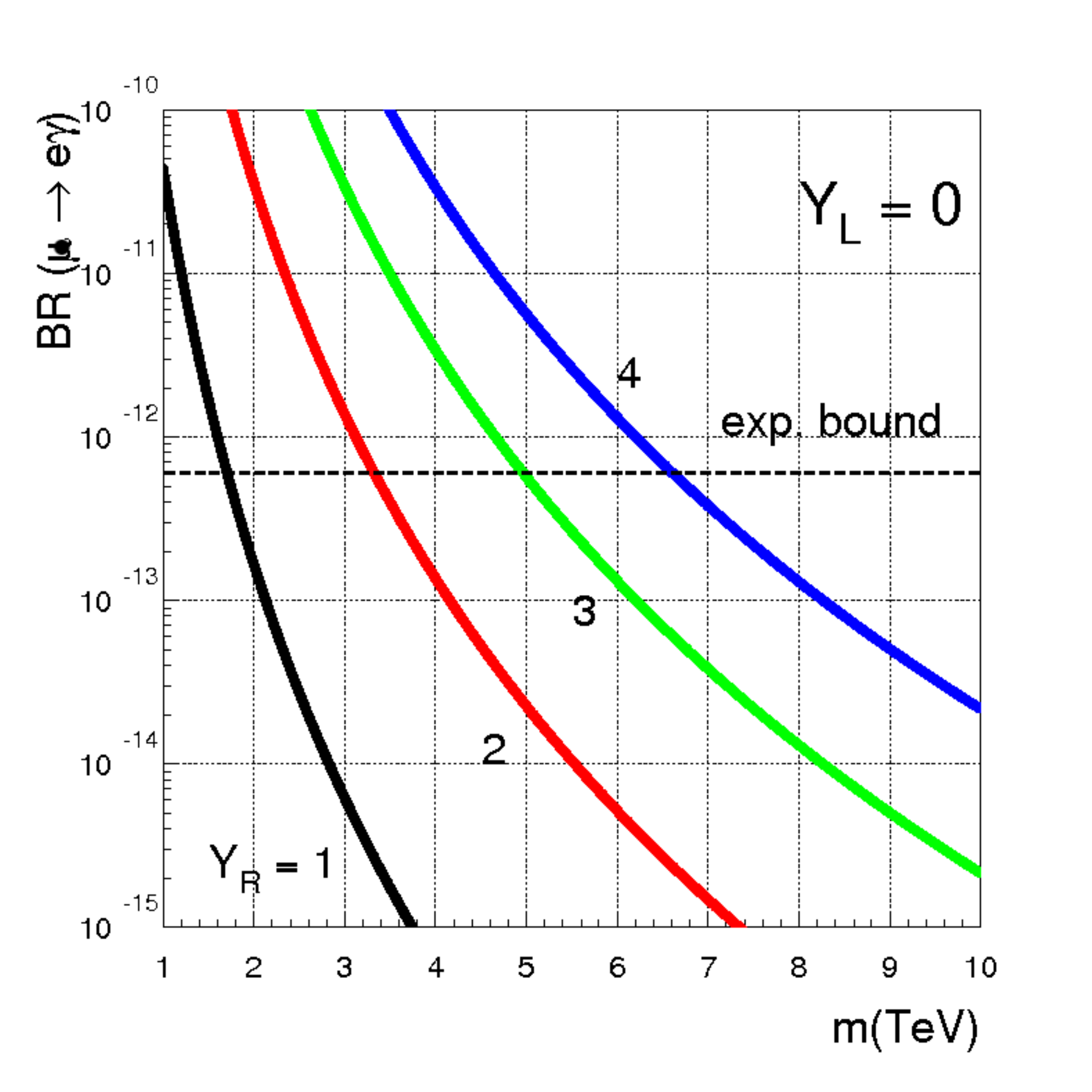}
\caption{Branching ratio of $\mu \to e\gamma$ as a function of the heavy fermion mass $m$.}
\label{fig:mueg_LO_vs_NLO}
\end{figure}
\begin{figure}
\centering
\includegraphics[scale=0.4]{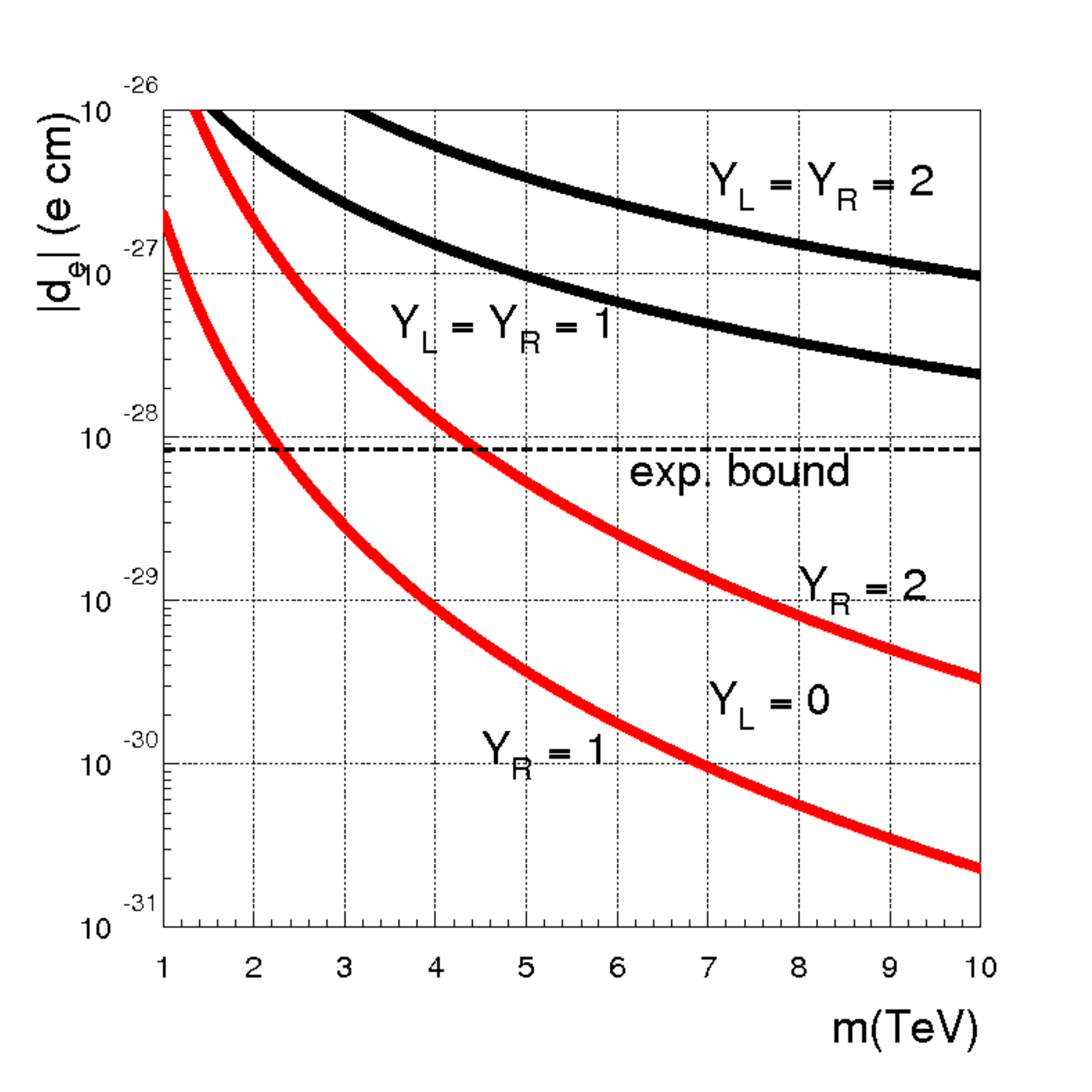}
\includegraphics[scale=0.4]{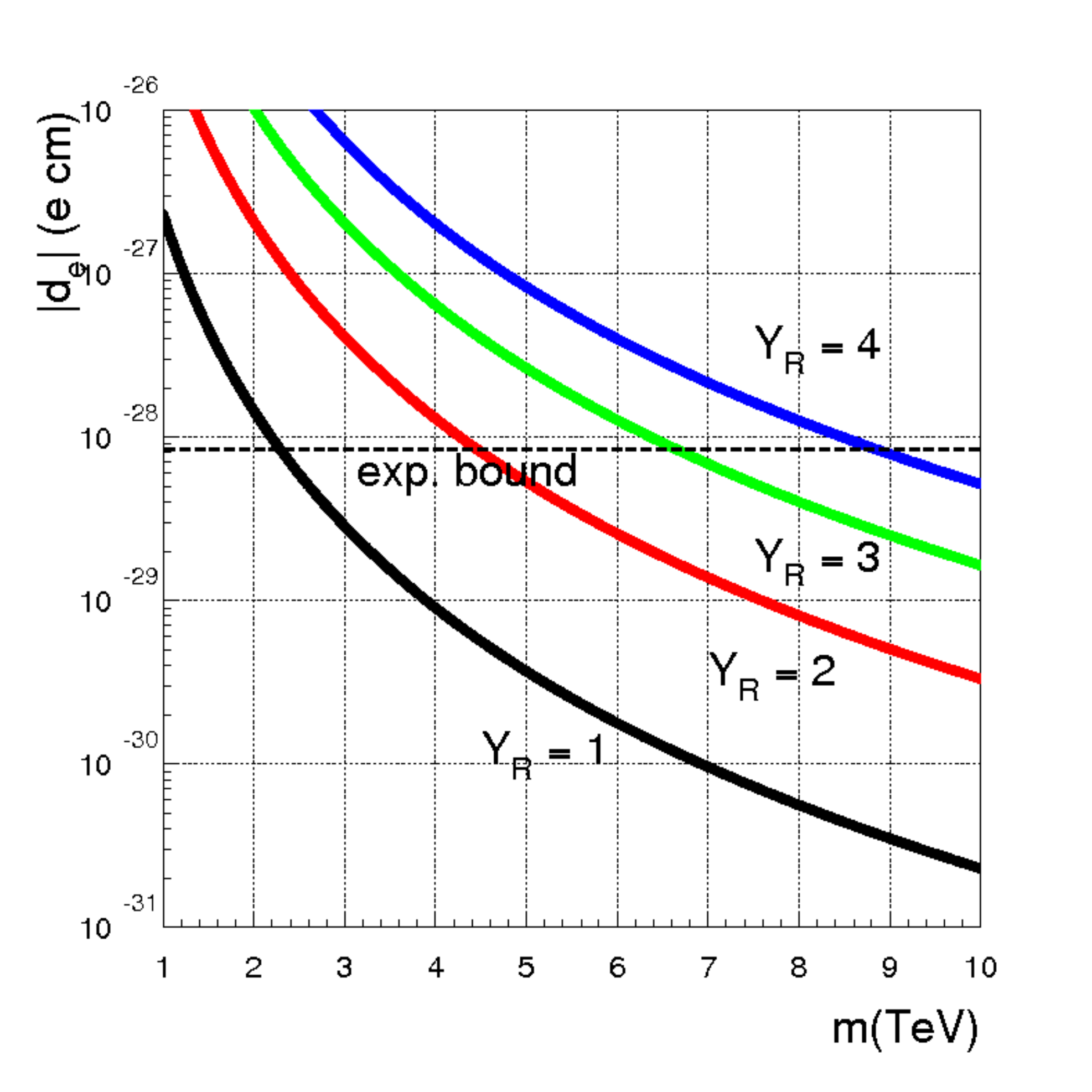}
\caption{Electron EDM $d_e$ as a function of the heavy fermion mass $m$.}
\label{fig:de_LO_vs_NLO}
\end{figure}

 A quite similar behavior is expected for the electron EDM. Indeed, if $\yl \neq 0$ and assuming $\mathcal{O}(1)$ CP-violating phases, 
it turns out that
\begin{align}
\frac{|d_e|}{e} & \approx \frac{m_e}{32\pi^2} \frac{\yr \yl}{m \mt} \approx 
8 \times 10^{-29}cm \left(\frac{20\,{\rm TeV}}{\sqrt{m \mt}}\right)^2 \yr\yl~,
\label{eq:edm}  
\end{align}
and therefore the electron EDM poses a severe constraint on the heavy fermion scale at the same level of $\mu \to e \gamma$.
Setting $\yl=0$, the EDM bound is satisfied with much lighter masses. In particular, we find that
\begin{align}
\frac{|d_e|}{e} & \approx 
\frac{m_e}{64\pi^2} 
\frac{v^2 \yr^4}{(m^2 - \mt^2)^2}  
\approx 
10^{-28}cm \left(\frac{3\,{\rm TeV}}{m}\right)^4 \yr^4~,
\end{align}			
where in the last equality we have assumed that $|m^2-\tilde m^2|\approx m^2$.

In fig.~\ref{fig:de_LO_vs_NLO} we show our numerical results for $d_e$. 
Indeed, fig.~\ref{fig:de_LO_vs_NLO} represents the analogous of fig.~\ref{fig:mueg_LO_vs_NLO} 
for the electron EDM case and similar conclusions drawn for $\mu\to e \gamma$ apply here too.

%%%%%%%%%%%%%%%%%%%%%%%%%%%%%%%%%%%%%%%%%%%%%%%%
\boldmath
\subsection{$\mu\to 3e$}
\unboldmath
%%%%%%%%%%%%%%%%%%%%%%%%%%%%%%%%%%%%%%%%%%%%%%%%

The process $\mu \to 3e$ receives more contributions than $\mu \to e \gamma$ as it is sensitive 
also to four-fermion operators. In particular, 
starting from the general expression of ref.~\cite{Kuno:1999jp,Brignole:2004ah} and neglecting 
scalar and non-dipole photonic contributions which are negligible in our model,
we find the following expression for its branching ratio
\begin{align}
{\rm BR}(\mu\to 3e ) ={}& 
\frac{1}{8 \Lambda^4 G^2_F} 
\bigg[
2|C_{VLL}|^2 + 2|C_{VRR}|^2 + |C_{VLR}|^2 + |C_{VRL}|^2 
\nonumber\\
&+\frac{8v e}{\sqrt{2}m_\mu}\mathrm{Re}
\bigg(\left(2 C_{VLL} + C_{VLR} \right) C_{e\gamma}^{e\mu \star} + 
\left( 2 C_{VRR} + C_{VRL} \right) C_{e\gamma}^{\mu e} \bigg) 
\nonumber\\
&+ 16 \,e^2 \left(\frac{v^2}{m_\mu^2}\right) \left(\log\frac{m_\mu^2}{m_e^2} - \frac{11}{4}\right)
(|C^{e\mu}_{e\gamma}|^2 + |C^{\mu e}_{e\gamma}|^2) 
\bigg]\;,
\label{eq:brg}
\end{align}
where 
\begin{align}
C_{VLL} &=  (2s_W^2-1) C_{\vp \ell}^{(1) e \mu } + C_{\ell\ell}^{ e \mu ee }\;,\qquad 
C_{VRR} = 2 s_W^2 C_{\vp e}^{ e\mu} + C_{ee}^{ e \mu ee } \;, \nonumber\\
C_{VLR} &=  2s_W^2 C_{\vp \ell}^{(1) e \mu } + C_{\ell e}^{ e \mu ee }\;,\qquad \qquad ~
C_{VRL} =  (2s_W^2-1) C_{\vp e}^{e\mu} + C_{\ell e}^{ eee\mu }  \;,
\end{align}
and the explicit expression for the he Wilson coefficients $C_{ab}$ can be found in sec.~\ref{sec.5}. 
In the above equations, we have neglected scalar four-fermion operators since they are very 
suppressed in our model.
We remind that, whenever the dipole operator is dominant in $\mu \to 3e$, there exists a model 
independent correlation between the branching ratio of $\mu \to 3e$ and $\mu \to e\gamma$ 
given by
\be
\frac{{\rm BR}(\mu \to 3e)}{{\rm BR}(\mu \to e \gamma)}
\simeq
\frac{\alpha_{el}}{3\pi}
\bigg(\log\frac{m^2_{\mu}}{m^2_{e}}- \frac{11}{4}\bigg)\approx 6.6\times 10^{-3}~.
\label{eq:mueg_mu3e_dipole}
%\nonumber\\
%B_{\mu N \to e N}
%&\simeq&
%\alpha_{\rm em} \times {\rm BR}(\mu \to e\gamma)~.
%\label{eq:dipole}
\ee
As a result, the current MEG bound ${\rm BR}(\mu\to e\gamma)\leq 5.7 \times 10^{-13}$
already implies that ${\rm BR}(\mu\to 3e)\lesssim 3.8\times 10^{-15}$.
However, as shown in our numerical analysis and illustrated in fig.~8, $\mu\to 3e$ turns out to be 
dominated by non-dipole operators and the correlation of eq.~(\ref{eq:mueg_mu3e_dipole})
is significantly violated. 
This is a  relevant result, as within composite Higgs models with $\yl \neq 0$, as well as in
supersymmetric scenarios, eq.~(\ref{eq:mueg_mu3e_dipole}) holds to an excellent approximation. 
Therefore, the two-site model with $\yl = 0$ can be disentangled among other models if 
both $\mu\to 3e$ and $\mu\to e\gamma$ will be observed.

In particular, in the HB scenario ($M\gg m,\tilde m$) with anarchic $\yr$, $m=\tilde m$ and $X=\tilde X$, we find 
\begin{align}
{\rm BR}(\mu \to 3e)
&\simeq
\frac{3}{16G_F^2} \frac{|\yr|^2}{m^4} \left(\frac{m_e m_{\mu}}{v^2}\right) 
\left(4s^4_W \!+\! (2s^2_W\!-\!1)^2 \right)
\nonumber\\
&\approx 5 \times 10^{-13} \left(\frac{1\,{\rm TeV}}{m}\right)^4 |\yr|^2~,
\label{eq:mu3e_approx}
\end{align}
implying that $\mu \to 3e$ saturates its current experimental bound for $m\approx 1$ TeV and $\yr \approx 1$.
This result is well reproduced by our numerical analysis shown in the left plot of fig.~\ref{fig:mu3e_mueconv}
where the behavior of ${\rm BR}(\mu \to 3e)$ as a function of the heavy fermion mass $m$ is shown for 
different values of $\yr$. Notice that ${\rm BR}(\mu \to 3e) \sim \yr^2/m^4$ and therefore it has a much milder growth 
with $\yr$ and a slower decoupling with $m$ compared to ${\rm BR}(\mu \to e\gamma)\sim \yr^8/m^8$ (in the case of $\yl=0$).
\begin{figure}[t]
\centering
\includegraphics[scale=0.4]{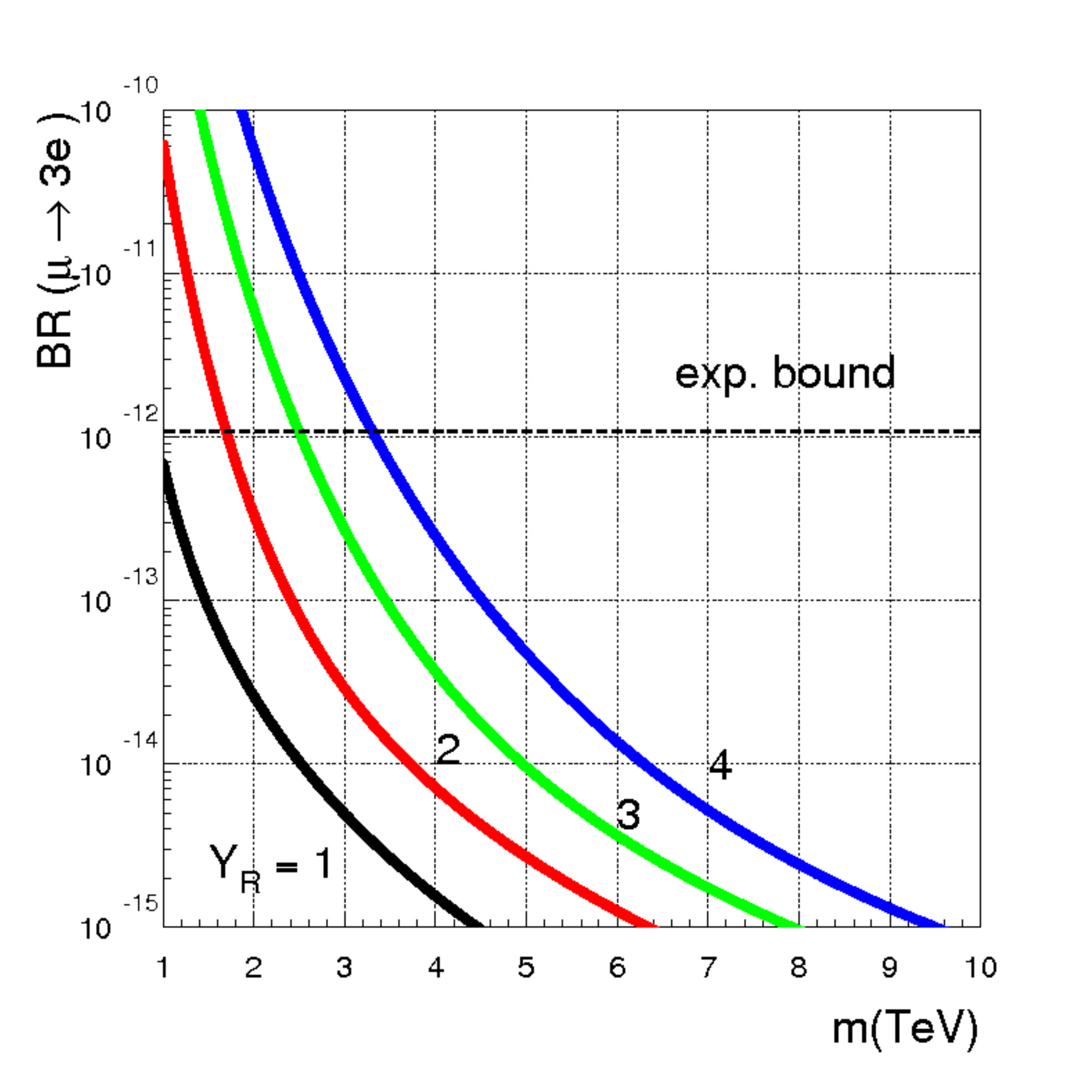}
\includegraphics[scale=0.4]{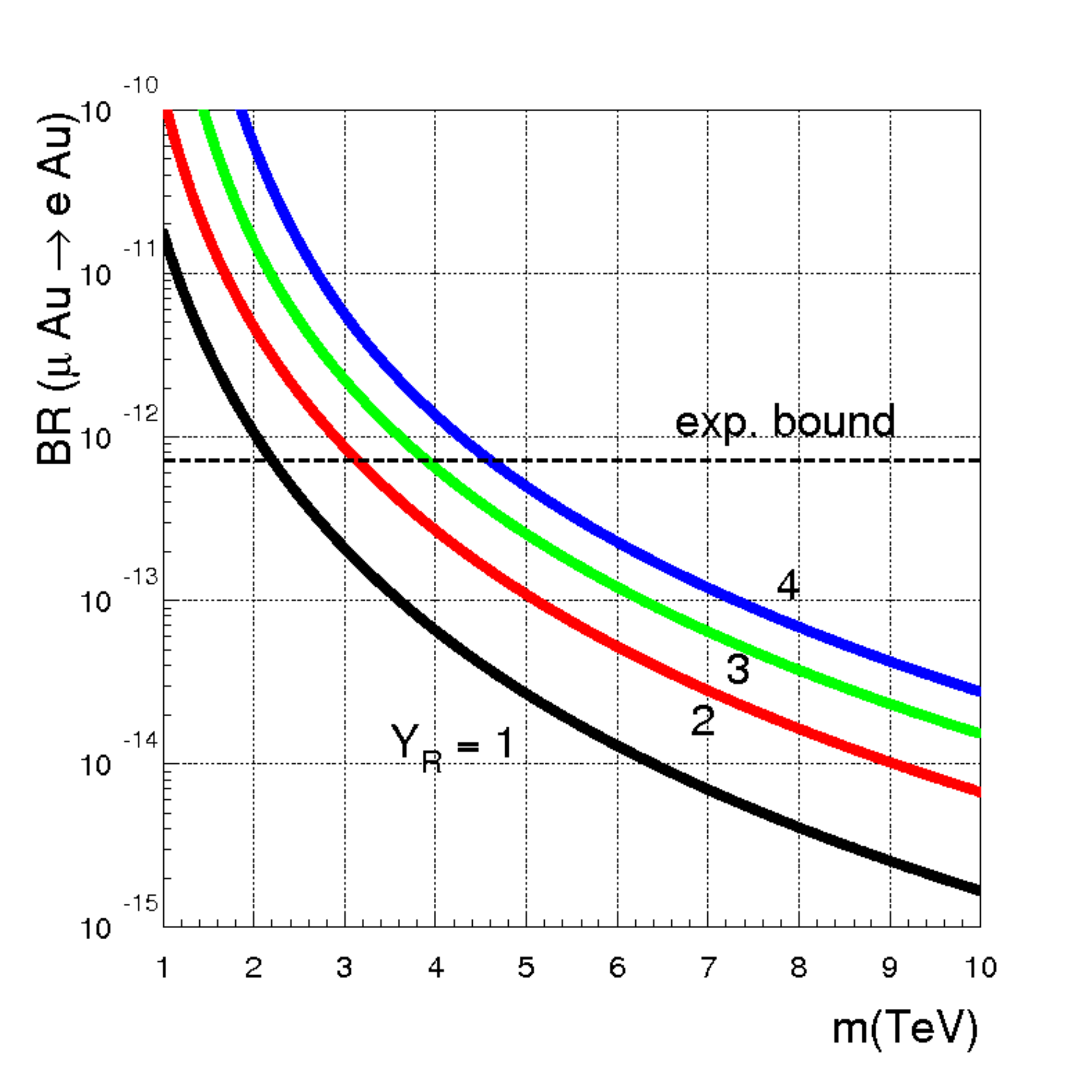}
\caption{Branching ratio of $\mu \to 3e$ (left) and $\mu^-Au\to e^-Au$ (right) as a function of the heavy fermion mass $m$.}
\label{fig:mu3e_mueconv}
\end{figure}
 %
 
%%%%%%%%%%%%%%%%%%%%%%%%%%%%%%%%%%%%%%%%%%%%%%%%
\boldmath
\subsection{$\mu N \to e N$}
\unboldmath
%%%%%%%%%%%%%%%%%%%%%%%%%%%%%%%%%%%%%%%%%%%%%%%%

As the last process governed by an underlying $\mu \to e$ transition we consider $\mu-e $ 
conversion in nuclei. As in the $\mu\to 3e$ case, this process receives effects from both dipole and 
four-fermion operators. In particular, the $\mu-e $ conversion branching ratio is defined as
\be
		{\rm BR}({\mu^- N \to e^- N})
		\equiv 
		\frac{\omega_{\rm conv}}{\omega_{\rm capt}}~,
		\label{eq:br_mueconv}
\ee
where $\omega_{\rm capt}$ is the muon capture rate while $\omega_{\rm conv}$ is given by the expression
\be
	\omega_{\rm conv} =
	\frac{m^5_\mu}{\Lambda^4}
	\left(
	\left|
	A_R^*  D
	+ g_{LV}^{(p)} V^{(p)}
	+ g_{LV}^{(n)} V^{(n)}
	\right|^2
	+ 
	\left|
	 A_L^*  D
	+ g_{RV}^{(p)} V^{(p)}
	+ g_{RV}^{(n)} V^{(n)}
	\right|^2
	\right)~,
	\label{eq:mueconv}
\ee
which we have obtained starting from the general expression of ref.~\cite{Kitano:2002mt} and setting to zero the scalar 
and non-dipole operator contributions which are negligible in our model.
The coupling constants $A_{L}$, $A_{R}$, $g_{LV}^{(p,n)}$ and $g_{RV}^{(p,n)}$ in eq.~(\ref{eq:mueconv}) are defined as
\begin{align}
	A_{R} & =  -\frac{1}{2\sqrt{2}} \frac{v}{m_\mu}~ C^{e\mu*}_{e\gamma},
	\\	
	A_{L} & =  -\frac{1}{2\sqrt{2}} \frac{v}{m_\mu}~  C^{\mu e}_{e\gamma},\\
         g_{LV}^{(p)}  & =  C^{(1) e\mu}_{\varphi\ell} \left( 4 s^2_W - 1 \right) - 2\left({C_{\ell q}^{(u)}}^{ e \mu}  + C_{\ell u}^{ e \mu}\right) - \left({C_{\ell q}^{(d)}}^{ e \mu} + C_{\ell d}^{ e \mu}\right),\\	
         g_{RV}^{(p)} &=C^{e\mu}_{\varphi e} \left( 4 s^2_W - 1 \right) - 2\left(C_{eq}^{e \mu} + C_{eu}^{ e \mu}\right) - \left(C_{eq}^{e \mu} + C_{ed}^{ e \mu}\right),\\
         g_{LV}^{(n)}  & =  C^{(1) e\mu}_{\varphi\ell}- \left({C_{\ell q}^{(u)}}^{ e \mu}  + C_{\ell u}^{ e \mu}\right) - 2\left({C_{\ell q}^{(d)}}^{ e \mu} + C_{\ell d}^{ e \mu}\right),\\	
         g_{RV}^{(n)} &= C^{e\mu}_{\varphi e} -\left(C_{eq}^{e \mu} + C_{eu}^{ e \mu}\right) - 2\left(C_{eq}^{e \mu} + C_{ed}^{ e \mu}\right)~.
%	g_{LV}^{(p)}  & =  C^{(1) e\mu}_{\varphi\ell} \left( 4 s^2_W - 1 \right) - 2\left(C_{\ell\ell}^{ e \mu uu } + C_{\ell e}^{ e \mu uu }\right) - \left(C_{\ell\ell}^{ e \mu dd } + C_{\ell e}^{ e \mu dd }\right),\\
%	g_{RV}^{(p)} &=
%  C^{e\mu}_{\varphi e} \left( 4 s^2_W - 1 \right) - 2\left(C_{\ell e}^{uu e \mu} + C_{ee}^{ e \mu uu }\right) - \left(C_{\ell e}^{dd e \mu} + C_{e e}^{ e \mu dd}\right),\\
%	g_{LV}^{(n)}  & =  C^{(1) e\mu}_{\varphi\ell} - \left(C_{\ell\ell}^{ e \mu uu } + C_{\ell e}^{ e \mu uu }\right) - 2\left(C_{\ell\ell}^{ e \mu dd } + C_{\ell e}^{ e \mu dd }\right),\\
%	g_{RV}^{(n)} &= C^{e\mu}_{\varphi e} - \left(C_{\ell e}^{uu e \mu} + C_{ee}^{ e \mu uu }\right) - 2\left(C_{\ell e}^{dd e \mu} + C_{e e}^{ e \mu dd}\right)\;,
\end{align}
The flavour-conserving interactions of the heavy vector resonances with the light quarks have been derived from the respective couplings with the 
light leptons simply rescaling them according to the different charges of the quarks under $Y$ and $T_3$.
Finally, the quantities $D$ and $V^{p,n}$ refer to the overlap integrals between the wave functions and the nucleon densities~\cite{Kitano:2002mt}
which we report for few relevant nuclei in table~\ref{table:overlap_integrals}. 
\begin{table}[htp]
    \begin{center}
	\begin{tabular}{|c|l|l|l|l|}
	    \hline
	    Nucleus &  $D$ & $V^{(p)}$  & $V^{(n)}$ & $\omega_{\rm capt}\,(10^{6}\,s^{-1})$
	    \\ 
	    \hline
	    Au & 0.189 & 0.0974 & 0.146 & 13.07\\
	     \hline
	     Ti & 0.0870 & 0.0399 & 0.0495 & 2.59 \\
	     \hline
	    Al & 0.0362 & 0.0161 & 0.0173 & 0.7054\\
	     \hline
	\end{tabular}
    \end{center}
    \caption{Overlap integrals from ref.~\cite{Kitano:2002mt}.}
    \label{table:overlap_integrals}
\end{table}
An inspection of eqs~(\ref{eq:br_mueconv}),(\ref{eq:mueconv}) and table~\ref{table:overlap_integrals} shows that at present 
$\mu^- Au \to e^- Au$ is the most sensitive probe of new physics among the various $\mu-e$ conversion in nuclei processes.
In scenarios with dipole dominance, the following model-independent relation holds
\begin{align}
\frac{{{\rm BR}}({\mu^- Au \to e^- Au})}{{{\rm BR}}({\mu \to e \gamma})}
= 
\frac{D^2}{192\pi^2}\frac{m^5_\mu G^2_F}{\omega_{capt}}
\approx 3.8 \times 10^{-3}
\label{eq:mueconv_dipole}
\end{align}
and therefore ${{\rm BR}}({\mu^- Au \to e^- Au})\lesssim 2.2\times 10^{-15}$
after imposing the bound ${\rm BR}(\mu\to e\gamma)\leq 5.7 \times 10^{-13}$.
However, as in the case of $\mu\to 3e$, in turns out that in our model ${{\rm BR}}({\mu^- Au \to e^- Au})$ 
is dominated by non-dipole operators and the correlation of eq.~(\ref{eq:mueconv_dipole}) is not at work.
This is shown in fig.~8.
%%%%%%%%%%%%%%%%%%%%%%%%%%%%%%%%%%%%%%%%%%%%%%%%%%%%%%%%%%%%%%%%%%%%%%%%%%%%%%
In particular, in the HB scenario ($M\gg m,\tilde m$) with anarchic $\yr$, $m=\tilde m$ and $X=\tilde X$, we find 
\begin{align}
{{\rm BR}}({\mu^- Au \to e^- Au})
&\simeq
\frac{m^5_\mu \, |\yr|^2}{m^4\, \omega_{capt}} \left(\frac{m_e m_\mu}{v^2}\right) |V^{(n)}|^2
\nonumber\\
&\approx 4 \times 10^{-13} \left(\frac{3\,{\rm TeV}}{m}\right)^4 |\yr|^2~,
\label{eq:mueconv_approx}
\end{align}
and therefore the current experimental bound is saturated for $m\approx 3~$TeV and $\yr \approx 1$.

These expectations are fully confirmed numerically as shown by the right plot of fig.~\ref{fig:mu3e_mueconv} where
we we report ${{\rm BR}}({\mu^- Au \to e^- Au})$ as a function of the heavy fermion mass $m$ for different values of $\yr$. 
Since both ${{\rm BR}}({\mu^- Au \to e^- Au})$ and ${\rm BR}(\mu \to 3e)$ are proportional to $\yr^2/m^4$, the consideration 
done for the case of $\mu\to 3e$ hold here too. However, we point out that in our model ${{\rm BR}}({\mu^- Au \to e^- Au})$ 
is much enhanced with respect to ${\rm BR}(\mu \to 3e)$ as opposite to scenarios with dipole-dominance where 
${{\rm BR}}({\mu^- Au \to e^- Au}) \approx 0.6\times{\rm BR}(\mu \to 3e)$. As a result, the simultaneous observation of these
processes would enable us to disentangle among the underlying theory at work. 
\begin{figure}[t]
\centering
\includegraphics[scale=0.4]{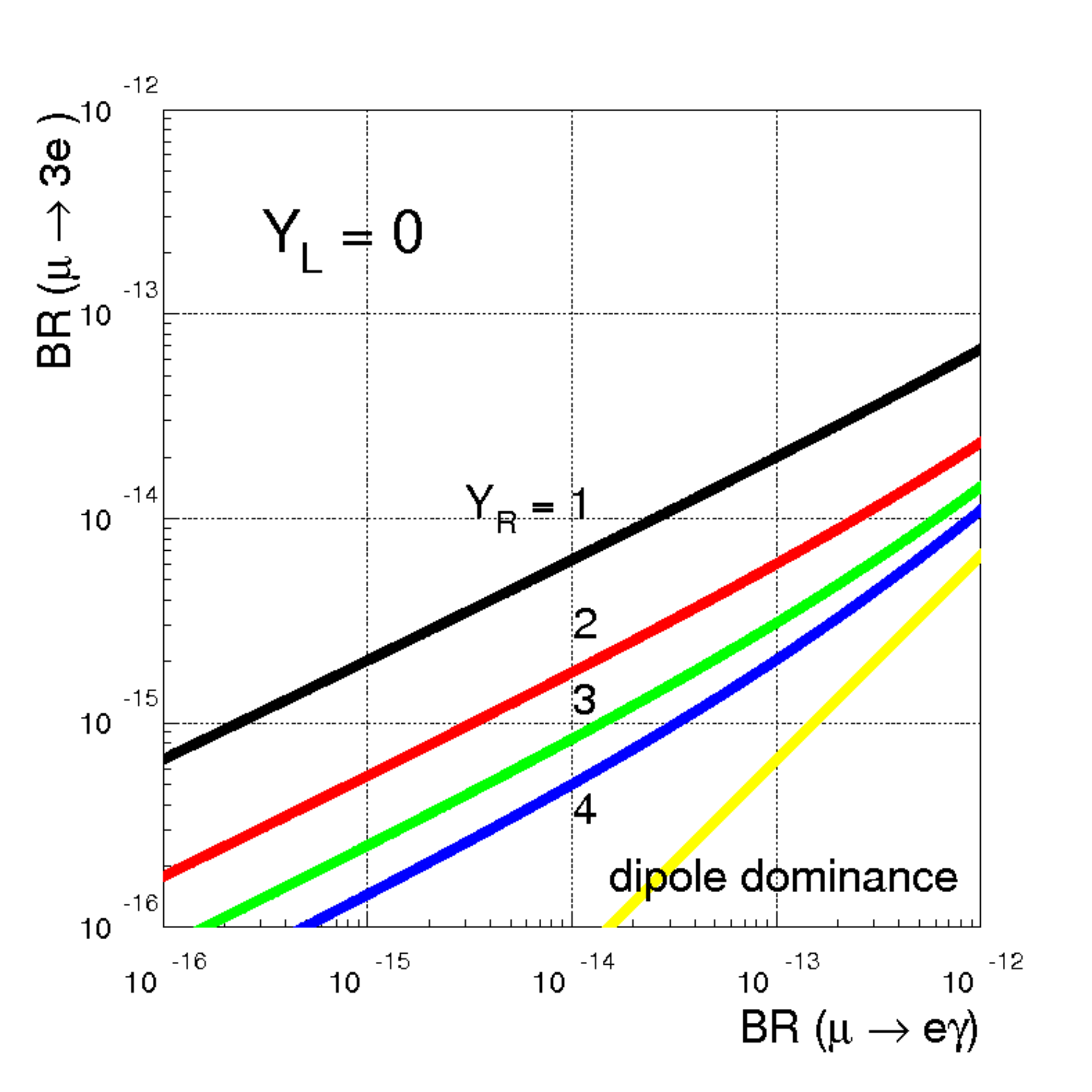}
\includegraphics[scale=0.4]{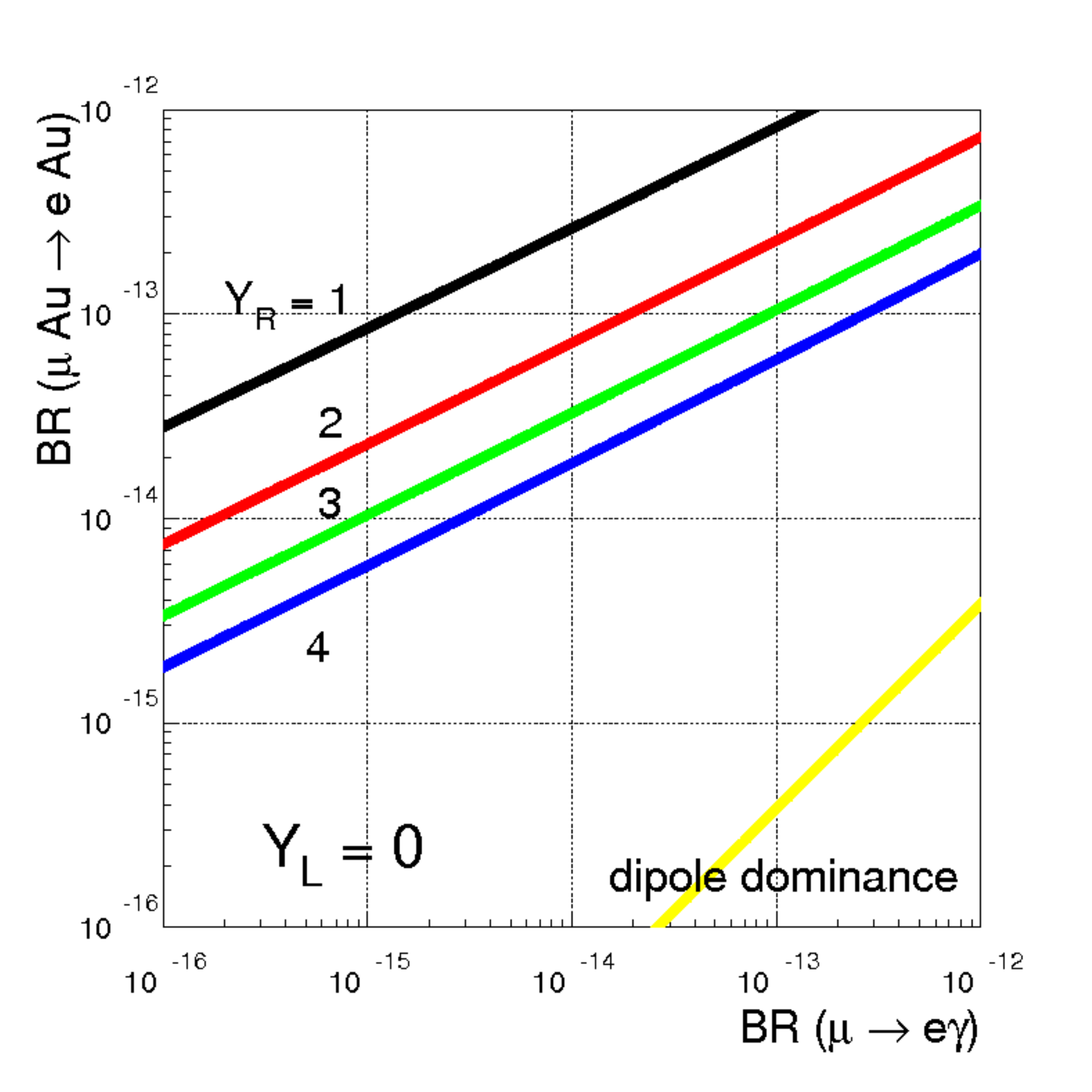}
\caption{Branching ratio of $\mu \to 3e$ (left) and $\mu^-Au\to e^-Au$ (right) versus the branching ratio of $\mu\to e \gamma$ for $\yl=0$. 
The case of dominance of the dipole operator is shown in yellow.}
\label{fig:mu3e_mueconv2}
\end{figure}
\section{Conclusion}
%%%%%%%%%%%%%%%%%%%%%%%%%%%%%%%%%%%%%%%%%%%%%%%%%%%%%%%%%%%%%%%%%%%%%%%%%%%%%%%
We have reanalysed the bounds on the compositeness scale derived from LFV and CP violation in the lepton sector, in the framework of PC.
In the generic case of anarchic Yukawa couplings, such bounds are known to be quite severe, stronger than the analogous bounds from the quark sector 
and requiring a compositeness scale above 30 TeV. In this work we focussed on the case of vanishing neutrino masses and vanishing ``wrong" Yukawa
couplings, in the hope of minimising FCNC and CP violating effects. We performed a general effective operator analysis, where the Wilson coefficients
of the relevant dimension-six operators are determined by a flavour symmetry group and by a set of spurions. We have considered all lowest dimensional 
operators leading to LFV violation and CP violation that can be constructed with the Higgs doublet and the SM leptons. We have formally expanded the Wilson coefficients 
in powers of the Yukawa couplings of the composite sector and in powers of the elementary-composite mixing terms. Our expansion includes, for the first time, terms quadrilinear 
in the elementary-composite mixing  $(X,\tilde X)$. By exploiting the known limits on the Wilson coefficients we have shown that, even in the case of vanishing ``wrong" Yukawa couplings,
the anarchic scenario is not compatible with a compositeness scale of 1 TeV, barring accidental cancellations not incorporated in our general spurion analysis. We have also shown that
there are schemes where a TeV scale can be accommodated, without necessarily reproducing the case of MFV. In the example we have provided
this is achieved in a semi-perturbative regime, where the universal Yukawa coupling  $y$ of the composite sector is close to unity. In this respect it is interesting to note that
while the bound on $\Lambda_{NP}$ coming from operators bilinear in  $(X,\tilde X)$ scales with $\sqrt{y}$, the limit from the operators quadrilinear in  $(X,\tilde X)$ scales with $1/y$,
thus making essentially impossible to lower $\Lambda_{NP}$ below the TeV. This also shows the importance of keeping terms quadrilinear in the mixing in the expansion of the Wilson coefficients.

We have also derived the low-energy effective Lagrangian relevant to LFV and CP violation in the lepton sector in the two-site model, that
includes heavy vector-like fermions as well as heavy spin-one particles. We focused on the case $\yl=0$, finding  a strong suppression of LFV effects compared to a generic composite Higgs scenario where $\yl\ne 0$. 
In a perturbative analysis where we evaluated the relevant amplitudes at the LO, we found that the $\mu\to e \gamma$ transition is dominated by
dimension 8 operators, at variance with our spurion analysis where the most important contribution comes from operators of dimension 6.
As a result the best probes of our scenario are $\mu\to e$ conversion in nuclei and the electron EDM. 
Al least in that portion of the parameter space where perturbation theory is applicable, LFV allows a compositeness scale close to the TeV,
even in the case of anarchic Yukawas. In this regime there are interesting relations among the various LFV transitions, which  
allow to disentangle the model from other possibilities.

\section*{Aknowledgements}
We would like to thank warmly Andrea Wulzer and Robert Ziegler for useful discussions. 
This work was supported in part by the MIUR-PRIN project 2010YJ2NYW and by the European Union network FP7 ITN INVISIBLES (Marie Curie Actions, PITN-GA-2011-289442).
The research of P.P. is supported by the ERC Advanced Grant No.  267985 (DaMeSyFla), by the research grant TAsP (Theoretical Astroparticle Physics), and by the Istituto Nazionale 
di Fisica Nucleare (INFN). P.P. thanks G.  Buchalla, G. Isidori, U. Nierste  and J. Zupan for the invitation to the MIAPP workshop Flavour 2015: New Physics at High Energy and High Precision,  
where part of his work was performed. 
The research of A.P. is supported by the Swiss National Science Foundation (SNF) under contract 200021-159720.

\appendix

\section{Contact operators}
 
In the following, we provide the full results for the coefficients $C^{(\rho)}_{\bar L L}$ and $C^{(\rho)}_{\bar R R}$ entering the interaction Lagrangian of heavy 
gauge bosons and SM leptons
\be
\mathcal{L} = \rho_\mu \, \bar \ell \gamma^\mu \left( C^{(\rho)}_{\bar L L}P_L +  C^{(\rho)}_{\bar R R}P_R \right) \ell\;,
\ee
with $\rho_\mu = \{B^*,\tilde B, W^*_3\}$.
For the $C_{\bar L L}$ coefficient, we find:
\begin{align}
C_{\bar L L}^{(B^*)} ={}&
\frac{g^*}{2}
\Big( \tan(\theta_2)^2 - ( \tan(\theta_2)^2 +1) X X^\dagger + ( \tan(\theta_2)^2 +1) X X^\dagger X X^\dagger 
\nonumber \\
&- \frac{1}{4} ( \tan(\theta_2)^2 +2) \frac{v^2}{\mt^2} X \yr \yr^\dagger X^\dagger
-  \frac{1}{2} ( \tan(\theta_2)^2 +1) \frac{v^2}{m \mt} X \yr \yl^\dagger X^\dagger + \hc \Big) \;, \\
C_{\bar L L}^{(\tilde B)} ={}& \frac{g^*}{2\cos\theta_2}
\left( \! X X^\dagger \!\left( 1 \!-\! X X^\dagger\right) \!+\! \frac{v^2}{\mt^2} X \yr \yr^\dagger X^\dagger \!+\!  
\left(  \frac{1}{2} \frac{v^2}{m \mt} X \yr \yl^\dagger X^\dagger \!+\! \hc \right)\! \right), \\
C_{\bar L L}^{(W^{*}_3)} ={}& -\frac{g^*}{2}\Big(
-\tan(\theta_1)^2 + ( \tan(\theta_1)^2 +1) X X^\dagger
		-  ( \tan(\theta_1)^2 +1) X X^\dagger X X^\dagger 
\nonumber \\
		&+ \!\!\! \frac{1}{4} \tan(\theta_1)^2 \frac{v^2}{\mt^2} X \yr \yr^\dagger X^\dagger
		+  \frac{1}{2} ( \tan(\theta_1)^2 +1) \frac{v^2}{m \mt} X \yr \yl^\dagger X^\dagger + \hc \Big)\;.
\end{align}
For the $C_{\bar R R}$ coefficient, we find:
\begin{align}
C_{\bar R R}^{(B^*)} ={}& g^*\Big( \tan(\theta_2)^2 -   ( \tan(\theta_2)^2 +1) \Xt \Xt^\dagger + ( \tan(\theta_2)^2 +1) \Xt \Xt^\dagger \Xt \Xt^\dagger
\nonumber \\
&-\!\!\! \frac{1}{8} ( 2\tan(\theta_2)^2 +1) \frac{v^2}{m^2} \Xt \yr^\dagger \yr \Xt^\dagger \!- \! \frac{1}{2} ( \tan(\theta_2)^2 +1) \frac{v^2}{m \mt} \Xt \yr^\dagger \yl \Xt^\dagger \!+ \! \hc \Big) ,\\
C_{\bar R R}^{(\tilde B)} ={}& \frac{g^*}{\cos\theta_2}\!
\left(\!
\Xt \Xt^\dagger \!-\! \Xt \Xt^\dagger \Xt \Xt^\dagger \!+\! \frac{1}{4}  \frac{v^2}{m^2} \Xt \yr^\dagger \yr \Xt^\dagger
\!+ \!\left( \frac{1}{2} \frac{v^2}{m \mt} X \yr \yl^\dagger X^\dagger \!+\! \hc \right) \!\!
\right)\!,\\
C^{(W^{*}_3)}_{\bar R R} ={}&  -\frac{g^*}{4} \frac{v^2}{m^2} \Xt \yr^\dagger \yr \Xt^\dagger .
\end{align}

\section{One loop contributions to the dipole operator}

In what follows, all lepton fields are meant to be in mass basis.
We introduce a compact notation to collectively denote such fields, using an upper case latin index to distinguish among them:

\begin{align}
(\mcE_L)_A &= \left\{ \ell_L , E_L , \tilde E_L \right\} \; ,
 &
(\mcE_R)_A &= \left\{ \tilde e_R , E_R , \tilde E_R \right\} \; ,
&
A &=1,2,3 \; ,
\\
(\mcN_L)_B &= \left\{ \nu_L , N_L \right\} \; , &
\mcN_R &= N_R \; ,
&
B &= 1,2 \; .
\end{align}

The flavor index, when necessary, will be explicited with a lower case latin letter. For most of the expressions however this index will be omitted to simplify the notation. 
In those cases one should interpret the expressions as $3{\rm x}3$ matrices in flavor space.

\begin{figure}[t]
\centering
\mbox{
\begin{minipage}{0.32\textwidth}
\centering
\includegraphics[height=0.45\textwidth]{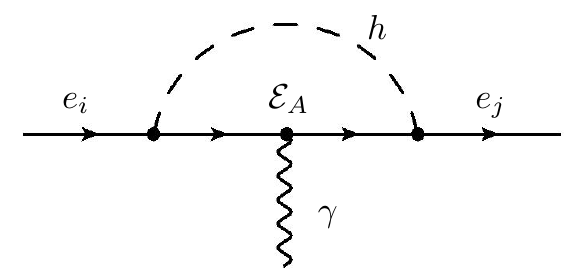}
\caption*{(1)}
\end{minipage}
\begin{minipage}{0.32\textwidth}
\centering
\includegraphics[height=0.45\textwidth]{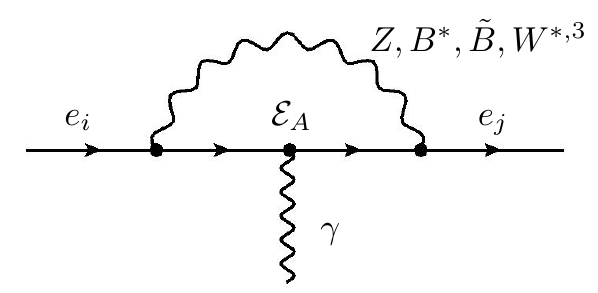}
\caption*{(2)}
\end{minipage}
\begin{minipage}{0.32\textwidth}
\centering
\includegraphics[height=0.45\textwidth]{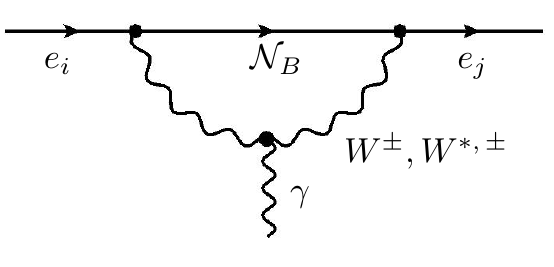}
\caption*{(3)}
\end{minipage}
}
\caption{{One-loop diagrams contributing to the dipole operator in our model. In diagrams (1) and (2) we have $A=1,2,3$, in diagram (3) we have $B=1,2$.}\label{fig: oneLoop diagrams}}
\end{figure}

\begin{figure}
\centering
\begin{minipage}{0.45\textwidth}
\centering
\includegraphics[height=0.32\textwidth]{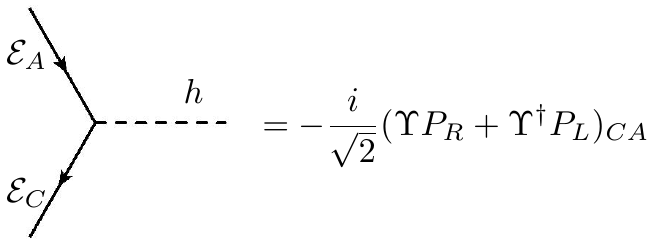}
\end{minipage} ,
\begin{minipage}{0.45\textwidth}
\centering
\includegraphics[height=0.32\textwidth]{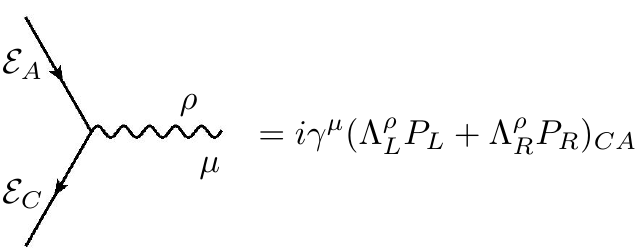}
\end{minipage} ,
%%%
\begin{minipage}{0.9\textwidth}  ~  \end{minipage}
%%%
\begin{minipage}{0.45\textwidth}
\centering
\includegraphics[height=0.32\textwidth]{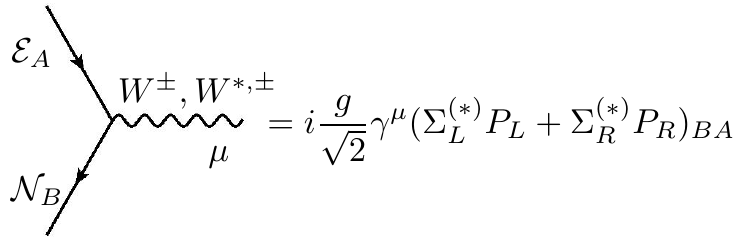}
\end{minipage} ,
\begin{minipage}{0.45\textwidth}
\centering
\includegraphics[height=0.32\textwidth]{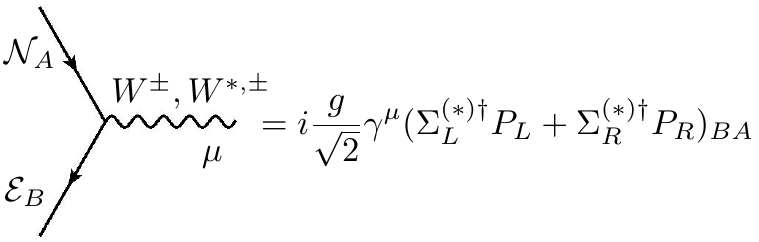}
\end{minipage} ,
%%%
\begin{minipage}{0.9\textwidth}  ~  \end{minipage}
%%%
\caption{{Feynman rules needed for the computations. Here $A,C=1,2,3$ and $B=1,2$, while $\rho=\{Z,B^*,\tilde B, W^{*,3}\}$. The expressions for 
$\Upsilon, \Lambda^\rho_{L,R},\Sigma^{(*)}_{L,R}$ can be found in the appendix.}\label{fig: FR for computations}}
\end{figure}

In figure \ref{fig: oneLoop diagrams} we see the different kinds of loop diagrams contributing to $C_{e\gamma}^{ij}$. From the Lagrangian (\ref{eq: CH lag tot}), after performing the rotations (\ref{rotations_mix_bos}), (\ref{eq: mass matrix leptons}) to the mass basis, one can read the Feynman rules of interest for these computations. We summarise them in figure (\ref{fig: FR for computations}), where the expressions for $\Upsilon, \Lambda^\rho_{L,R},\Sigma^{(*)}_{L,R}$ are:
\begin{align}
\Sigma_R & = \left[ 
	\begin{array}{c c c}
	0 & 0  \\
	0 & U_R^\dagger 
	\end{array} \right]  \left[ 
	\begin{array}{c c c}
	0 & 0 & 0 \\
	0 & 1 & 0
	\end{array} \right] V_R \; ,
	&
	\Sigma_L & = U^\dagger_L \left[ 
	\begin{array}{c c c}
	1 & 0 & 0 \\
	0 & 1 & 0
	\end{array} \right] V_L \; ,
\\
{\Sigma^*}_R & = \left[ 
	\begin{array}{c c c}
	0 & 0  \\
	0 & U_R^\dagger 
	\end{array} \right]  \left[ 
	\begin{array}{c c c}
	0 & 0 & 0 \\
	0 & \cot\theta_1 & 0
	\end{array} \right] V_R \; ,
	&
	{\Sigma^*}_L & = U^\dagger_L \left[ 
	\begin{array}{c c c}
	-\tan\theta_1 & 0 & 0 \\
	0 & \cot\theta_1 & 0
	\end{array} \right] V_L \; ,
	\\
\Lambda^Z_R &= 
	\frac{g}{c_W} s_W^2 
	- V^\dagger_R \left[ 
	\begin{array}{c c c}
	0 &  &  \\
	 & \frac{g}{2c_W} &  \\
	 &  & 0
	\end{array} \right] V_R  \; ,
	&
		\Lambda^Z_L &  = 
	\frac{g}{c_W}  \left( s_W^2 - \frac{1}{2} \right)
	+ V^\dagger_L \left[ 
	\begin{array}{c c c}
	0 &  &  \\
	& 0 &  \\
	 &  & \frac{g}{2c_W}
	\end{array} \right] V_L  \; ,
\\
    \Lambda^{B^*}_R &= g_*' V_R^\dagger \left[ 
    	\begin{array}{c c c}
    	\tan^2 \theta_2 &  &  \\
    	 & - \frac{1}{2} &  \\
    	 &  & -1
    	\end{array} \right] V_R \; ,
  &
  \Lambda^{B^*}_L &= g_*' V_L^\dagger \left[ 
    	\begin{array}{c c c}
    	\frac{1}{2} \tan^2 \theta_2 &  &  \\
    	 & - \frac{1}{2} &  \\
    	 &  & -1
    	\end{array} \right] V_L \; ,
  \\
      \Lambda^{\tilde B}_R &= g_*' \sec \theta_2 \;V_R^\dagger \left[ 
    	\begin{array}{c c c}
    	0 &  &  \\
    	 &  \frac{1}{2} &  \\
    	 &  & 1
    	\end{array} \right] V_R \; ,
    	&
    \Lambda^{\tilde B}_L &= g_*' \sec \theta_2 \, V_L^\dagger \left[ 
    	\begin{array}{c c c}
    	0 &  &  \\
    	 & \frac{1}{2} &  \\
    	 &  & 1
    	\end{array} \right] V_L \; ,  
  \\
    \Lambda^{W^{*,3}}_R &= \frac{g_*}{2} \, V_R^\dagger \left[ 
    	\begin{array}{c c c}
    	0 &  &  \\
    	 & -1 &  \\
    	 &  & 0
    	\end{array} \right] V_R \; ,
    	&
    	   \Lambda^{W^{*,3}}_L &= \frac{g_*}{2} \, V_L^\dagger \left[ 
    	\begin{array}{c c c}
    	 \tan^2 \theta_1 &  &  \\
    	 & -1 &  \\
    	 &  & 0
    	\end{array} \right] V_L \; ,
    	\\
    	\Upsilon &= V^\dagger_L \left[ 
	\begin{array}{c c c}
	0 & 0 & 0 \\
	0 & 0 & \yr \\
	0 & \yl^\dagger & 0
	\end{array} \right] V_R \; .
\end{align}
where $g_*=g \cot\theta_1$ and $g'_*=g'\cot\theta_2$.
Remember that this is a compact notation in which the flavor indices are omitted. Then each element of the above matrices is really a $3{\rm x}3$ matrix in flavor space (such as $\yr,\yl$).

We can now summarise the leading results for the different loop diagrams. In these expressions, terms of order $\frac{m_\ell^2}{v^2}$ or  $\frac{m_\ell^2}{\Lambda^2}$ (where $\Lambda$ 
is either an heavy lepton or boson mass) have been neglected. The final results are:
\begin{align} 
\frac{1}{\Lambda^2}(C_{e\gamma}^h)_{ij} = {} & -\frac{e}{32\sqrt{2} \pi^2 v} \bigg( \frac{1}{2} \Upsilon_{12} \frac{1}{m} \Upsilon_{21} 
					+ \frac{1}{2} \Upsilon_{13} \frac{1}{\tilde m} \Upsilon_{31}
					+ \frac{m^{SM}_i}{12} \Upsilon_{12} \frac{1}{m^2} \Upsilon_{21} ^\dagger
					+ \frac{m^{SM}_i}{12} \Upsilon_{13} \frac{1}{\tilde m^2} \Upsilon_{31}^\dagger 
	\nonumber \\
	& 
					+ \frac{m^{SM}_j}{12} \Upsilon_{12}^\dagger  \frac{1}{m^2}  \Upsilon_{21}
					+ \frac{m^{SM}_j}{12} \Upsilon_{13}^\dagger  \frac{1}{\tilde m^2}  \Upsilon_{31}
\bigg)_{ij}\; ,\label{eq:SR}
\\
\frac{1}{\Lambda^2}(C_{e\gamma}^Z)_{ij} = & - \frac{e}{16\sqrt{2}\pi^2 v} \frac{g^2}{ c_W^2} \frac{1}{M_Z^2} \times
	\nonumber \\*
	&
					\times \bigg[ 2 \Lambda^Z_{L \,11} m^{SM}  \Lambda^Z_{R \, 11}
					+ \frac{1}{2} \Lambda^Z_{L \,12} 
								\left( m + 4 \frac{M_Z^2}{m} \right)  \Lambda^Z_{R \, 21}
					+ \frac{1}{2} \Lambda^Z_{L \,13} 
								\left( \tilde m + 4 \frac{M_Z^2}{\tilde m} \right) \Lambda^Z_{R \, 31}
	\nonumber \\
	&
					- m^{SM}_i \left( \frac{2}{3} \Lambda^Z_{L \,11} \Lambda^Z_{L \, 11}
					+ \frac{5}{6} \Lambda^Z_{L \,12} 
								\left( \frac{1}{2}+  \frac{M_Z^2}{m^2} \right) \Lambda^Z_{L \, 21}
					+ \frac{5}{6} \Lambda^Z_{L \,13} 
								\left( \frac{1}{2}+  \frac{M_Z^2}{\tilde m^2} \right) \Lambda^Z_{L \, 31} \right)
	\nonumber \\*
	&
					- m^{SM}_j \left( \frac{2}{3} \Lambda^Z_{R \,11} \Lambda^Z_{R \, 11}
					+ \frac{5}{6} \Lambda^Z_{R \,12} 
								\left(  \frac{1}{2}+  \frac{M_Z^2}{m^2} \right) \Lambda^Z_{R \, 21}
					+ \frac{5}{6} \Lambda^Z_{R \,13} 
								\left( \frac{1}{2}+ \frac{M_Z^2}{\tilde m^2} \right) \Lambda^Z_{R \, 31} \right)
	\bigg]_{ij} \; ,
\\
\frac{1}{\Lambda^2}(C_{e\gamma}^W)_{ij} = & - \frac{e}{32\sqrt{2}\pi^2 v} \frac{g^2}{ M_W^2} \times
	\nonumber \\*
	&
					\times \bigg[ -2 \, \Sigma^\dagger_{L \,11} m^{SM}  \Sigma_{R \, 11}
					+ \frac{1}{2} \Sigma^\dagger_{L \,12} 
								\left( - m + 3\frac{M_W^2}{m} \left( 3+ 2 \log\frac{M_W^2}{m} \right)  \right)  \Sigma_{R \, 21}
	\nonumber \\
	&
					+ m^{SM}_i \left( \frac{5}{6} \Sigma^\dagger_{L \,11} \Sigma_{L \, 11}
					+ \Sigma^\dagger_{L \,12} 
								\left(  \frac{1}{3} - \frac{M_W^2}{m^2} \left(\frac{11}{4} + \frac{3}{2}\log\frac{M_W^2}{m} \right) \right) \Sigma_{L \, 21}\right)
	\nonumber \\*
	&
					+ m^{SM}_j \left( \frac{5}{6} \Sigma^\dagger_{R \,11} \Sigma_{R \, 11}
					+ \Sigma^\dagger_{R \,12} 
								\left(  \frac{1}{3} - \frac{M_W^2}{m^2} \left(\frac{11}{4} + \frac{3}{2}\log\frac{M_W^2}{m} \right) \right) \Sigma_{R \, 21} \right)
	\bigg]_{ij} \; ,
\\
\frac{1}{\Lambda^2}(C_{e\gamma}^{\rho'})_{ij} = {} & -\frac{1}{16 \sqrt{2} \pi^2 v} \frac{1}{M_{\rho'}^2} \bigg(
							2 \Lambda^{\rho'}_{L \,11} m^{SM} \Lambda^{\rho'}_{R \, 11}
							+ \Lambda^{\rho'}_{L \,12} m f^H_1(y) \Lambda^{\rho'}_{R \, 21}
					   		+ \Lambda^{\rho'}_{L \,13} \tilde m f^H_1(z) \Lambda^{\rho'}_{R \, 31}
	\nonumber \\*
	& 
			+ m^{SM}_i \left(
							- \frac{2}{3}  \Lambda^{\rho'}_{L \,11} \Lambda^{\rho'}_{L \, 11}
							+ \Lambda^{\rho'}_{L \,12} m f^H_2(y) \Lambda^{\rho'}_{L \, 21}
							+ \Lambda^{\rho'}_{L \,13} \tilde m f^H_2(z) \Lambda^{\rho'}_{L \, 31}
			\right)
	\nonumber \\*
	&
			+ m^{SM}_j \left(
							- \frac{2}{3}  \Lambda^{\rho'}_{L \,11} \Lambda^{\rho'}_{L \, 11}
							+ \Lambda^{\rho'}_{L \,12} m f^H_2(y) \Lambda^{\rho'}_{L \, 21}
							+ \Lambda^{\rho'}_{L \,13} \tilde m f^H_2(z) \Lambda^{\rho'}_{L \, 31}
			\right)
	\bigg)_{ij} \; ,\label{eq:SR1}
	\\
\frac{1}{\Lambda^2}(C_{e\gamma}^{W^*})_{ij} ={} & - \frac{1}{32 \sqrt{2} \pi^2 v}
						\frac{g_*^2}{M_H^2} 
					\bigg[ -2 \, \Sigma^{* \,\dagger}_{L \,11} m^{SM}  \Sigma^*_{R \, 11}
					+ \frac{1}{2} \Sigma^{* \,\dagger}_{L \,12} 
								m f^H_3(y)  \Sigma^*_{R \, 21}
	\nonumber \\
	&
					+ m^{SM}_i \left( \frac{5}{6} \Sigma^{* \,\dagger}_{L \,11} \Sigma^*_{L \, 11}
					+ \Sigma^{* \,\dagger}_{L \,12} 
								m f^H_4(y) \Sigma^*_{L \, 21}\right)
	\nonumber \\*
	&
					+ m^{SM}_j \left( \frac{5}{6} \Sigma^{* \,\dagger}_{R \,11} \Sigma^*_{R \, 11}
					+ \Sigma^{* \,\dagger}_{R \,12} 
								m f^H_4(y) \Sigma^*_{R \, 21} \right)
	\bigg]_{ij} \; ,\label{eq:SR2}
&
\end{align}
where $y=\frac{m^2}{M_{\rho'}^2}, z=\frac{\tilde m^2}{M_{\rho'}^2}$, while in eq.~(\ref{eq:SR1}) $\rho'=\{ B^*,\tilde B,W^{*,3} \}$ and the expression is the same for the three heavy bosons. Notice that the result is expressed in a block matrix notation. For example, taking eq.~(\ref{eq:SR}) and recalling that $\Upsilon$ is a $9 \textrm{x} 9$ matrix, $\Upsilon_{12}$ should be read as the $3 \textrm{x} 3$ block of $\Upsilon$ in position $(1,2)$. In this notation, the expression 
($\Upsilon_{12} \frac{1}{m} \Upsilon_{21})_{ij}$ stands for the matrix element: $\sum_{k=1}^3 (\Upsilon_{12})_{ik} (1/m_k)(\Upsilon_{12})_{kj}$.

The expressions for the loop functions used in eq.~(\ref{eq:SR1}), (\ref{eq:SR2}) are:

\begin{align}
f^H_1(x) & = \frac{4 - 3x - x^3 + 6x \log x}{(1-x)^3} \; , \\
f^H_2(x) & = \frac{-8 + 38 x - 39 x^2 + 14 x^3 - 5 x^4 + 18 x^2 \log x}{(1-x)^4} \; , \\
f^H_3(x) & = \frac{-4 + 15 x - 12 x^2 + x^3 + 6 x^2 \log x}{2(1-x)^3} \; , \\
f^H_4(x) & = \frac{10 - 43 x + 78 x^2 - 49 x^3 + 4 x^4 + 18 x^3 \log x}{12(1-x)^4} \; .
\end{align}
%

%%%%%%%%%%%%%%%%%%%%%%%%%%%%%%%%%%%%%%%%%%%%%%%%%%%%%%%%%%%%%%%%%%%%%%%%%%%%%%%
\section{Loop functions}
%%%%%%%%%%%%%%%%%%%%%%%%%%%%%%%%%%%%%%%%%%%%%%%%%%%%%%%%%%%%%%%%%%%%%%%%%%%%%%%

The loop functions for dipoles at the LO read:
\begin{align}
f^{B}_1(y,z) = {}& \frac{2 (y + z) (-4 + 3 y + y^3 - 6 y \log y)}{(1 - y)^3 (y - z)} + \frac{4 (z - 2 y) (-4 + 3 z + z^3 - 6 z \log z)}{(1 - z)^3 (y - z)} \; ,
\\
f^{B}_2(y,z) ={}& \frac{\sqrt{yz}}{(y-z)}\left(\frac{4 - 3 z - z^3 + 6 z \log z}{(1 - z)^3} - (y \leftrightarrow z) \right) \; ,
\\
f^{W^*}_1(y) ={}&  \frac{4 - 27 y + 24 y^2 - y^3 - 6 y (1 + 2 y) \log y}{(1-y)^3}  \; .
\end{align}
The loop functions for dipoles at the next-to-leading order read:
\begin{align}
f^h_1(x) ={}& \frac{4-7x}{3(1-x)} \\
f^Z_1(x) ={}& \frac{-8 + 16 s_W^2 (-1 + x) + 11 x}{3(1-x)} \\
f^Z_2(x) ={}& \frac{19 + 16 s_W^2 (-1 + x) - 16 x}{3(1-x)} \\
f^B_3(y,z) =&-\frac{ y^4 (79 + 58 z + 7 z^2) + 2 z (32 - 37 z + 23 z^2) + 
   3 y^2 (74 + 41 z + 14 z^2 + 3 z^3)}{3(1 -y)^3 (y - z) (1 - z)^2} 
\nonumber \\*  
&   +\frac{2 y (20 + 53 z + 5 z^2 + 12 z^3) + y^3 (279 + 85 z + 19 z^2 + 13 z^3)}{3(1 - y)^3 (y - z) (-1 + z)^2} 
\nonumber \\*  
 &		+\frac{6 y (5 y^3 + 2 y^2 (-2 + z) + 4 z^2 - y z (4 + 3 z))}{(1-y)^4(y-z)^2} \log y
	+\frac{12 z (4 y^2 - 3 y z + z^2)}{(1-z)^3(y-z)^2} \log z
 \\
f^B_4(y,z) =& 
			 -\frac{5 y^4 (1-z)^2 + 4 z^2 (1 + 7 z - 2 z^2) - 
   y^3 (-5 + 52 z + 19 z^2 + 6 z^3)}{2(1-y)^2 (y - z)^3 (1-z)^2}
\nonumber \\*
			& + \frac{-y z (24 + 95 z - 10 z^2 + 11 z^3) + 
  y^2 (20 + 62 z + 45 z^2 + 40 z^3 + z^4)}{2(1-y)^2 (y - z)^3 (1-z)^2}
\nonumber \\*
&		-\frac{3 y (5 y^2 + 6 y z - 3 z^2)}{(1-y)^3(y-z)^3} \log y
	+\frac{12 (3 y - z) z^2}{(1-z)^3(y-z)^3} \log z
 \\
f^B_5(y,z) ={}& 
			 -\frac{4 z (31 - 33 z + 36 z^2 - 16 z^3) + 
 y^3 -157 + 15 z + 45 z^2 + 25 z^3}{3 (1 - y)^2 (y - z) (1 - z)^3}
\nonumber \\*
			& - \frac{y^2 (323 + 58 z - 60 z^2 - 86 z^3 - 19 z^4) + 
 y (-148 - 233 z + 75 z^2 + 97 z^3 - 7 z^4)}{3 (1 - y)^2 (y - z) (1 - z)^3}
\nonumber \\*
&		-\frac{6 (y^3 - 5 y z^2)}{(1-y)^3(y-z)^2} \log y
	-\frac{24 z (y (7 z- 5 z^2) + (-2 + z) z^2 + y^2 (-4 + 3 z))}{(1-z)^4(y-z)^2} \log z
& \\
f^{W^*}_2(y,z) =& - \frac{3 (4 - 36 y + 55 y^2 - 24 y^3 + y^4 + 2 y (-4 - 5 y + 6 y^2) \log y)}{2(1 - y)^4}  \\
f^{W^*}_3(y,z) =& \frac{ -4 + 27 y - 24 y^2 + y^3 + 6 y (1 + 2 y) \log y}{2(1 - y)^3}  \\
f^{W^*}_4(y,z) ={}& \frac{4 + y^2 (-1 + z)^2 + 10 z - 2 z^2 + y (1 - 8 z - 5 z^2)}{4 (1 - y)^2 (y - z) (1 - z)^2}  
\nonumber \\*
&
				 +\frac{ 3 y (y + z)}{2 (1 - y)^2 (y - z)^2 } \log(y)
				- \frac{ 3 z^2}{ (y - z)^2 (1 - z)^3}  \log(z)
\end{align}
%
%%%%%%%%%%%%%%%%%%%%%%%%%%%%%%%%%%%%%%%%%%%%%%%%%%%%%%%%%%%%%%%%%%%%%%%%%%%%%%%%%%%%%%%%%%%%%

%%%%%%%%%%%%%%%%%%%%%%%%%%%%%%%%%%%%%%%%%%%%%%%%%%%%%%%%%%%%%%%

\end{document}